\documentclass[11pt]{article}

\usepackage[a4paper, margin=2.5cm]{geometry}
\usepackage{setspace}
\usepackage{graphicx}
\usepackage{dcolumn}
\usepackage{bm}
\usepackage[version=4]{mhchem}
\usepackage{braket}
\usepackage{nicefrac}
\usepackage[dvipsnames,svgnames]{xcolor}
\usepackage{amsmath}
\usepackage{amssymb}
\usepackage{booktabs}
\usepackage{pifont}
\usepackage[colorlinks=true,linkcolor=blue,citecolor=blue,urlcolor=blue]{hyperref}

\usepackage[super,sort&compress,comma]{natbib}

\usepackage{pdfpages}

\begin{document}

\title{\textbf{Kohn-Sham density encoding rescues coupled cluster theory for strongly correlated molecules}}

\author{
Abdulrahman Y. Zamani$^{1}$, Barbaro Zulueta$^{2}$, Andrew M. Ricciuti$^{1}$,\\ John A. Keith$^{2,*}$, Kevin Carter-Fenk$^{1,*}$
}

\date{}

\maketitle

\noindent $^{1}$Department of Chemistry, University of Pittsburgh, Pittsburgh, Pennsylvania 15260, USA\\
$^{2}$Department of Chemical and Petroleum Engineering, University of Pittsburgh, Pittsburgh, Pennsylvania, 15213, USA\\[0.5em]
$^{*}$Corresponding authors: jakeith@pitt.edu (J.A.K.); kay.carter-fenk@pitt.edu (K.C.F.)

\vspace{1em}

\begin{abstract}
\noindent 
Coupled cluster theory with a Kohn-Sham reference (KS-CC) can dramatically outperform its Hartree-Fock counterpart for strongly correlated systems, but the origin of these improvements has remained unclear. Here we demonstrate that these improvements arise from differences in the one-particle density matrix that are encoded into the non-canonical Fock matrix and not from the nature of the KS orbitals, as is commonly assumed. Equipped with this insight, KS-CCSD(T) can be leveraged to achieve near-chemical-accuracy for electronic and thermochemical properties of transition-metal dimers and main-group compounds. Most strikingly, KS-CCSD(T) qualitatively recovers the entire \ce{Cr2} potential energy surface, a notorious failure case for HF-CCSD(T) and single-reference density functional theory. We further introduce a density difference diagnostic that identifies multireference character and guides practitioners toward rational selections of optimal references at mean-field cost. These results establish KS-CCSD(T) as a practical route to treat strong correlation within the ``gold standard'' framework, and this has immediate implications for machine learning potential development and materials research, areas that heavily rely on KS-DFT for model-parameter fitting.
\end{abstract}

\newpage

\section*{Introduction}

Chemical catalysis by transition-metal compounds (TMCs) underpins renewable energy technologies,\cite{Johnson2023,Zhang2024,Guo2024,Li2024,Pei2024} yet predicting their thermochemistry and kinetics from first principles remains challenging. Heightened interest in replacing precious metals with Earth-abundant alternatives\cite{Bullock2020} amplifies this challenge since the same first-row transition metals prized for sustainability (e.g. iron, cobalt, nickel, copper, manganese) often exhibit the multiconfigurational electronic structures that confound standard quantum-chemical methods. Coupled cluster (CC) theory with singles, doubles, and perturbative triples [CCSD(T)]\cite{Raghavachari1989,Stanton1997} achieves ``gold standard" accuracy for main-group thermochemistry, but its reliability for TMCs can be compromised by strong correlation arising from the d-orbital manifold.\cite{Xu2015,Cheng2017,Vogiatzis2018} This limitation has profound consequences: CCSD(T) reference data train machine learning potentials\cite{Keith2021} and parameterize semi-empirical methods\cite{Thiel2014,Bannwarth2021} that enable exascale simulations,\cite{Smith2019,Unke2021,Kulichenko2024} yet systematic errors for TMCs propagate through this entire computational ecosystem.

Replacing the Hartree-Fock (HF) reference with one from Kohn-Sham density functional theory\cite{Hohenberg1964,Kohn1965} (KS-DFT) offers a tantalizing solution. KS-CC methods have yielded accurate results for specific transition-metal thermochemistry,\cite{Fang2016,Fang2017} spin-state energetics,\cite{Radon2019,Mallick2021,Drosou2022,Drabik2024} barrier heights, and reaction kinetics.\cite{Rettig2020,Bertels2021} However, improvements over HF-CC are maddeningly inconsistent: comprehensive assessments reveal no systematic advantage, and the choice of density functional has minimal impact when improvements do occur.\cite{Benedek2022} The prevailing explanation has been that KS orbitals resemble Brueckner orbitals\cite{Brueckner1954a,Nesbet1958,Stolarczyk1984} and therefore accelerate the convergence of truncated CC to the full configuration interaction limit, but this explanation fails to account for the fact that the KS orbitals are not actually the orbitals used in a KS-CC calculation. Orbital-based arguments are therefore inconsistent with the reality of practical KS-CC calculations and thus fail to account for these erratic outcomes.\cite{Jankowski2004,Wasilewski2009}

The present work resolves this puzzle. We demonstrate that improvements from KS-CC arise not from the orbitals themselves, but actually from differences in the one-particle density matrix that are encoded into the non-canonical Fock operator.\cite{Purvis1982} In practical KS-CC calculations, the KS density is iterated to self-consistency, then the Hartree-Fock potential is used to build the one-electron Fock matrix and the orbitals are semi-canonicalized to ensure that their energies are meaningful.
This procedure has the critical consequence of transforming the KS orbitals toward the HF result while preserving the density, meaning the orbitals that enter CC calculations bear little resemblance to the converged KS solution. Thus, the KS density matrix itself acts as the operative quantity by etching information about electron correlation into the Fock matrix. This Fock matrix is then used to compute the CC wavefunction and, under the right conditions, can drive the CC equations toward a more accurate solution.

With this insight, we systematically assess KS-CCSD(T) for first-row transition-metal diatomics (M–O, M–Cl, M–H$^+$, M–H, and M–M; where M = Sc–Zn)\cite{Wiedner2016,Moltved2018,Moltved2019a} and main-group species, benchmarking against experiment and higher-level wavefunction methods across Rungs 1–4 of Jacob's ladder.\cite{Perdew2001} We achieve substantial improvements for systems with purported multireference character.\cite{Jiang2012,Bao2017a,Bao2017b} Most notably, KS-CC qualitatively recovers the entire \ce{Cr2} potential energy surface, a notorious failure case normally assumed to require multireference methods since methods such as HF-CCSD(T) and hybrid DFT functionals generally perform less well than their lower-rung counterparts.\cite{Perdew1996,Perdew1997} Finally, we propose a density difference metric that introduces a new multireference diagnostic at mean-field cost and provides a rational basis for selecting optimal reference densities in KS-CC calculations.
 
\section*{Results}

\subsection*{Preconditioning coupled cluster with Kohn-Sham densities}

To identify the physical origins of differences in CC results when beginning with a KS solution, we first examine the effect of the semi-canonicalization\cite{Handy1989,Raghavachari1990,Lauderdale1991} (SC) of initial reference single-particle quantities that are used in post-SCF methods. SC involves diagonalizing the occupied-occupied and virtual-virtual blocks of ${\widetilde{F}}$, and then rotating the molecular orbitals to provide a set of single-particle energies. This transformation does not generally result in zero-entries for occupied-virtual elements, ${\widetilde{F}}_\mathrm{ov}$.

Generalized CC implementations can accommodate arbitrary single reference determinants, but when the Fock operator is not strictly diagonal, obtaining physically meaningful orbital energies and MOs requires an SC step. This is especially important for KS-CC calculations, where the physicality of KS eigenvalues remains ambiguous at best\cite{Stowasser1999,Perdew1985} unless recast into a correlated orbital theory.\cite{Bartlett2009,Kim2025,AraujoMendes2025,Shigeta2005,Bartlett2005a,Bartlett2005b,Grabowski2002,Grabowski2007} When using KS references in correlated wave function calculations, the ${\widetilde{F}}_\mathrm{ov}$ elements carry essential information about the KS density, appearing as non-Brillouin singles (NBS) corrections in many-body perturbation theory (MBPT) and as singles clusters within the CC amplitude equations.\cite{Bartlett1995} 

\begin{figure}[htbp]
\centering
\includegraphics{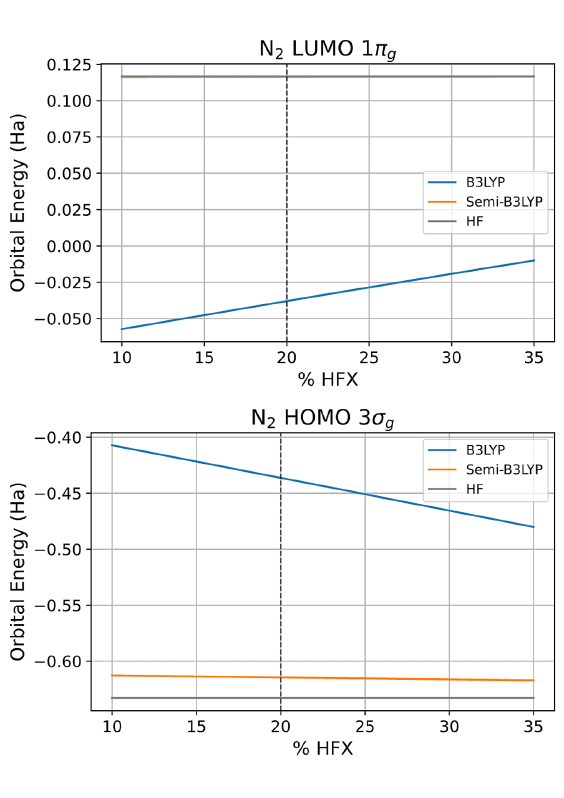}
\caption{\textbf{MO eigenvalue changes with increasing \%HFX for HF, B3LYP, and B3LYP after semi-canonicalization.} The percent exchange of the DFA is adjusted in proportion to the exact exchange. The vertical dashed lines represent the standard  \%HFX in the DFA. Highest occupied (HOMO) and lowest unoccupied (LUMO) MO energies are obtained with the cc-pVTZ and aug-cc-pVTZ basis sets respectively.}  
     \label{Fig:n2-b3lyp} 
\end{figure}

We illustrate this point by computing MO energies for \ce{N2} with HF and density functional approximations (DFAs) that contain different amounts of exact exchange (HFX). Figure \ref{Fig:n2-b3lyp} shows that, upon SC of the KS orbitals, their energies will closely resemble the HF eigenvalues. This effect remains stable with respect to increases in \%HFX.  Conversely, canonical occupied KS eigenvalues are vastly different from those of HF with the differences between the two vanishing as more HFX is included in the DFA. The SC effect appears even stronger on virtual MOs; the semi-canonical ``KS" virtual MO energies align with their HF counterparts and are stable against increasing \%HFX. Larger molecules exhibit similar behavior when semi-empirical functionals are employed (see Supplementary Table 8).

Modifying the single-particle quantities significantly affects the energy denominator magnitudes in CC and M{\o}ller-Plesset (MP) theory amplitude expressions (see Supplementary Figs. 46 and 47), including those with NBS contributions that would otherwise vanish for a canonical HF reference. Since orbital energies after SC ($\widetilde{\epsilon_{p}}$) become similar (but not necessarily identical) to their canonical HF values under the action of the same Fock operator, their eigenvectors must be similar to the canonical HF MOs. The eigenvectors after SC are also involved in the transformation of the two-electron integrals into the semi-canonical basis: $\widetilde{\langle pq|| rs \rangle}$.\cite{RodrguezMayorga2021} This suggests that   \textit{neither} the initial KS MOs \textit{nor} the KS MO energies enter correlated interaction terms of post-SCF methods. Instead, the SC procedure exactly preserves the KS density matrix, which is invariant under unitary transformations, and encodes this density information into the Fock matrix elements that enter the MBPT/NBS and CC equations.

\subsection*{Response properties using KS references}

Previous work leveraged SC KS references with Green's function (GF) methods to obtain correlated orbital energies.\cite{Shigeta2001} More recently, the renormalized singles (RS) approach for one- and two-shot $GW$ methods adopted this step.\cite{Jin2019,li2021,Li2022} These methods incorporate essential NBS terms and yield accurate ionization potentials (IPs), electron affinities (EAs), and dipole moments. Benchmark studies show that RS $GW$ methods with GGA or hybrid functional reference densities outperform their HF-based counterparts.

Here, we observe how KS densities influence diagonal 2nd and 3rd order self-energy approximations that are derived from electron propagator theory (EPT) for IPs and EAs (Supplementary Fig. 7 and Table 7). The SC eigenvalues obtained under the HF potential $\widetilde{\epsilon_p}$ either closely restore\cite{Bartlett2006} or slightly improve Koopmans' theorem for IPs. However, MAEs of self-energy corrected orbital energies from KS-EPT deviate from HF-EPT by up to $\sim$0.17~eV. An exception arises for EAs computed with EPT@B3LYP for BeO, $\mathrm{C}_{2}$, and BN---molecules noted to contain appreciable MR character. Clearly, low-order perturbation theory does not fully alleviate the dependence of EPT methods on the initial single-determinant reference. Thus, methods that optimize or renormalize the reference state,\cite{Ortiz1991,Rishi2015} such as CC, can be beneficial for non-self-consistent EPT approaches.\cite{Schilfgaarde2006,Rostgaard2010}

Finally, we test reference-state dependence for response properties by computing single excitation energies with equation-of-motion CC (EOM-CCSD) on the QUEST \#1 dataset (see Supplementary Figure 6).\cite{Loos2018,veril2021,loos2025}  These molecules are presumed to be dominated by the initial (i.e.\ single reference) ground-state configuration. Here, semi-canonicalized KS references closely approach or reproduce HF-based results, and this indicates that CC solutions are qualitatively the same.

\subsection*{Bond dissociation energies}

\begin{figure}[htbp]
\centering
\includegraphics[width=0.9\linewidth]{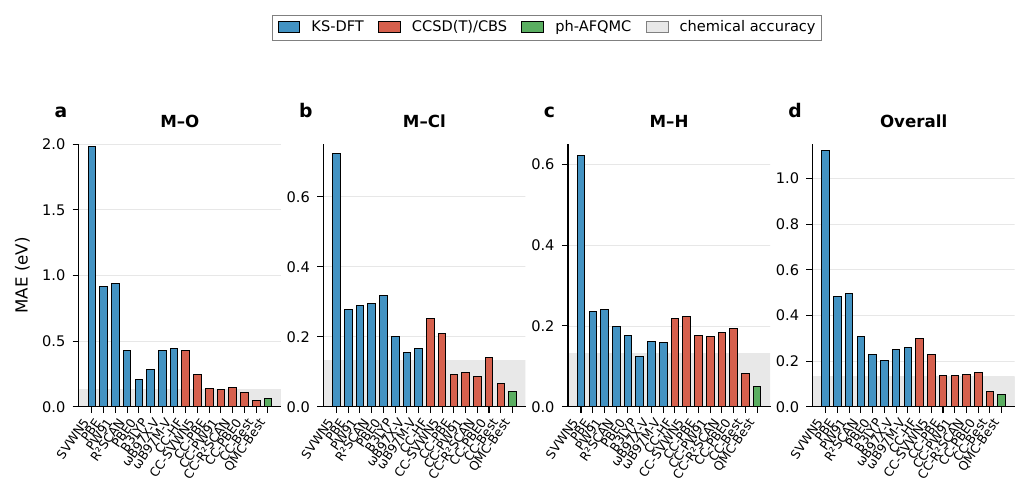}
\caption{\label{fig:dimer_fig1}  \textbf{Benchmark comparison of density functional and coupled cluster methods for transition metal–ligand bond dissociation energies.} Mean absolute errors (MAE) in bond dissociation energies for first-row transition metal compounds across four bond classes: a, metal–oxygen (\ce{M-O}); b, metal–chlorine (\ce{M-Cl}); c, metal–hydrogen (\ce{M-H}); and d, over all bonds. Blue bars represent density functional methods (KS-DFT), orange bars represent CCSD(T)/CBS calculations using different reference determinants (CC-X denotes CCSD(T)/CBS with densities from method X), and the green bar represents phaseless auxiliary-field quantum Monte Carlo (ph-AFQMC) results from Ref.\citenum{Shee2019}. CC-Best denotes the optimal reference selection either from HF or DFT for each system, and QMC-Best denotes the optimal reference when using MP2, CC, or TZ/QZ extrapolation in ph-AFQMC. The subtle gray shading indicates the chemical accuracy threshold (0--3 kcal/mol $\approx$ $0$--$0.13$ eV).}
\end{figure}

\begin{figure}[htbp]
\centering
\includegraphics{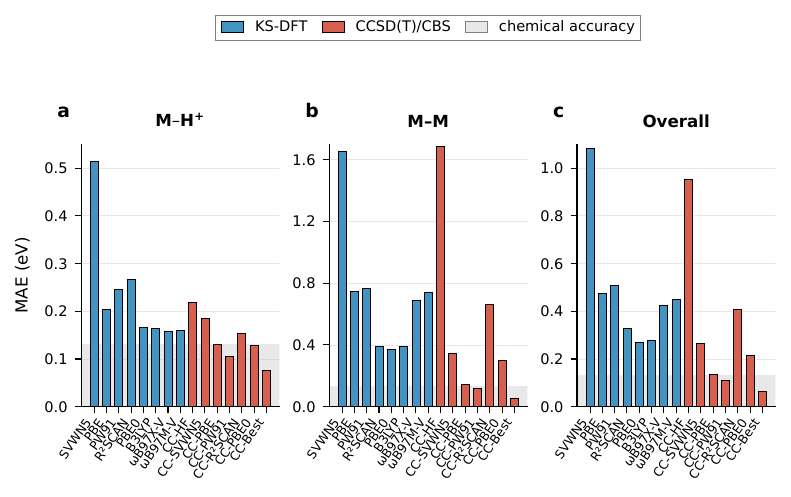}
\caption{\label{fig:dimer_fig2}  \textbf{Performance of density functional and coupled cluster methods for metal hydride cations and bimetallic bond dissociation energies.} Mean absolute errors (MAE) in bond dissociation energies for: a, metal hydride cations (\ce{M-H+}); b, homonuclear metal–metal dimers (\ce{M-M}); and c, overall performance. Blue bars represent density functional methods (KS-DFT) and orange bars represent CCSD(T)/CBS with different reference determinants. CC-Best denotes the optimal reference selection (HF or DFT) for each system. The subtle gray shading indicates the chemical accuracy threshold (0--3 kcal/mol $\approx$ $0$--$0.13$ eV).}
\end{figure}

Having established how SC transforms single-particle quantities, we now assess whether KS-CC improves \textit{total energy differences} for strongly correlated metal systems.\cite{Jiang2015,Xu2015,Aoto2017,Hait2019,Shee2019,Shee2021} We benchmark KS-DFT, HF-CCSD(T), and KS-CCSD(T) for BDEs across transition-metal diatomics. Figure \ref{fig:dimer_fig1} compares error measures against ph-AFQMC for \ce{M-H}, \ce{M-O}, and \ce{M-Cl} species. CCSD(T) with optimally selected DFA or HF references achieves ph-AFQMC accuracy across all bond types (MAE = 0.07~eV vs.\ 0.05~eV for ph-AFQMC). KS-CCSD(T) outperforms standalone KS-DFT for these BDEs; in particular, $\omega$B97X-V, $\omega$B97M-V, and B3LYP references yield KS-CCSD(T) results that match or slightly surpass HF-CC accuracy. For \ce{M-H+} and \ce{M-M} species (Figure~\ref{fig:dimer_fig2}), the differences widen. Conventional CCSD(T)/CBS with HF density fails dramatically for metal--metal bonds (MAE = 0.95 eV) and performs no better than the local (spin) density approximation (LDA) functional. However, CCSD(T)/CBS with GGA references (PBE, PW91) approaches chemical accuracy (MAE $\approx$ 0.11--0.14~eV).\cite{DeYonker2007,Neugebauer2023}

Accurately modeling reactivity and thermochemistry requires accurate potential energy curves (PECs). We therefore turn to the infamous \ce{Cr2}\cite{Larsson2022,Feng2025} as a stringent test for our approach. Previous work applied KS-DFT and HF-CC independently to \ce{Cr2}, finding that certain functionals yield qualitatively reliable PECs and accurate $r_{eq}$.\cite{Goodgame1982,Delley1983,Salahub1985,Edgecombe1995,Johnson2017} Such success stories have also been born out for other qualitative descriptions of transition-metal bonding, albeit strategies to guide the optimal choice of DFA remain unclear.\cite{Schultz2005,RuizDiaz2010,Hongo2012,Siegbahn2010} Curiously, Bauschlicher \& Partridge\cite{Bauschlicher1994} showed that at $r_{eq}$, UCCSD(T) underbinds, RCCSD(T) overbinds, and UBLYP provides the best comparison to experiment.

\begin{figure}[htbp]
\centering
\includegraphics{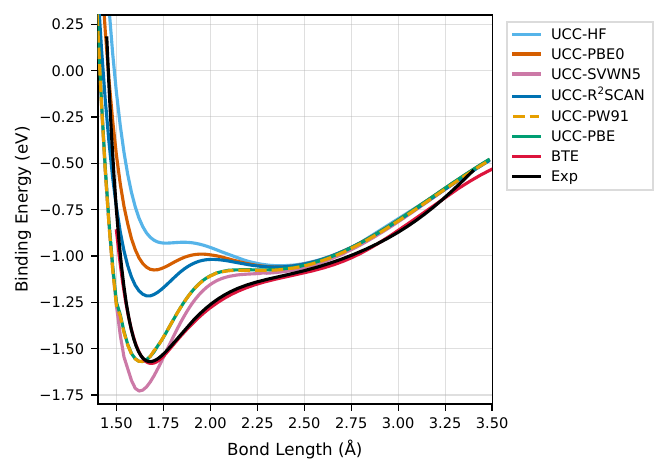}
\caption{\label{fig:Cr2_dimer}  \textbf{Potential energy curve of the antiferromagnetic state of \ce{Cr2} dimer computed with UCCSD(T)/CBS using different reference densities.} Binding energy as a function of bond length for \ce{Cr2}, the prototypical multireference transition metal dimer. Plot depicts UCCSD(T)/CBS calculations with reference orbitals from Hartree–Fock (UCC-HF), SVWN5 (UCC-SVWN5), PBE (UCC-PBE), PW91 (UCC-PW91), R²SCAN (UCC-R²SCAN), and PBE0 (UCC-PBE0). The experimental curve (Exp, black) and the best theoretical estimate from state-of-the-art multireference calculations (BTE, crimson) from Ref.\citenum{Larsson2022} provide comparison.}
\end{figure}

Figure~\ref{fig:Cr2_dimer} shows PECs from KS-UCCSD(T) with spin-symmetry constraints removed to describe the antiferromagnetic state alongside a best theoretical estimate (BTE) and experimental data. Again, conventional UCCSD(T)@HF/CBS produces a qualitatively incorrect potential that has a shallow minimum shifted to longer bond lengths. 

Remarkably, KS-UCCSD(T)/CBS with PW91 or PBE qualitatively reproduces the PEC topography of both experiment and high-level MR theory, particularly in the characteristic shelf region. We find that KS-UCCSD(T) with these DFAs attains greater accuracy than HF-UCCSD(T) and even rivals the accuracy of sophisticated MR methods\cite{Larsson2022,Celani2004,Vancoillie2016,Muller2009} at much lower cost.  Most strikingly, both DFAs deliver chemically accurate BDEs consistent with experiment.\cite{Kant1966,Hilpert1987,Simard1998} Thus, the decades-long failure of CCSD(T) for \ce{Cr2} shown in Refs~\citenum{Larsson2022}, \citenum{Bauschlicher1994}, and \citenum{Purwanto2015} is not a fundamental limitation of single-reference CC, as it can largely be remedied by simply employing alternatives to the canonical HF reference density. PBE and PW91 references reduce the UCCSD(T)/CBS errors from 0.52~eV to 0.001~eV for $D_e$ and from 0.70~\AA\ to 0.04~\AA\ for $r_{eq}$ (Supplementary Table 6).

\subsection*{Spin-state energetics}

Functional choice and exact exchange content significantly impact KS-DFT spin-state energetics.\cite{Harvey2004,Harvey2006,Ghosh2006,Siegbahn2000,Swart2016,Radon2023} Systematically improvable methods such as CC should provide a more consistent route to accurate spin-dependent properties. We therefore examine singlet-triplet gaps (STGs) for two classes of species: neutral transition-metal diatomics (denoted as the TinySpins25 set) and main-group molecules with established multireference character. Supplementary Sections 14 and 15 provide error measures, Frobenius norms of $\widetilde{F}_{ov}$, $T_1$ diagnostics, $\langle\mathcal{S}^{2}\rangle$ values, and additional qualitative analysis.

By comparing STGs computed with CCSD(T)/def2-QZVPPD against CBS HF-CCSDT(Q)$_{\Lambda}$ estimates, we find that hybrid functionals, notably SCAN0, closely reproduce HF-CC results (Supplementary Figure 15). LDA and (meta-)GGA references perform less well but remain within 0.25~kcal/mol of the HF-CCSD(T) MAE. For the TinySpins25 set, KS densities bring no added benefit. Average $T_{1}^\mathrm{diag}$ values for singlet and triplet states exceed the heuristic threshold\cite{Lee1989} ($\sim$0.02), suggesting MR character. However, standard amplitude-based diagnostics may inadequately categorize molecules with pronounced nondynamical correlation; alternative thresholds (e.g.\ $T_{1}^\mathrm{diag} > 0.05$) have been proposed for inorganic systems.\cite{Jiang2012} Consistent with previous analyses,\cite{Harvey2003,Fang2016,Bartlett2020,Benedek2022} KS references suppress singlet $T_{1}^\mathrm{diag}$ values below HF-CC thresholds. For triplets, only hybrid DFAs produce this suppression. We return to these observations in the section on classifying MR character.

\begin{figure}[htbp]
\centering
\includegraphics{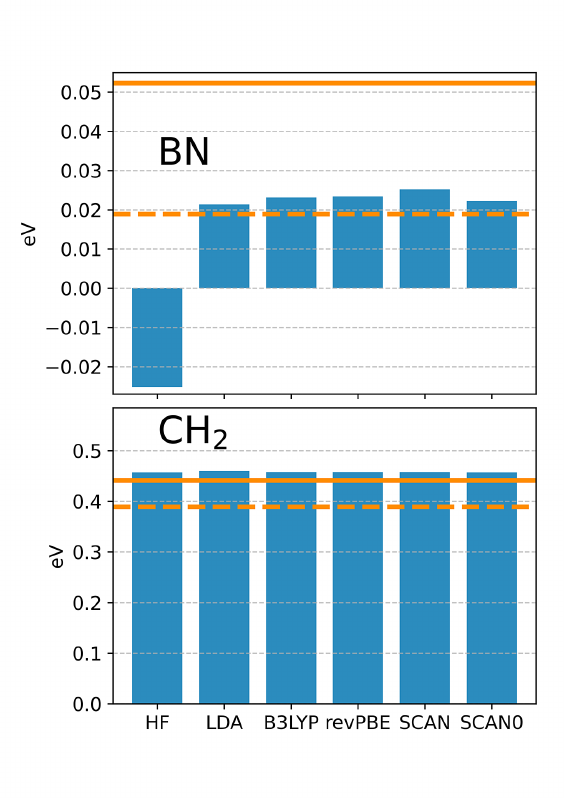} 
\caption{\textbf{Singlet-triplet gaps (STGs) for BN and {$\mathbf{{CH}_{2}}$} computed with CCSD(T)/cc-pVTZ with HF and KS-DFT determinants.} The solid and dashed horizontal orange line represent the reference theoretical estimate and experimental value, respectively.}  
\label{Fig:examples} 
\end{figure}

Next, we assess KS-CC performance on STGs using main-group molecules that higher-level theories predict to have MR ground states. Specifically, we consider an isoelectronic series (BN, $\mathrm{C}_2$, $\mathrm{BO}^{+}$, $\mathrm{CN}^{+}$) and a diradical series ($\mathrm{CH}_2$, $\mathrm{O}_{2}$, NF, NH). Figure \ref{Fig:examples} compares STGs for two prototypical MR species against experiment and high-level theoretical estimates and Supplementary Figure 25 presents the remaining species. BN and $\mathrm{CN}^{+}$ show clear improvements over HF-CC with various DFA references. Notably, HF-CC predicts a negative STG for BN, which implies an incorrect ground-state assignment, while KS-CC correctly identifies the singlet ground state, consistent with Li and Paldus.\cite{Li2006} However, $\mathrm{CH}_2$, whose closed-shell singlet is partially MR,\cite{Perera2014,Slipchenko2002} shows no improvement with KS-CC. These examples illustrate both the promise and the limits of KS-CC for states with pronounced MR character.

\subsection*{Classifying Multireference Character}

KS references can treat some strong correlation effects\cite{Cremer2002,Boyn2022,Gao2016} that arise from MR character, but $T_{1}^\mathrm{diag}$ alone cannot predict whether KS-CC will improve results (Ref.~\citenum{Benedek2022}, see Supplementary Fig. 48). Many other MR diagnostics exist,\cite{Coe2015,Chan2024,Bartlett2020,Duan2020,RamosCordoba2017,Xu2025,Ganyecz2025,Weflen2025} but none reliably predicts KS-CC performance. Comparing KS-CC densities to near-FCI densities via distance or entropic metrics could gauge reference quality, but generating CI-quality densities incurs substantial computational cost. Moreover, Benedek et al.\cite{Benedek2022} showed that KS-CC density errors relative to approximate FCI are more inconsistent than HF-CC errors. Overall, most metrics require the calculation of a wavefunction (CC or multi-configurational) which is suboptimal for the user that is interested in screening DFAs for their potential to improve CC calculations. It would therefore be highly desirable to define a proactive metric that can predict MR character before a computationally demanding wavefunction calculation is run.

\begin{figure}[htbp]
\centering
\includegraphics{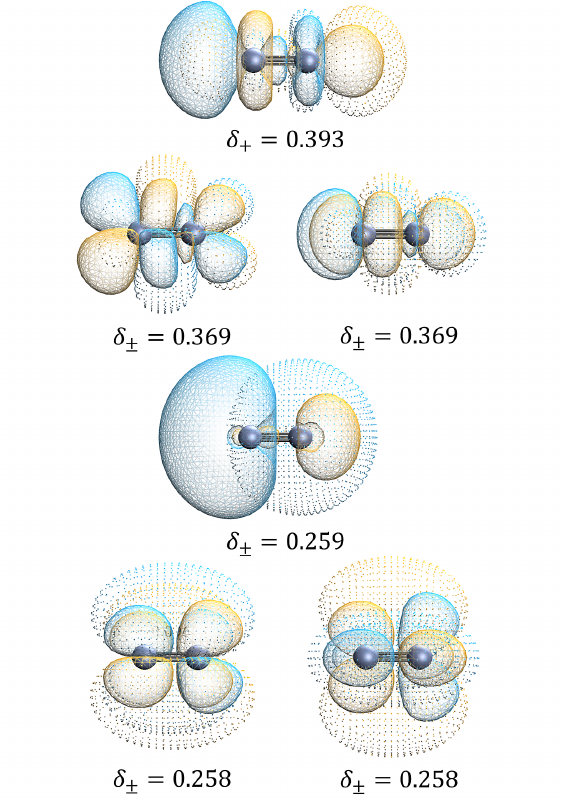} 
\caption{\textbf{Difference density natural orbitals for singlet $\mathrm{Cr}_{2}$ computed between HF and PW91 with def2-QZVPP.} The six largest electron displacement eigenpairs $\delta_{\pm}$ are reported. The orbitals corresponding to $\delta_{-}$ and $\delta_{+}$ are represented by radial dots and line meshes, respectively. An isosurface value of 0.02 is used except for the fourth orbital from the top, where a value of 0.04 was used. The NNED and $||\Delta_P||_{F}$ values are 0.083 and 2.240, respectively.}  
\label{Fig:DDNO-Cr2} 
\end{figure}

Our results imply that the KS density matrix used to construct $\widetilde{F}$ must differ sufficiently from its HF counterpart for KS-CC to yield a different solution than HF-CC. This raises a practical question: can we determine \textit{a priori} at the SCF level whether a given DFA will produce significantly different CC results? Drawing on recent interpretive tools for $\Delta$SCF methods, we develop a difference-density analysis based on the natural orbital decomposition of $\Delta_P$ to quantify the charge density shift between HF and KS-DFT (see Methods). From this we define the normalized number of electrons displaced (NNED) between SCF solutions. Figure \ref{Fig:DDNO-Cr2} illustrates this analysis for \ce{Cr2}, supplemented by Frobenius norms $||\Delta_P||_F$. Supplementary Fig. 14 presents main-group species. The NNED serves as an economical proxy for $T_{1}^\mathrm{diag}$, capturing the magnitude of reference density changes. Applying NNED to assess MR character, we obtain classification thresholds proportionate to HF-CC $T_{1}^\mathrm{diag}$ limits. Supplementary Section 12 details the classification protocols. When NNED approaches or exceeds a $T_1$-equivalent threshold, this suggests an MR treatment may be needed, and so testing alternative DFAs for improved CC results is warranted.

Applying $T_1$-equivalent NNED thresholds to metal diatomics and main-group species reveals that strong MR character can diminish under KS-CC (Table~\ref{table:mr-character}). This indicates that strong correlations in MR species, including TMCs, can be captured by an alternative single-reference density as long as it differs qualitatively from the canonical HF solution.\cite{Rishi2015} While no simple metric can guarantee improved thermochemical or electronic properties, the density deformation between HF and KS-DFT provides a useful descriptor for identifying functionals that may enhance single-reference CC methods for MR systems. However, when a given DFA yields a much more single-reference picture at the CC level (measured by a sufficiently large NNED), it is safe to presume that the changes imparted by the KS density may beget improvements over HF-CC, as the KS density will be a more amenable starting point to converge the single-reference CC equations. At the very least, the choice of KS-CC will not have a deleterious effect on the results if this is indeed the case.

\section*{Discussion}

This study clarifies how KS-DFT references can improve correlated many-body calculations relative to HF references. We demonstrate significant BDE improvements over HF-CC for a variety of TMCs with notable MR character. In particular, employing KS-DFT references with UCCSD(T) alone drastically improves the \ce{Cr2} ground-state PES topography. We also show improved spin-state energetics for \textit{bona fide} MR main-group species. The concept of MR character is sometimes nebulous\cite{Bartlett2020} because this facet of strong correlation diminishes with more optimal reference densities.

KS-DFT incorporates some correlation and resists symmetry breaking\cite{Sherrill1999,Perdew2003,Perdew2021,Pople1995,Sonnenberg2005,Schattenberg2018,Baker1993} more than HF, so we posit that enhancements from KS-CC may stem from improved descriptions of electronic pair densities.\cite{Perdew2021} Crucially, the on-top pair density relates directly to the total electronic energy. In spin-polarized KS-DFT, broken-symmetry LSDA and GGA solutions yield realistic on-top pair densities and total energies.\cite{Perdew1995,Perdew1997b} For antiferromagnetic systems like \ce{Cr2}, LSDA and GGA energies remain viable at the SCF level despite the symmetry dilemma from lifting variational constraints on $\mathcal{S}^2$. We believe this artifact permeates the CC amplitude expressions through $\widetilde{F}_{ov}$ terms primed by $P_\mathrm{KS}$, as reflected in our \ce{Cr2} PEC results. Self-interaction error (SIE)\cite{Perdew1981,Lundberg2005,Polo2002,MoriSanchez2006} always affects DFT accuracy,\cite{Maniar2024} but the SC step considerably suppresses functional-driven SIE because KS and HF eigenquantities become similar. However, we find that the density and its square, proportional to the on-top density of a single determinant, must be sufficiently accurate for KS-CC to improve upon HF-CC, complicating DFA selection beyond SIE considerations.

To simplify this choice, we also introduce a new difference-density metric, NNED, that addresses deficiencies in conventional MR diagnostics like $T_{1}^\mathrm{diag}$ and directly gauges the viability of KS densities. NNED captures features of $P_\mathrm{KS}$ that precondition CC and MBPT methods, so NNED should guide practitioners toward DFAs that may evolve the correlated reference closer to the exact solution. NNED can be evaluated at mean-field cost while revealing \textit{a priori} that certain DFAs can alleviate MR character by preconditioning the CC problem with a single-reference density. For practitioners, we offer the following guidance: GGA references (PBE, PW91) consistently achieve chemical accuracy for metal--metal bonds, and we recommend them as default choices for systems with suspected MR character. For metal--ligand bonds (\ce{M-O}, \ce{M-Cl}, \ce{M-H}, and \ce{M-H+}), the choice of reference matters less, though hybrid functionals may offer marginal improvements for metal oxides. Closed-shell or $d^{10}$ systems (e.g., \ce{Cu2}, \ce{Zn2}) show minimal sensitivity to reference choice, consistent with their negligible MR character.

In general, single-reference KS-CC can handle cases of moderate MR character that challenge HF-CC. The variability in KS-CC results reflects a complicated interplay among non-Brillouin terms,\cite{Brillouin1932,Engel2006,Bour2000,Koren2001,Cramer1995,Robinson2007} the transformed integrals $\widetilde{\langle pq|| rs \rangle}$, and the semi-canonical orbital-energy denominators. However, the epicenter of all these changes is the KS density matrix and not the KS orbitals, as the density remains invariant under semi-canonicalization.

Examining the difference density for qualitative changes in reference determinants also underscores the importance of information-theoretic concepts and precise density reconstruction in quantum chemistry. We have clarified the physical origin of changes that KS densities induce in many-body methods. These changes allow \textit{useful} Thouless rotations that drive KS-CC toward the exact solution. We have also shown that KS-CCSD(T) achieves near chemical accuracy for metal--metal BDEs at a fraction of the cost. Notably, GGA-referenced CCSD(T) produces no errors exceeding 0.51~eV across 49 transition-metal diatomics, demonstrating robust transferability across diverse bonding environments (Supplementary Fig.\ 2). Current protocols for generating coupled cluster training data typically exclude systems that $T_1^{\mathrm{diag}}$ flags as multireference, introducing systematic bias toward single-reference chemistry. The NNED diagnostic offers a path to recover these systems: when NNED indicates sufficient density difference, KS-CCSD(T) can provide benchmark-quality thermochemistry data without recourse to expensive multireference methods, thereby enabling unbiased training sets that span the full spectrum of electronic structure. Finally, this work carries crucial implications for solid-state materials simulations, where HF-CCSD(T) is often dismissed due to poor performance relative to DFT. We anticipate that the NNED metric will guide DFA selection in future KS-CCSD(T) applications to solids.

\section*{Methods}
 
\subsection*{Electronic Structure Methods}

We computed metal diatomic BDEs at CCSD(T)/CBS using ORCA 6.0.\cite{Neese2012,Neese2018,Neese2020,Neese2023,Neese2025} We calculated orbital energy changes with varying HFX, STGs, excitation energies, NNED metrics, $T_{1}^\mathrm{diag}$, and Frobenius norms of $\widetilde{F}_{ov}$/$\Delta_{P}$ using a development version of Q-Chem v6.2.\cite{qchem} EPT calculations used a modified version of UQUANTCHEM.\cite{uppsala} We visualized DDNOs with IQmol.\cite{gilbert2012iqmol} The Supplementary Information provides additional computational details, including reference data, relativistic effects, zero-point energy corrections, and molecular geometries.
 
\subsection*{Difference Density Analysis}

We formulate a density-difference metric\cite{Ortiz2020,Bovill2026} to better assess MR character in non-canonical CC references. We define the difference between the one-electron density matrices from KS-DFT ($P_\mathrm{KS}$) and HF ($P_\mathrm{HF}$) as an approximate density cumulant:
\begin{align}
\Delta_{P} = P_\mathrm{KS} - P_\mathrm{HF}.
\end{align}
Analogous to L\"{o}wdin's natural orbital decomposition, we solve:
\begin{align}
U^{\dagger}S^{\nicefrac{1}{2}}\Delta_{P}S^{\nicefrac{1}{2}}U = \delta. 
\end{align}

This yields symmetrically orthogonalized corresponding orbitals\cite{Amos1961,Lowdin1950} and eigenvalues representing electron occupation number shifts in the diagonal matrix $\delta$. The eigenvalue sum gives the number of electrons displaced (NED) between SCF solutions. Normalizing by the total electron count yields NNED, our criterion for gauging reference density similarity and \textit{a priori} MR character in KS-CC. We performed this analysis using a development version of Q-Chem v6.2.\cite{qchem}

\section*{Acknowledgements}

B.Z. acknowledges support from the National Science Foundation Graduate Research Fellowship. J.A.K. acknowledges support from the U.S. Naval Research Lab (N00173-25-1-0040). A.M.R. acknowledges support from the Wass Undergraduate Research Fellowship. This research was supported in part by the University of Pittsburgh and the University of Pittsburgh Center for Research Computing and Data through the resources provided. Specifically, this work used the H2P cluster, which is supported by NSF award number OAC-2117681.

\section*{Author Contributions}
J.A.K. and K.C.F. conceptualized the study. B.Z., A.M.R., and A.Y.Z. collected numerical data to support the study. A.Y.Z. and K.C.F. performed the theoretical analysis to support the study. A.Y.Z. and B.Z. wrote the manuscript and all authors provided input on revisions.

\section*{Competing Interests}
The authors declare no competing interests.

\section*{Additional Information}
Supplementary Information The online version contains
supplementary material available at...

\section*{Data Availability}
All ORCA output files, including KS-DFT optimized geometries, CCSD(T)/CBS single-point energies, atomic energies, and \ce{Cr2} potential energy surface data, are freely available on Zenodo at \url{https://doi.org/10.5281/zenodo.17958091}. Source data for all figures are provided with the Supplementary Information.

\section*{Code Availability}
Jupyter notebooks for parsing ORCA outputs and generating figures are available on GitHub at \url{https://github.com/BLZ11/cc-dft}. The NNED analysis was performed using a development version of Q-Chem v6.2; the relevant routines will be included in a future public release.

\clearpage

\section*{Figure Legends}
 
\textbf{Figure 1} $\vert$ \textbf{MO eigenvalue changes with increasing \%HFX for HF, B3LYP, and B3LYP after semi-canonicalization.} The percent exchange of the DFA is adjusted in proportion to the exact exchange. The vertical dashed lines represent the standard  \%HFX in the DFA. Highest occupied (HOMO) and lowest unoccupied (LUMO) MO energies are obtained with the cc-pVTZ and aug-cc-pVTZ basis sets respectively.

\vspace{1em}

\textbf{Figure 2} $\vert$ \textbf{Benchmark comparison of density functional and coupled cluster methods for transition metal–ligand bond dissociation energies.} Mean absolute errors (MAE) in bond dissociation energies for first-row transition metal compounds across four bond classes: a, metal–oxygen (\ce{M-O}); b, metal–chlorine (\ce{M-Cl}); c, metal–hydrogen (\ce{M-H}); and d, over all bonds. Blue bars represent density functional methods (KS-DFT), orange bars represent CCSD(T)/CBS calculations using different reference determinants (CC-X denotes CCSD(T)/CBS with densities from method X), and the green bar represents phaseless auxiliary-field quantum Monte Carlo (ph-AFQMC) results from Ref.\citenum{Shee2019}. CC-Best denote the optimal reference selection either from HF or DFT for each system and QMC-Best denote the optimal reference when MP2, CC, or TZ/QZ extrapolation was used in ph-AFQMC. The subtle gray shading  indicates chemical accuracy threshold (0--3 kcal/mol $\approx$ $0$--$0.13$ eV).

\vspace{1em}

\textbf{Figure 3} $\vert$ \textbf{Performance of density functional and coupled cluster methods for metal hydride cations and bimetallic bond dissociation energies.} Mean absolute errors (MAE) in bond dissociation energies for: a, metal hydride cations (\ce{M-H+}); b, homonuclear metal–metal dimers (\ce{M-M}); and c, overall performance. Blue bars represent density functional methods (KS-DFT) and orange bars represent CCSD(T)/CBS with different reference determinants. CC-Best denotes the optimal reference selection (HF or DFT) for each system. The subtle gray shading  indicates chemical accuracy threshold (0--3 kcal/mol $\approx$ $0$--$0.13$ eV).

\vspace{1em}

\textbf{Figure 4} $\vert$ \textbf{Potential energy curve of the anti-ferromagnetic state of \ce{Cr2} dimer computed with UCCSD(T)/CBS using different reference densities.} Binding energy as a function of bond length for \ce{Cr2}, the prototypical multireference transition metal dimer. Solid lines show UCCSD(T)/CBS calculations with reference orbitals from Hartree–Fock (UCC-HF), SVWN5 (UCC-SVWN5), PBE (UCC-PBE), PW91 (UCC-PW91), R²SCAN (UCC-R²SCAN), and PBE0 (UCC-PBE0). The experimental curve (Exp, black) and best theoretical estimate from state-of-the-art multireference calculations (BTE, crimson) are shown for comparison obtained from Ref.\citenum{Larsson2022}.

\vspace{1em}

\textbf{Figure 5} $\vert$ \textbf{Singlet triplet gaps (STGs) for BN and {$\mathbf{{CH}_{2}}$} computed with CCSD(T)/cc-pVTZ with HF and KS-DFT determinants.} The solid and dashed horizontal orange line represent the reference theoretical estimate and experimental value, respectively.

\vspace{1em}

\textbf{Figure 6} $\vert$ \textbf{Difference density natural orbitals for singlet $\mathrm{Cr}_{2}$ computed between HF and PW91 with def2-QZVPP.} The six largest electron displacement eigenpairs $\delta_{\pm}$ are reported. The orbitals corresponding to $\delta_{-}$ and $\delta_{+}$ are represented by radial dots and line meshes, respectively. An isosurface value of 0.02 is used except for the fourth orbital from the top, where a value of 0.04 was used. The NNED and $||\Delta_P||_{F}$ values are 0.083 and 2.240, respectively.

\clearpage


\clearpage

\newcommand{\greenCheck}{{{\color{ForestGreen}{{\ding{51}}}}}}
\newcommand{\redEx}{{{\color{BrickRed}{{\ding{55}}}}}}

\begin{table}[htbp]
\caption{\textbf{Designations$^{a}$ of MR character based on HF-CC $T_1$ diagnotics and $T_1$-equivalent KS-based NNED metrics.}}
\label{table:mr-character}
\setlength{\tabcolsep}{3.5pt}
\renewcommand{\arraystretch}{1.1} 
\begin{tabular}{lcccccc}
\toprule
 & \small{$T_1$} &  &  & \small{NNED}  &  &  \\ 
 \cline{3-7}
 & \small{HF}&  \small{LDA}&  \small{B3LYP}&  \small{revPBE}&  \small{SCAN}&  \small{SCAN0}\\
 
\midrule
\footnotesize{I} &  &  & &  &  &  \\
\cline{1-1}
\addlinespace
\small{BN} & \greenCheck  & \greenCheck & \greenCheck & \greenCheck & \greenCheck & \greenCheck \\

\addlinespace
\small{$\mathrm{C}_2$} & \greenCheck  & \greenCheck & \greenCheck & \greenCheck & \greenCheck & \redEx  \\

\addlinespace
\small{$\mathrm{BO}^{+}$} & \greenCheck  & \greenCheck & \greenCheck & \greenCheck & \greenCheck & \redEx    \\

\addlinespace
\small{$\mathrm{CN}^{+}$} & \greenCheck  & \greenCheck & \greenCheck & \greenCheck & \greenCheck & \redEx \\
\addlinespace
\small{II}$^{b}$ &  &  & &  &  &  \\
\cline{1-1}
\addlinespace
\small{$\mathrm{CH}_{2}$}  &  \greenCheck    & \greenCheck   &  \redEx      &    \redEx   &  \redEx  &  \redEx   \\

\addlinespace
\small{$\mathrm{O}_{2}$} &   \greenCheck       &  \greenCheck$^{c}$    & \redEx   & \redEx   & \redEx   & \redEx   \\

\addlinespace
\small{NF} &   \greenCheck   & \greenCheck   & \redEx &  \greenCheck    & \redEx   &  \redEx    \\

\addlinespace
\small{NH} &   \redEx  & \greenCheck   &  \redEx    &  \redEx  &  \redEx   &  \redEx   \\
\addlinespace

 & \small{HF}&  \small{LDA}&  \small{PW91}&  \small{r$^{2}$SCAN}&  \small{PBE}&  \small{PBE0}\\
 
\midrule
\small{III}$^{b}$ &  &  & &  &  &  \\
\cline{1-1}
\addlinespace
\small{$\mathrm{Mn}_{2}$}  &  \redEx     & \redEx    & \greenCheck  &  \greenCheck   &  \redEx  &  \redEx   \\

\addlinespace
\small{NiO} &   \greenCheck       &  \greenCheck    &  \greenCheck$^{c}$   & \redEx   & \redEx   &  \greenCheck   \\

\addlinespace
\small{CoCl} &   \greenCheck   & \greenCheck   &  \greenCheck & \redEx  & \greenCheck   &  \redEx    \\

\addlinespace
\small{VH} &   \greenCheck   &  \redEx   &  \redEx    &  \redEx  &  \redEx   &  \redEx   \\
\bottomrule
\end{tabular}\\ \vspace{0.5em}
\footnotesize{Two classes of main-group singlets (I \& II) and one class of assorted metal diatomics (III) are examined. General NNED classifier limits are trained on TinySpins25 and BDE20.$^{a}$Markers \ding{51} and \ding{55} denote MR and not MR, respectively. $^{b}$When applicable, broken-symmetry references are used. $^{c}$At the general T$_1$-equivalent NNED threshold.}
\end{table}

\clearpage

\bibliography{references}

@ARTICLE{Bullock2020,
  title     = "Using nature's blueprint to expand catalysis with Earth-abundant
               metals",
  author    = "Bullock, R Morris and Chen, Jingguang G and Gagliardi, Laura and
               Chirik, Paul J and Farha, Omar K and Hendon, Christopher H and
               Jones, Christopher W and Keith, John A and Klosin, Jerzy and
               Minteer, Shelley D and Morris, Robert H and Radosevich,
               Alexander T and Rauchfuss, Thomas B and Strotman, Neil A and
               Vojvodic, Aleksandra and Ward, Thomas R and Yang, Jenny Y and
               Surendranath, Yogesh",
  journal   = "Science",
  publisher = "American Association for the Advancement of Science (AAAS)",
  volume    =  369,
  number    =  6505,
  pages     = "eabc3183",
  month     =  aug,
  year      =  2020,
  copyright = "https://www.sciencemag.org/about/science-licenses-journal-article-reuse",
  language  = "en"
}

@article{Drabik2024,
    title = {{Approaching the Complete Basis Set Limit for Spin-State Energetics of Mononuclear First-Row Transition Metal Complexes}},
    year = {2024},
    journal = {J. Chem. Theory Comput.},
    author = {Drabik, Gabriela and Rado{\'{n}}, Mariusz},
    number = {8},
    month = {4},
    pages = {3199--3217},
    volume = {20},
    doi = {10.1021/acs.jctc.4c00092},
    issn = {1549-9618}
}

@article{Radon2019,
    title = {{Benchmarking quantum chemistry methods for spin-state energetics of iron complexes against quantitative experimental data}},
    year = {2019},
    journal = {Phys. Chem. Chem. Phys.},
    author = {Rado{\'{n}}, Mariusz},
    number = {9},
    pages = {4854--4870},
    volume = {21},
    doi = {10.1039/C9CP00105K},
    issn = {1463-9076}
}

@article{Keith2021,
    title = {{Combining Machine Learning and Computational Chemistry for Predictive Insights Into Chemical Systems}},
    year = {2021},
    journal = {Chem. Rev. },
    author = {Keith, John A. and Vassilev-Galindo, Valentin and Cheng, Bingqing and Chmiela, Stefan and Gastegger, Michael and M{\"{u}}ller, Klaus-Robert and Tkatchenko, Alexandre},
    number = {16},
    month = {8},
    pages = {9816--9872},
    volume = {121},
    doi = {10.1021/acs.chemrev.1c00107},
    issn = {0009-2665}
}

@article{Vogiatzis2018,
    title = {{Computational Approach to Molecular Catalysis by 3d Transition Metals: Challenges and Opportunities}},
    year = {2019},
    journal = {Chem. Rev.},
    author = {Vogiatzis, Konstantinos D. and Polynski, Mikhail V. and Kirkland, Justin K. and Townsend, Jacob and Hashemi, Ali and Liu, Chong and Pidko, Evgeny A.},
    number = {4},
    month = {2},
    pages = {2453--2523},
    volume = {119},
    doi = {10.1021/acs.chemrev.8b00361},
    issn = {0009-2665}
}

@article{Bauschlicher1994,
    title = {{$Cr_{2}$ revisited}},
    year = {1994},
    journal = {Chem. Phys. Lett.},
    author = {Bauschlicher, Charles W. and Partridge, Harry},
    number = {2-3},
    month = {12},
    pages = {277--282},
    volume = {231},
    doi = {10.1016/0009-2614(94)01243-1},
    issn = {00092614}
}

@article{Schultz2005,
author = {Schultz, Nathan E. and Zhao, Yan and Truhlar, Donald G.},
title = {Databases for Transition Element Bonding: Metal-Metal Bond Energies and Bond Lengths and Their Use To Test Hybrid, Hybrid Meta, and Meta Density Functionals and Generalized Gradient Approximations},
journal = {J. Phys. Chem. A},
volume = {109},
number = {19},
pages = {4388-4403},
year = {2005},
doi = {10.1021/jp0504468},
}

@article{Kant1966,
    author = {Kant, Arthur and Strauss, Bernard},
    title = {Dissociation Energy of $Cr_{2}$},
    journal = {J. Chem. Phys.},
    volume = {45},
    number = {8},
    pages = {3161-3162},
    year = {1966},
    month = {10},
    issn = {0021-9606},
    doi = {10.1063/1.1728082},
}

@article{Hilpert1987,
author = {Hilpert, K. and Ruthardt, R.},
title = {Determination of the Dissociation Energy of the $Cr_{2}$ Molecule},
journal = {Berichte der Bunsengesellschaft für physikalische Chemie},
volume = {91},
number = {7},
pages = {724-731},
doi = {https://doi.org/10.1002/bbpc.19870910707},
year = {1987}
}

@article{Simard1998,
    author = {Simard, Benoit and Lebeault-Dorget, Marie-Ange and Marijnissen, Adrian and ter Meulen, J. J.},
    title = {Photoionization spectroscopy of dichromium and dimolybdenum: {Ionization} potentials and bond energies},
    journal = {J. Chem. Phys.},
    volume = {108},
    number = {23},
    pages = {9668-9674},
    year = {1998},
    month = {06},
    issn = {0021-9606},
    doi = {10.1063/1.476442},
}

@article{Xu2015,
    title = {{Do Practical Standard Coupled Cluster Calculations Agree Better than Kohn–Sham Calculations with Currently Available Functionals When Compared to the Best Available Experimental Data for Dissociation Energies of Bonds to 3d Transition Metals?}},
    year = {2015},
    journal = {J. Chem. Theory Comput.},
    author = {Xu, Xuefei and Zhang, Wenjing and Tang, Mingsheng and Truhlar, Donald G.},
    number = {5},
    month = {5},
    pages = {2036--2052},
    volume = {11},
    doi = {10.1021/acs.jctc.5b00081},
    issn = {1549-9618}
}

@article{Cheng2017,
  title = {Bond Dissociation Energies for Diatomic Molecules Containing 3d Transition Metals: Benchmark Scalar-Relativistic Coupled-Cluster Calculations for 20 Molecules},
  volume = {13},
  ISSN = {1549-9626},
  DOI = {10.1021/acs.jctc.6b00970},
  number = {3},
  journal = {J. Chem. Theory Comput.},
  publisher = {American Chemical Society (ACS)},
  author = {Cheng,  Lan and Gauss,  J\"{u}rgen and Ruscic,  Branko and Armentrout,  Peter B. and Stanton,  John F.},
  year = {2017},
  month = feb,
  pages = {1044–1056}
}

@article{Bannwarth2021,
    title = {{Extended tight‐binding quantum chemistry methods}},
    year = {2021},
    journal = {WIREs Comput Mol Sci},
    author = {Bannwarth, Christoph and Caldeweyher, Eike and Ehlert, Sebastian and Hansen, Andreas and Pracht, Philipp and Seibert, Jakob and Spicher, Sebastian and Grimme, Stefan},
    number = {2},
    month = {3},
    volume = {11},
    doi = {10.1002/wcms.1493},
    issn = {1759-0876}
}

@article{Perdew1996,
    title = {{Generalized Gradient Approximation Made Simple}},
    year = {1996},
    journal = {Phys. Rev. Lett.},
    author = {Perdew, John P. and Burke, Kieron and Ernzerhof, Matthias},
    number = {18},
    month = {10},
    pages = {3865--3868},
    volume = {77},
    doi = {10.1103/PhysRevLett.77.3865},
    issn = {0031-9007}
}

@article{Perdew1997,
    title = {{Generalized Gradient Approximation Made Simple [Phys. Rev. Lett. 77, 3865 (1996)]}},
    year = {1997},
    journal = {Phys. Rev. Lett.},
    author = {Perdew, John P and Burke, Kieron and Ernzerhof, Matthias},
    number = {7},
    pages = {1396--1396},
    volume = {78},
    url = {https://link.aps.org/doi/10.1103/PhysRevLett.78.1396},
    doi = {10.1103/PhysRevLett.78.1396}
}

@article{Aoto2017,
    title = {{How To Arrive at Accurate Benchmark Values for Transition Metal Compounds: Computation or Experiment?}},
    year = {2017},
    journal = {J. Chem. Theory Comput.},
    author = {Aoto, Yuri A. and de Lima Batista, Ana Paula and K{\"{o}}hn, Andreas and de Oliveira-Filho, Antonio G. S.},
    number = {11},
    month = {11},
    pages = {5291--5316},
    volume = {13},
    doi = {10.1021/acs.jctc.7b00688},
    issn = {1549-9618}
}

@article{Hohenberg1964,
    title = {{Inhomogeneous Electron Gas}},
    year = {1964},
    journal = {Phys. Rev.},
    author = {Hohenberg, P. and Kohn, W.},
    number = {3B},
    pages = {B864--B871},
    volume = {136},
    doi = {10.1103/PhysRev.136.B864},
    issn = {0378620X}
}

@article{Perdew2021,
    title = {{Interpretations of ground-state symmetry breaking and strong correlation in wavefunction and density functional theories}},
    year = {2021},
    journal = {Proc. Natl. Acad. Sci. U.S.A.},
    author = {Perdew, John P. and Ruzsinszky, Adrienn and Sun, Jianwei and Nepal, Niraj K. and Kaplan, Aaron D.},
    number = {4},
    month = {1},
    volume = {118},
    doi = {10.1073/pnas.2017850118},
    issn = {0027-8424}
}

@article{Jiang2012,
    title = {{Multireference Character for 3d Transition-Metal-Containing Molecules}},
    year = {2012},
    journal = {J. Chem. Theory Comput.},
    author = {Jiang, Wanyi and DeYonker, Nathan J. and Wilson, Angela K.},
    number = {2},
    month = {2},
    pages = {460--468},
    volume = {8},
    doi = {10.1021/ct2006852},
    issn = {1549-9618}
}

@article{Shee2019,
    title = {{On Achieving High Accuracy in Quantum Chemical Calculations of 3d Transition Metal-Containing Systems: A Comparison of Auxiliary-Field Quantum Monte Carlo with Coupled Cluster, Density Functional Theory, and Experiment for Diatomic Molecules}},
    year = {2019},
    journal = {J. Chem. Theory Comput.},
    author = {Shee, James and Rudshteyn, Benjamin and Arthur, Evan J. and Zhang, Shiwei and Reichman, David R. and Friesner, Richard A.},
    number = {4},
    month = {4},
    pages = {2346--2358},
    volume = {15},
    doi = {10.1021/acs.jctc.9b00083},
    issn = {1549-9618}
}

@article{Sherrill1999,
    title = {{On the performance of density functional theory for symmetry-breaking problems}},
    year = {1999},
    journal = {Chem. Phys. Lett.},
    author = {Sherrill, C.David and Lee, Michael S. and Head-Gordon, Martin},
    number = {5-6},
    month = {3},
    pages = {425--430},
    volume = {302},
    doi = {10.1016/S0009-2614(99)00206-7},
    issn = {00092614}
}

@article{Bertels2021,
    title = {{Polishing the Gold Standard: The Role of Orbital Choice in CCSD(T) Vibrational Frequency Prediction}},
    year = {2021},
    journal = {J. Chem. Theory Comput.},
    author = {Bertels, Luke W. and Lee, Joonho and Head-Gordon, Martin},
    number = {2},
    month = {2},
    pages = {742--755},
    volume = {17},
    doi = {10.1021/acs.jctc.0c00746},
    issn = {1549-9618}
}

@article{Bao2017b,
    title = {{Predicting Bond Dissociation Energies of Transition-Metal Compounds by Multiconfiguration Pair-Density Functional Theory and Second-Order Perturbation Theory Based on Correlated Participating Orbitals and Separated Pairs}},
    year = {2017},
    journal = {J. Chem. Theory Comput.},
    author = {Bao, Junwei Lucas and Odoh, Samuel O. and Gagliardi, Laura and Truhlar, Donald G.},
    number = {2},
    month = {2},
    pages = {616--626},
    volume = {13},
    doi = {10.1021/acs.jctc.6b01102},
    issn = {1549-9618}
}

@article{Bao2017a,
    title = {{Predicting bond dissociation energy and bond length for bimetallic diatomic molecules: a challenge for electronic structure theory}},
    year = {2017},
    journal = {Phys. Chem. Chem. Phys.},
    author = {Bao, Junwei Lucas and Zhang, Xin and Xu, Xuefei and Truhlar, Donald G.},
    number = {8},
    pages = {5839--5854},
    volume = {19},
    doi = {10.1039/C6CP08896A},
    issn = {1463-9076}
}

@article{Fang2017,
    title = {{Prediction of Bond Dissociation Energies/Heats of Formation for Diatomic Transition Metal Compounds: CCSD(T) Works}},
    year = {2017},
    journal = {J. Chem. Theory Comput.},
    author = {Fang, Zongtang and Vasiliu, Monica and Peterson, Kirk A. and Dixon, David A.},
    number = {3},
    month = {3},
    pages = {1057--1066},
    volume = {13},
    doi = {10.1021/acs.jctc.6b00971},
    issn = {1549-9618}
}

@article{Drosou2022,
    title = {{Reconciling Local Coupled Cluster with Multireference Approaches for Transition Metal Spin-State Energetics}},
    year = {2022},
    journal = {J. Chem. Theory Comput.},
    author = {Drosou, Maria and Mitsopoulou, Christiana A. and Pantazis, Dimitrios A.},
    number = {6},
    month = {6},
    pages = {3538--3548},
    volume = {18},
    doi = {10.1021/acs.jctc.2c00265},
    issn = {1549-9618}
}

@article{Kohn1965,
    title = {{Self-Consistent Equations Including Exchange and Correlation Effects}},
    year = {1965},
    journal = {Phys. Rev.},
    author = {Kohn, W and Sham, L. J.},
    number = {4A},
    pages = {A1133-A1138},
    volume = {140},
    url = {https://link.aps.org/doi/10.1103/PhysRev.140.A1133},
    doi = {10.1103/PhysRev.140.A1133}
}

@article{Thiel2014,
    title = {{Semiempirical quantum-chemical methods}},
    year = {2014},
    journal = {WIREs Comput Mol Sci},
    author = {Thiel, Walter},
    number = {2},
    month = {3},
    pages = {145--157},
    volume = {4},
    doi = {10.1002/wcms.1161},
    issn = {17590876}
}

@article{Benedek2022,
    title = {{Sensitivity of coupled cluster electronic properties on the reference determinant: Can Kohn–Sham orbitals be more beneficial than Hartree–Fock orbitals?}},
    year = {2022},
    journal = {J. Comput. Chem.},
    author = {Benedek, Zsolt and Tim{\'{a}}r, Paula and Szilv{\'{a}}si, Tibor and Barcza, Gergely},
    number = {32},
    month = {12},
    pages = {2103--2120},
    volume = {43},
    doi = {10.1002/jcc.26996},
    issn = {0192-8651}
}

@article{Neese2025,
    title = {{Software Update: The {ORCA} Program System—Version 6.0}},
    year = {2025},
    journal = {WIREs Comput. Mol. Sci.},
    author = {Neese, Frank},
    number = {2},
    month = {3},
    volume = {15},
    doi = {10.1002/wcms.70019},
    issn = {1759-0876}
}

@article{Neese2018,
    title = {{Software update: the ORCA program system, version 4.0}},
    year = {2018},
    journal = {WIREs Comput. Mol. Sci.},
    author = {Neese, Frank},
    number = {1},
    month = {1},
    volume = {8},
    doi = {10.1002/wcms.1327},
    issn = {1759-0876}
}

@article{Maniar2024,
    title = {{Symmetry breaking and self-interaction correction in the chromium atom and dimer}},
    year = {2024},
    journal = {J. Chem. Phys.},
    author = {Maniar, Rohan and Withanage, Kushantha P. K. and Shahi, Chandra and Kaplan, Aaron D. and Perdew, John P. and Pederson, Mark R.},
    number = {14},
    month = {4},
    volume = {160},
    doi = {10.1063/5.0180863},
    issn = {0021-9606}
}

@article{Larsson2022,
    title = {{The Chromium Dimer: Closing a Chapter of Quantum Chemistry}},
    year = {2022},
    journal = {J. Am. Chem. Soc.},
    author = {Larsson, Henrik R. and Zhai, Huanchen and Umrigar, C. J. and Chan, Garnet Kin-Lic},
    number = {35},
    month = {9},
    pages = {15932--15937},
    volume = {144},
    doi = {10.1021/jacs.2c06357},
    issn = {0002-7863}
}

@article{Moltved2019a,
    title = {{The Metal Hydride Problem of Computational Chemistry: Origins and Consequences}},
    year = {2019},
    journal = {J. Phys. Chem. A},
    author = {Moltved, Klaus A. and Kepp, Kasper P.},
    number = {13},
    month = {4},
    pages = {2888--2900},
    volume = {123},
    doi = {10.1021/acs.jpca.9b02367},
    issn = {1089-5639}
}

@article{Neese2012,
    title = {{The ORCA program system}},
    year = {2012},
    journal = {WIREs Comput. Mol. Sci.},
    author = {Neese, Frank},
    number = {1},
    month = {1},
    pages = {73--78},
    volume = {2},
    doi = {10.1002/wcms.81},
    issn = {1759-0876}
}

@article{Neese2020,
    title = {{The ORCA quantum chemistry program package}},
    year = {2020},
    journal = {J. Chem. Phys.},
    author = {Neese, Frank and Wennmohs, Frank and Becker, Ute and Riplinger, Christoph},
    number = {22},
    month = {6},
    volume = {152},
    doi = {10.1063/5.0004608},
    issn = {0021-9606}
}

@article{Neese2023,
    title = {{The SHARK integral generation and digestion system}},
    year = {2023},
    journal = {J. Comp. Chem.},
    author = {Neese, Frank},
    number = {3},
    month = {1},
    pages = {381--396},
    volume = {44},
    doi = {10.1002/jcc.26942},
    issn = {0192-8651}
}

@article{Fang2016,
    title = {{Use of Improved Orbitals for CCSD(T) Calculations for Predicting Heats of Formation of Group IV and Group VI Metal Oxide Monomers and Dimers and UCl <sub>6</sub>}},
    year = {2016},
    journal = {J. Chem. Theory Comput.},
    author = {Fang, Zongtang and Lee, Zachary and Peterson, Kirk A. and Dixon, David A.},
    number = {8},
    month = {8},
    pages = {3583--3592},
    volume = {12},
    doi = {10.1021/acs.jctc.6b00327},
    issn = {1549-9618}
}

@article{Hait2019,
    title = {{What Levels of Coupled Cluster Theory Are Appropriate for Transition Metal Systems? A Study Using Near-Exact Quantum Chemical Values for 3d Transition Metal Binary Compounds}},
    year = {2019},
    journal = {J. Chem. Theory Comput.},
    author = {Hait, Diptarka and Tubman, Norman M. and Levine, Daniel S. and Whaley, K. Birgitta and Head-Gordon, Martin},
    number = {10},
    month = {10},
    pages = {5370--5385},
    volume = {15},
    doi = {10.1021/acs.jctc.9b00674},
    issn = {1549-9618}
}

@article{Johnson2023,
  title     = "Review and perspective on transition metal electrocatalysts
               toward carbon-neutral energy",
  author    = "Johnson, Denis and Pranada, Eugenie and Yoo, Ray and Uwadiunor,
               Ekenedilichukwu and Ngozichukwu, Bright and Djire, Abdoulaye",
  journal   = "Energy Fuels",
  publisher = "American Chemical Society (ACS)",
  volume    =  37,
  number    =  3,
  pages     = "1545--1576",
  month     =  feb,
  year      =  2023
}

@article{Zhang2024,
  title     = "Strategies for designing advanced transition metal-based
               electrocatalysts for alkaline water/seawater splitting at
               ampere-level current densities",
  author    = "Zhang, Xian and Zuo, Ziteng and Liao, Chengzhu and Jia, Feifei
               and Cheng, Chun and Guo, Zhiguang",
  journal   = "ACS Catal.",
  publisher = "American Chemical Society (ACS)",
  volume    =  14,
  number    =  23,
  pages     = "18055--18071",
  month     =  dec,
  year      =  2024
}

@article{Guo2024,
  title    = "Electronic structure design of transition metal-based catalysts
              for electrochemical carbon dioxide reduction",
  author   = "Guo, Liang and Zhou, Jingwen and Liu, Fu and Meng, Xiang and Ma,
              Yangbo and Hao, Fengkun and Xiong, Yuecheng and Fan, Zhanxi",
  journal  = "ACS Nano",
  volume   =  18,
  number   =  14,
  pages    = "9823--9851",
  month    =  apr,
  year     =  2024
}

@article{Li2024,
  title     = "Engineering interfacial sulfur migration in transition-metal
               sulfide enables low overpotential for durable hydrogen evolution
               in seawater",
  author    = "Li, Min and Li, Hong and Fan, Hefei and Liu, Qianfeng and Yan,
               Zhao and Wang, Aiqin and Yang, Bing and Wang, Erdong",
  journal   = "Nat. Commun.",
  publisher = "Springer Science and Business Media LLC",
  volume    =  15,
  number    =  1,
  pages     = "6154",
  month     =  jul,
  year      =  2024,
  copyright = "https://creativecommons.org/licenses/by/4.0"
}

@ARTICLE{Pei2024,
  title    = "A replacement strategy for regulating local environment of
              single-atom {Co-SxN4-x} catalysts to facilitate {CO2}
              electroreduction",
  author   = "Pei, Jiajing and Shang, Huishan and Mao, Junjie and Chen, Zhe and
              Sui, Rui and Zhang, Xuejiang and Zhou, Danni and Wang, Yu and
              Zhang, Fang and Zhu, Wei and Wang, Tao and Chen, Wenxing and
              Zhuang, Zhongbin",
  journal  = "Nat. Commun.",
  volume   =  15,
  number   =  1,
  pages    = "416",
  month    =  jan,
  year     =  2024
}

@article{Kulichenko2024,
  title     = "Data generation for machine learning interatomic potentials and
               beyond",
  author    = "Kulichenko, Maksim and Nebgen, Benjamin and Lubbers, Nicholas
               and Smith, Justin S and Barros, Kipton and Allen, Alice E A and
               Habib, Adela and Shinkle, Emily and Fedik, Nikita and Li, Ying
               Wai and Messerly, Richard A and Tretiak, Sergei",
  journal   = "Chem. Rev.",
  publisher = "American Chemical Society (ACS)",
  volume    =  124,
  number    =  24,
  pages     = "13681--13714",
  month     =  dec,
  year      =  2024,
  copyright = "https://creativecommons.org/licenses/by-nc-nd/4.0/"
}

@article{Unke2021,
  title     = "Machine learning force fields",
  author    = "Unke, Oliver T and Chmiela, Stefan and Sauceda, Huziel E and
               Gastegger, Michael and Poltavsky, Igor and Sch{\"u}tt, Kristof T
               and Tkatchenko, Alexandre and M{\"u}ller, Klaus-Robert",
  journal   = "Chem. Rev.",
  publisher = "American Chemical Society (ACS)",
  volume    =  121,
  number    =  16,
  pages     = "10142--10186",
  month     =  aug,
  year      =  2021,
  copyright = "https://creativecommons.org/licenses/by-nc-nd/4.0/"
}

@article{Smith2019,
  title     = "Approaching coupled cluster accuracy with a general-purpose
               neural network potential through transfer learning",
  author    = "Smith, Justin S and Nebgen, Benjamin T and Zubatyuk, Roman and
               Lubbers, Nicholas and Devereux, Christian and Barros, Kipton and
               Tretiak, Sergei and Isayev, Olexandr and Roitberg, Adrian E",
  journal   = "Nat. Commun.",
  publisher = "Springer Science and Business Media LLC",
  volume    =  10,
  number    =  1,
  pages     = "2903",
  month     =  jul,
  year      =  2019,
  copyright = "https://creativecommons.org/licenses/by/4.0"
}

@article{Raghavachari1989,
  title     = "A fifth-order perturbation comparison of electron correlation
               theories",
  author    = "Raghavachari, Krishnan and Trucks, Gary W and Pople, John A and
               Head-Gordon, Martin",
  journal   = "Chem. Phys. Lett.",
  publisher = "Elsevier BV",
  volume    =  157,
  number    =  6,
  pages     = "479--483",
  month     =  may,
  year      =  1989
}

@article{Stanton1997,
  title     = "Why {CCSD(T}) works: a different perspective",
  author    = "Stanton, John F",
  journal   = "Chem. Phys. Lett.",
  publisher = "Elsevier BV",
  volume    =  281,
  number    = "1-3",
  pages     = "130--134",
  month     =  dec,
  year      =  1997
}

@article{Moltved2018,
  title     = "Chemical bond energies of 3d transition metals studied by
               density functional theory",
  author    = "Moltved, Klaus A and Kepp, Kasper P",
  journal   = "J. Chem. Theory Comput.",
  publisher = "American Chemical Society (ACS)",
  volume    =  14,
  number    =  7,
  pages     = "3479--3492",
  month     =  jul,
  year      =  2018
}

@article{Wiedner2016,
  title     = "Thermodynamic hydricity of transition metal hydrides",
  author    = "Wiedner, Eric S and Chambers, Matthew B and Pitman, Catherine L
               and Bullock, R Morris and Miller, Alexander J M and Appel, Aaron
               M",
  journal   = "Chem. Rev.",
  publisher = "American Chemical Society (ACS)",
  volume    =  116,
  number    =  15,
  pages     = "8655--8692",
  month     =  aug,
  year      =  2016,
  copyright = "http://pubs.acs.org/page/policy/authorchoice\_termsofuse.html"
}

@article{Bovill2026,
author = {Bovill, Andrew J. and Abou Taka, Ali and Harb, Hassan and Hratchian, Hrant P.},
title = {Excitation/Relaxation Analysis of Electronic Transitions Using Difference Density Natural Orbitals},
journal = {J. Chem. Theory Comput.},
volume = { },
number = { },
pages = { },
year = {2026},
doi = {10.1021/acs.jctc.5c01792},
}

@article{Ortiz2020,
    author = {Ortiz, J. V. and Zalik, R. A.},
    title = {Eigenvalues of uncorrelated, density-difference matrices and the interpretation of {$\Delta$}-self-consistent-field calculations},
    journal = {J. Chem. Phys.},
    volume = {153},
    number = {11},
    pages = {114122},
    year = {2020},
    month = {09},
    doi = {10.1063/5.0019542},
}

@article{Amos1961,
author = {Amos, A. T.  and Hall, G. G. },
title = {Single determinant wave functions},
journal = {Proc. R. Soc. Lond. A},
volume = {263},
number = {1315},
pages = {483-493},
year = {1961},
doi = {10.1098/rspa.1961.0175},
}

@article{Lowdin1950,
  title = {On the Non-Orthogonality Problem Connected with the Use of Atomic Wave Functions in the Theory of Molecules and Crystals},
  volume = {18},
  ISSN = {1089-7690},
  DOI = {10.1063/1.1747632},
  number = {3},
  journal = {J. Chem. Phys.},
  publisher = {AIP Publishing},
  author = {L\"{o}wdin,  Per-Olov},
  year = {1950},
  month = mar,
  pages = {365–375}
}

@Article{Harvey2003,
author ="Harvey, Jeremy N. and Aschi, Massimiliano",
title  ="Modelling spin-forbidden reactions: recombination of carbon monoxide with iron tetracarbonyl",
journal  ="Faraday Discuss.",
year  ="2003",
volume  ="124",
issue  ="0",
pages  ="129-143",
publisher  ="The Royal Society of Chemistry",
doi  ="10.1039/B211871H",
}

@article{Jankowski2004,
  title = {A comparative study of Kohn–Sham,  Brueckner and Hartree–Fock orbitals},
  volume = {389},
  ISSN = {0009-2614},
  DOI = {10.1016/j.cplett.2004.03.114},
  number = {4–6},
  journal = {Chem. Phys. Lett.},
  publisher = {Elsevier BV},
  author = {Jankowski,  K. and Nowakowski,  K. and Wasilewski,  J.},
  year = {2004},
  month = may,
  pages = {393–399}
}

@article{Wasilewski2009,
  title = {Evolution of orbital spaces along potential curves for diatomic molecules. {A} comparative study of {Hartree–Fock},  {Kohn–Sham},  {Brueckner} and multi-configurational orbital spaces},
  volume = {905},
  ISSN = {0166-1280},
  DOI = {10.1016/j.theochem.2009.03.004},
  number = {1–3},
  journal = {J. Mol. Struct.: THEOCHEM},
  publisher = {Elsevier BV},
  author = {Wasilewski,  Jan and Zelek,  Sławomir},
  year = {2009},
  month = jul,
  pages = {24–33}
}

@article{Shigeta2005,
author = {Shigeta, Yasuteru},
title = {Optimized effective potential method at finite temperature: {An} application to superconductivity},
journal = {Int. J. Quantum Chem.},
volume = {101},
number = {6},
pages = {774-782},
doi = {https://doi.org/10.1002/qua.20337},
year = {2005}
}

@article{Bartlett2005a,
  title = {The exchange-correlation potential in ab initio density functional theory},
  volume = {122},
  ISSN = {1089-7690},
  DOI = {10.1063/1.1809605},
  number = {3},
  journal = {J. Chem. Phys.},
  publisher = {AIP Publishing},
  author = {Bartlett,  Rodney J. and Grabowski,  Ireneusz and Hirata,  So and Ivanov,  Stanislav},
  year = {2005},
  month = dec 
}

@article{Bartlett2005b,
    author = {Bartlett, Rodney J. and Lotrich, Victor F. and Schweigert, Igor V.},
    title = {Ab initio density functional theory: {The} best of both worlds?},
  journal = {J. Chem. Phys.},
    volume = {123},
    number = {6},
    pages = {062205},
    year = {2005},
    month = {08},
    issn = {0021-9606},
    doi = {10.1063/1.1904585},
}

@article{Bartlett2006,
title = {Ab initio {DFT}: Getting the right answer for the right reason},
journal = {J. Mol. Struct.: THEOCHEM},
volume = {771},
number = {1},
pages = {1-8},
year = {2006},
note = {Modelling Structure and Reactivity: the 7th triennial conference of the World Association of Theoritical and Computational Chemists (WATOC 2005)},
issn = {0166-1280},
doi = {https://doi.org/10.1016/j.theochem.2006.02.004},
author = {Rodney J. Bartlett and Igor V. Schweigert and Victor F. Lotrich},
}

@article{Grabowski2002,
    author = {Grabowski, Ireneusz and Hirata, So and Ivanov, Stanislav and Bartlett, Rodney J.},
    title = {Ab initio density functional theory: {OEP-MBPT(2)}. {A} new orbital-dependent correlation functional},
  journal = {J. Chem. Phys.},
    volume = {116},
    number = {11},
    pages = {4415-4425},
    year = {2002},
    month = {03},
    issn = {0021-9606},
    doi = {10.1063/1.1445117},
}

@article{Grabowski2007,
    author = {Grabowski, Ireneusz and Lotrich, Victor and Bartlett, Rodney J.},
    title = {Ab initio density functional theory applied to quasidegenerate problems},
  journal = {J. Chem. Phys.},
    volume = {127},
    number = {15},
    pages = {154111},
    year = {2007},
    month = {10},
    issn = {0021-9606},
    doi = {10.1063/1.2790013},
}

@article{Stowasser1999,
author = {Stowasser, Ralf and Hoffmann, Roald},
title = {What Do the Kohn-Sham Orbitals and Eigenvalues Mean?},
journal = {J. Am. Chem. Soc.},
volume = {121},
number = {14},
pages = {3414-3420},
year = {1999},
doi = {10.1021/ja9826892}
}

@Inbook{Perdew1985,
author="Perdew, John P.",
title="What do the Kohn-Sham Orbital Energies Mean? How do Atoms Dissociate?",
bookTitle="Density Functional Methods In Physics",
year="1985",
publisher="Springer US",
address="Boston, MA",
pages="265--308",
isbn="978-1-4757-0818-9",
doi="10.1007/978-1-4757-0818-9_10",
}

@article{Bartlett2009,
  title = {Towards an exact correlated orbital theory for electrons},
  volume = {484},
  ISSN = {0009-2614},
  DOI = {10.1016/j.cplett.2009.10.053},
  number = {1–3},
  journal = {Chem. Phys. Lett.},
  publisher = {Elsevier BV},
  author = {Bartlett,  Rodney J.},
  year = {2009},
  month = dec,
  pages = {1–9}
}

@article{Kim2025,
    author = {Kim, Hyunsik and Perera, Ajith and Mendes, Rodrigo A. and Bartlett, Rodney J.},
    title = {Benchmarking ionization potentials and electron affinities of potential photovoltaic molecules using DFT/QTP functionals and EOM-CC},
    journal = {J. Chem. Phys.},
    volume = {163},
    number = {17},
    pages = {174703},
    year = {2025},
    month = {11},
    issn = {0021-9606},
    doi = {10.1063/5.0293131},
}

@article{AraujoMendes2025,
  title = {Does Correlated Orbital Theory improve PBE-like functionals?},
  url = {http://dx.doi.org/10.26434/chemrxiv-2025-8wvfp-v2},
  DOI = {10.26434/chemrxiv-2025-8wvfp-v2},
  journal = {ChemRxiv},
  author = {Araujo Mendes,  Rodrigo and Windom,  Zachary W. and Haiduke,  Roberto L. A. and Bartlett,  Rodney J.},
  year = {2025},
  month = oct 
}

@article{Lee1989,
author = {Lee, Timothy J. and Taylor, Peter R.},
title = {A diagnostic for determining the quality of single-reference electron correlation methods},
journal = {Int. J. Quantum Chem.},
volume = {36},
number = {S23},
pages = {199-207},
doi = {https://doi.org/10.1002/qua.560360824},
year = {1989}
}

@article{Rettig2020,
author = {Rettig, Adam and Hait, Diptarka and Bertels, Luke W. and Head-Gordon, Martin},
title = {Third-Order {Møller–Plesset} Theory Made More Useful? The Role of Density Functional Theory Orbitals},
journal = {J. Chem. Theory Comput.},
volume = {16},
number = {12},
pages = {7473-7489},
year = {2020},
doi = {10.1021/acs.jctc.0c00986},
}

@article{Mallick2021,
title = {Accurate estimation of singlet-triplet gap of strongly correlated systems by CCSD(T) method using improved orbitals},
journal = {Comput. Theor. Chem.},
volume = {1202},
pages = {113326},
year = {2021},
issn = {2210-271X},
doi = {https://doi.org/10.1016/j.comptc.2021.113326},
author = {Subhasish Mallick and Philips Kumar Rai and Pradeep Kumar},
}

@article{Perdew2001,
    author = {Perdew, John P. and Schmidt, Karla},
    title = {Jacob’s ladder of density functional approximations for the exchange-correlation energy},
    journal = {AIP Conference Proceedings},
    volume = {577},
    number = {1},
    pages = {1-20},
    year = {2001},
    month = {07},
    issn = {0094-243X},
    doi = {10.1063/1.1390175},
}

@article{RodrguezMayorga2021,
  title = {Coupling Natural Orbital Functional Theory and Many-Body Perturbation Theory by Using Nondynamically Correlated Canonical Orbitals},
  volume = {17},
  ISSN = {1549-9626},
  url = {http://dx.doi.org/10.1021/acs.jctc.1c00858},
  DOI = {10.1021/acs.jctc.1c00858},
  number = {12},
  journal = {J. Chem. Theory Comput.},
  publisher = {American Chemical Society (ACS)},
  author = {Rodríguez-Mayorga,  Mauricio and Mitxelena,  Ion and Bruneval,  Fabien and Piris,  Mario},
  year = {2021},
  pages = {7562–7574}
}

@article{Shigeta2001,
author = {Shigeta, Y. and Ferreira, A. M. and Zakrzewski, V. G. and Ortiz, J. V.},
title = {Electron propagator calculations with {Kohn–Sham} reference states},
journal = {Int. J. Quantum Chem.},
volume = {85},
number = {4-5},
pages = {411-420},
doi = {https://doi.org/10.1002/qua.1543},
year = {2001}
}

@article{Li2022,
author = {Li, Jiachen and Yang, Weitao},
title = {Renormalized Singles with Correlation in {GW} Green’s Function Theory for Accurate Quasiparticle Energies},
journal = {J. Phys. Chem. Lett. },
volume = {13},
number = {40},
pages = {9372-9380},
year = {2022},
doi = {10.1021/acs.jpclett.2c02051},
}

@article{Jin2019,
author = {Jin, Ye and Su, Neil Qiang and Yang, Weitao},
title = {Renormalized Singles {Green’s} Function for Quasi-Particle Calculations beyond the $G_{0}W_{0}$ Approximation},
journal = {J. Phys. Chem. Lett. },
volume = {10},
number = {3},
pages = {447-452},
year = {2019},
doi = {10.1021/acs.jpclett.8b03337},
}

@article{Li2021,
author = {Li, Jiachen and Chen, Zehua and Yang, Weitao},
title = {Renormalized Singles {Green’s} Function in the {T}-Matrix Approximation for Accurate Quasiparticle Energy Calculation},
journal = {J. Phys. Chem. Lett. },
volume = {12},
number = {26},
pages = {6203-6210},
year = {2021},
doi = {10.1021/acs.jpclett.1c01723},
}

@article{Rostgaard2010,
  title = {Fully self-consistent {GW} calculations for molecules},
  author = {Rostgaard, C. and Jacobsen, K. W. and Thygesen, K. S.},
  journal = {Phys. Rev. B},
  volume = {81},
  issue = {8},
  pages = {085103},
  numpages = {10},
  year = {2010},
  month = {Feb},
  publisher = {American Physical Society},
  doi = {10.1103/PhysRevB.81.085103},
}

@article{Schilfgaarde2006,
  title = {Quasiparticle Self-Consistent $GW$ Theory},
  author = {van Schilfgaarde, M. and Kotani, Takao and Faleev, S.},
  journal = {Phys. Rev. Lett.},
  volume = {96},
  issue = {22},
  pages = {226402},
  numpages = {4},
  year = {2006},
  month = {Jun},
  publisher = {American Physical Society},
  doi = {10.1103/PhysRevLett.96.226402},
}

@article{Brillouin1932,
  title={Les probl{\`e}mes de perturbations et les champs self-consistents},
  author={Brillouin, L},
  journal={Journal de Physique et le Radium},
  volume={3},
  number={9},
  pages={373--389},
  year={1932}
}

@article{Jiang2015,
author = {Jiang, Hong},
title = {First-principles approaches for strongly correlated materials: A theoretical chemistry perspective},
journal = {Int. J. Quantum Chem.},
volume = {115},
number = {11},
pages = {722-730},
doi = {https://doi.org/10.1002/qua.24905},
year = {2015}
}

@Inbook{Perdew2003,
author="Perdew, John P.",
editor="Gonis, A.
and Kioussis, N.
and Ciftan, M.",
title="Can Density Functional Theory Describe Strongly Correlated Electronic Systems?",
bookTitle="Electron Correlations and Materials Properties 2",
year="2003",
publisher="Springer US",
address="Boston, MA",
pages="237--252",
isbn="978-1-4757-3760-8",
doi="10.1007/978-1-4757-3760-8_13",
}

@article{Shee2021,
    author = {Shee, James and Loipersberger, Matthias and Hait, Diptarka and Lee, Joonho and Head-Gordon, Martin},
    title = {Revealing the nature of electron correlation in transition metal complexes with symmetry breaking and chemical intuition},
    journal = {J. Chem. Phys.},
    volume = {154},
    number = {19},
    pages = {194109},
    year = {2021},
    month = {05},
    issn = {0021-9606},
    doi = {10.1063/5.0047386},
}

@article{Johnson2017,
    author = {Johnson, Erin R. and Becke, Axel D.},
    title = {Communication: {DFT} treatment of strong correlation in 3d transition-metal diatomics},
    journal = {J. Chem. Phys.},
    volume = {146},
    number = {21},
    pages = {211105},
    year = {2017},
    month = {06},
    issn = {0021-9606},
    doi = {10.1063/1.4985084},
}

@article{DeYonker2007,
author = {DeYonker, Nathan J. and Peterson, Kirk A. and Steyl, Gideon and Wilson, Angela K. and Cundari, Thomas R.},
title = {Quantitative Computational Thermochemistry of Transition Metal Species},
journal = {J. Phys. Chem. A},
volume = {111},
number = {44},
pages = {11269-11277},
year = {2007},
doi = {10.1021/jp0715023},
}

@article{Neugebauer2023,
author = {Neugebauer, Hagen and Vuong, Hung T. and Weber, John L. and Friesner, Richard A. and Shee, James and Hansen, Andreas},
title = {Toward Benchmark-Quality \textit{Ab Initio} Predictions for 3d Transition Metal Electrocatalysts: A Comparison of {CCSD(T)} and {ph-AFQMC}},
journal = {J. Chem. Theory Comput.},
volume = {19},
number = {18},
pages = {6208-6225},
year = {2023},
doi = {10.1021/acs.jctc.3c00617},
}

@article{Edgecombe1995,
title = {$Cr_{2}$ in density-functional theory: approximate spin projection},
journal = {Chem. Phys. Lett.},
volume = {244},
number = {5},
pages = {427-432},
year = {1995},
issn = {0009-2614},
doi = {https://doi.org/10.1016/0009-2614(95)00945-Z},
author = {Kenneth E. Edgecombe and Axel D. Becke},
}

@article{Siegbahn2010,
author = {Siegbahn, Per E. M. and Blomberg, Margareta R. A.},
title = {Bond-dissociation using hybrid DFT},
journal = {Int. J. Quantum Chem.},
volume = {110},
number = {2},
pages = {317-322},
doi = {https://doi.org/10.1002/qua.22204},
year = {2010}
}

@article{Siegbahn2000,
author = {Siegbahn, Per E. M. and Blomberg, Margareta R. A.},
title = {Transition-Metal Systems in Biochemistry Studied by High-Accuracy Quantum Chemical Methods},
journal = {Chem. Rev.},
volume = {100},
number = {2},
pages = {421-438},
year = {2000},
doi = {10.1021/cr980390w},
}

@article{Salahub1985,
title = {{LCAO}-local-spin-density calculations for $V_{2}$ and $Mn_{2}$},
journal = {Surf. Sci.},
volume = {156},
pages = {605-614},
year = {1985},
issn = {0039-6028},
doi = {0.1016/0039-6028(85)90231-6},
author = {D.R. Salahub and N.A. Baykara},
}

@article{Delley1983,
  title = {Metal-Metal Bonding in {Cr-Cr} and {Mo-Mo} Dimers: {Another} Success of Local Spin-Density Theory},
  author = {Delley, B. and Freeman, A. J. and Ellis, D. E.},
  journal = {Phys. Rev. Lett.},
  volume = {50},
  issue = {7},
  pages = {488--491},
  numpages = {0},
  year = {1983},
  month = {Feb},
  publisher = {American Physical Society},
  doi = {10.1103/PhysRevLett.50.488},
}

@article{Purwanto2015,
    author = {Purwanto, Wirawan and Zhang, Shiwei and Krakauer, Henry},
    title = {An auxiliary-field quantum {Monte Carlo} study of the chromium dimer},
    journal = {J. Chem. Phys.},
    volume = {142},
    number = {6},
    pages = {064302},
    year = {2015},
    month = {02},
    issn = {0021-9606},
    doi = {10.1063/1.4906829},
}

@article{RuizDiaz2010,
  title = {Magnetism of small Cr clusters: Interplay between structure, magnetic order, and electron correlations},
  author = {Ruiz-D\'{\i}az, P. and Ricardo-Ch\'avez, J. L. and Dorantes-D\'avila, J. and Pastor, G. M.},
  journal = {Phys. Rev. B},
  volume = {81},
  issue = {22},
  pages = {224431},
  numpages = {11},
  year = {2010},
  month = {Jun},
  publisher = {American Physical Society},
  doi = {10.1103/PhysRevB.81.224431},
}

@article{Goodgame1982,
  title = {Nature of Mo-Mo and Cr-Cr Multiple Bonds: A Challenge for the Local-Density Approximation},
  author = {Goodgame, Marvin M. and Goddard, William A.},
  journal = {Phys. Rev. Lett.},
  volume = {48},
  issue = {3},
  pages = {135--138},
  numpages = {0},
  year = {1982},
  month = {Jan},
  publisher = {American Physical Society},
  doi = {10.1103/PhysRevLett.48.135},
}

@article{Hongo2012,
author = {Hongo, Kenta and Maezono, Ryo},
title = {A benchmark quantum {Monte Carlo} study of the ground state chromium dimer},
journal = {Int. J. Quantum Chem.},
volume = {112},
number = {5},
pages = {1243-1255},
doi = {https://doi.org/10.1002/qua.23113},
year = {2012}
}

@article{Muller2009,
author = {M{\"u}ller, Thomas},
title = {Large-Scale Parallel Uncontracted Multireference-Averaged Quadratic Coupled Cluster: The Ground State of the Chromium Dimer Revisited},
journal = {J. Phys. Chem. A },
volume = {113},
number = {45},
pages = {12729-12740},
year = {2009},
doi = {10.1021/jp905254u},
}

@article{Celani2004,
author = {P. Celani and H. Stoll and H.-J. Werner and P.J. Knowles *},
title = {The {CIPT2} method: Coupling of multi-reference configuration interaction and multi-reference perturbation theory. Application to the chromium dimer},
journal = {Mol. Phys.},
volume = {102},
number = {21-22},
pages = {2369--2379},
year = {2004},
publisher = {Taylor \& Francis},
doi = {10.1080/00268970412331317788},
}

@article{Vancoillie2016,
author = {Vancoillie, Steven and Malmqvist, Per Ake and Veryazov, Valera},
title = {Potential Energy Surface of the Chromium Dimer Re-re-revisited with Multiconfigurational Perturbation Theory},
journal = {J. Chem. Theory Comput. },
volume = {12},
number = {4},
pages = {1647-1655},
year = {2016},
doi = {10.1021/acs.jctc.6b00034},
}

@article{Weflen2025,
author = {Weflen, Kaila E. and Bentley, Megan R. and Thorpe, James H. and Franke, Peter R. and Martin, Jan M. L. and Matthews, Devin A. and Stanton, John F.},
title = {Exploiting a Shortcoming of Coupled-Cluster Theory: The Extent of Non-Hermiticity as a Diagnostic Indicator of Computational Accuracy},
journal = {J. Phys. Chem. Lett.},
volume = {16},
number = {20},
pages = {5121-5127},
year = {2025},
doi = {10.1021/acs.jpclett.5c00885},
}

@Inbook{Harvey2004,
author="Harvey, Jeremy N.",
title="{DFT} Computation of Relative Spin-State Energetics of Transition Metal Compounds",
bookTitle="Principles and Applications of Density Functional Theory in Inorganic Chemistry I",
year="2004",
publisher="Springer Berlin Heidelberg",
address="Berlin, Heidelberg",
pages="151--184",
doi="10.1007/b97939",
}

@Article{Harvey2006,
author ="Harvey, Jeremy N.",
title  ="On the accuracy of density functional theory in transition metal chemistry",
journal  ="Annu. Rep. Prog. Chem.{,} Sect. C: Phys. Chem.",
year  ="2006",
volume  ="102",
issue  ="0",
pages  ="203-226",
publisher  ="The Royal Society of Chemistry",
doi  ="10.1039/B419105F",
}

@article{Swart2016,
author = {Swart, Marcel and Gruden, Maja},
title = {Spinning around in Transition-Metal Chemistry},
journal = {Acc. Chem. Res.},
volume = {49},
number = {12},
pages = {2690-2697},
year = {2016},
doi = {10.1021/acs.accounts.6b00271},
}
\bibliographystyle{naturemag}



\section*{Supplementary material}

# Supporting Information: "Kohn-Sham density encoding rescues coupled cluster theory for strongly correlated molecules"

Abdulrahman Y. Zamani[1], Barbaro Zulueta[2], Andrew M. Ricciuti[1], John A. Keith[2,*], Kevin Carter-Fenk[1,*]

[1]Department of Chemistry, University of Pittsburgh, Pittsburgh, Pennsylvania 15260, USA
[2]Department of Chemical and Petroleum Engineering, University of Pittsburgh, Pittsburgh, Pennsylvania, 15213, USA

*E-mail: jakeith@pitt.edu; kay.carter-fenk@pitt.edu

In the following sections, we provide additional data and computational details for obtaining the quantities discussed in this manuscript. Unless otherwise stated, DFT and other electronic structure calculations are performed as implemented in the codes employed.

## S1  General Computational Details

We performed approximate KS-DFT and CCSD(T) calculations with ORCA 6.0.[1–5] Geometry optimizations employed SVWN5,[6] PBE,[7, 8] PW91,[9, 10] r$^2$SCAN,[11, 12] PBE0,[13] B3LYP,[14] $\omega$B97X-V,[15] and $\omega$B97M-V[16] functionals for all dimers. For first-row transition-metal systems, we used the exact-two-component (X2C) scalar relativistic approximation[17, 18] with the x2c-TZVPall[19] basis set.

We computed CCSD(T) single-point energies at experimental equilibrium distances with HF, SVWN5, PBE, PW91, r$^2$SCAN, and PBE0 reference densities. We obtained CCSD(T)/CBS energies via two-point extrapolation of SCF[20, 21] and correlation energies[22, 23] using def2-TZVPP[24] and def2-QZVPP[25] basis sets (*vide infra*). For all dimers except Mn$_2$, we used the X2C relativistic approximation with the segmented all-electron relativistically contracted (SARC) basis sets for Coulomb fitting,[26] decontracted auxiliary sets for Coulomb and exchange fitting, and x2c-TZVPall for correlation fitting.[27] For Mn$_2$, we followed the same procedure but with the ZORA[28, 29] scalar relativistic approximation and the def2-TZVP basis set for correlation fitting. For Mn, we performed two single-point energy calculations with CC-PBE0 using the ZORA and X2C relativistic approximations and the same basis sets and procedures described above. The ZORA-based energy was used for calculating the BDE of Mn$_2$, while the X2C-based energy was used for Mn−H, Mn−H$^+$, Mn−Cl, and Mn−O. All calculations used tight SCF convergence and the finite Gaussian-nucleus model for relativistic treatments.[30]

We also performed potential energy surface (PES) scans for Cr$_2$ using the same CCSD(T)/CBS and KS-DFT protocols (*vide supra*), except under spin-polarization. We evaluated KS-DFT PES data with def2-QZVPP and compared them directly to CCSD(T)/CBS results. The Cr$_2$ PES contained 84–124 points and used the same convergence criteria and auxiliary basis sets as in the CCSD(T) and KS-DFT calculations.

1

We define the dissociation energy for each dimer ($D_e$) as

$$D_e = (E_{\text{M·}} + E_{\text{L·}}) - E_{\text{M-L}} + \Delta E_{\text{SO}}, \tag{S1}$$

where $E_{\text{M·}}$ and $E_{\text{L·}}$ are the ground-state atomic energies, $E_{\text{M-L}}$ is the ground-state dimer energy, and $\Delta E_{\text{SO}}$ is the spin–orbit (SO) correction. $\Delta E_{\text{SO}}$ can be obtained from experiment[31, 32] or first-principles calculations,[33] where

$$\Delta E_{\text{SO}} = (E_{\text{M·,SO}} + E_{\text{L·,SO}}) - E_{\text{M-L,SO}}, \tag{S2}$$

with $E_{\text{M·,SO}}$ and $E_{\text{L·,SO}}$ the atomic SO correction energies and $E_{\text{M-L,SO}}$ the dimer SO correction energy. Note that when calculating the $D_e$ of M–H and M–H$^+$, we took the H energy from the two-point extrapolated UHF energy using the procedures described above.

We treated several dimers with spin-polarized KS-DFT and CCSD(T)/CBS. These include Cr$_2$ and Mn$_2$, which have singlet antiferromagnetic ground states, and Ni$_2$, which exhibits mixed singlet-triplet character; spin-polarization improves the accuracy of the BDEs for these singlet states (*vide infra*).

All ORCA output files, including KS-DFT optimized geometries, CCSD(T)/CBS single points at reference bond lengths, atomic energies, and Cr$_2$ PES scans, are freely available on Zenodo. Jupyter notebooks that parse the ORCA outputs and generate the plots shown in this study are available on GitHub. Spectroscopic constants, reference bond lengths, reference bond dissociation energies, and reference SO values are provided below.

## S2 Details of the Complete Basis Set (CBS) Extrapolation for CCSD(T)

To obtain CCSD(T)/CBS energies, we employ a two-point extrapolation scheme for the self-consistent energy ($E_{\text{SCF}}$) and correlation energy ($E_{\text{corr}}$) using triple and quadruple zeta basis sets. We outline the relevant equations and parameters below. The extrapolated self-consistent energy ($E_{\text{SCF}}^\infty$) follows the approaches of Martin[20] and Petersson:[21]

$$E_{\text{SCF}}^\infty = \frac{E_{\text{SCF},X} e^{-\alpha\sqrt{Y}} - E_{\text{SCF},Y} e^{-\alpha\sqrt{X}}}{e^{-\alpha\sqrt{Y}} - e^{-\alpha\sqrt{X}}}, \tag{S3}$$

where $X$ and $Y$ are the cardinal numbers of the basis sets (e.g. TZ = 3, QZ = 4); $E_{\text{SCF},X}$ and $E_{\text{SCF},Y}$ are the self-consistent energies for each basis set; and $\alpha$ is an empirical parameter dependent on the family of basis functions and the cardinal numbers of the basis set pair. The extrapolated correlation energy ($E_{\text{corr}}^\infty$) follows Truhlar:[22]

$$E_{\text{corr}}^\infty = \frac{X^\beta E_{\text{corr},X} - Y^\beta E_{\text{corr},Y}}{X^\beta - Y^\beta}, \tag{S4}$$

where $\beta$ is an empirical parameter dependent on the family of basis functions and the cardinal numbers of the basis set pair, and $E_{\text{corr},X}$ and $E_{\text{corr},Y}$ are the correlation energies for each basis set. Finally, the total CCSD(T)/CBS energy ($E_{\text{CCSD(T)}}^\infty$) is

$$E_{\text{CCSD(T)}}^\infty = E_{\text{SCF}}^\infty + E_{\text{corr}}^\infty. \tag{S5}$$

We employed the Ahlrichs def2-TZVPP[24] and def2-QZVPP[25] basis sets for all CCSD(T) calculations, corresponding to cardinal numbers $X = 3$ and $Y = 4$, respectively. For this basis set pair, the values of $\alpha$ and $\beta$ are 7.880 and 2.970, respectively, as given in Ref. 23.

## S3 Spectroscopic Term Symbols of the Dimers and Atoms

Table S1 lists the ground-state spectroscopic term symbols for the M–H, M–Cl, M–O, M–M, and M–H$^+$ dimers. We took term symbols for M–H, M–O, and M–Cl from Ref. 33, and those for M–H$^+$ from Refs. 34, 35. For M–M, we determined several ground-state term symbols from previous work by Wilson's group[36], except for Ti$_2$, Mn$_2$, Fe$_2$, and Ni$_2$.

*Ab initio* methods with sufficiently large active spaces have established the ground states of Ti$_2$[37] and Fe$_2$[38] to be $^3\Delta_g$ and $^9\Sigma_u^-$, respectively. For Mn$_2$, the ground state is an antiferromagnetic $^1\Sigma_g^+$ state based on experiments and *ab initio* methods.[39–41] For Ni$_2$, the ground state is difficult to assign, as different *ab initio* methods produce mixed singlet-triplet character.[42] In our work, we assigned the ground state for each method as either $^1\Sigma_g^+$ or $^3\Sigma_g^-$. Spin-polarized r$^2$SCAN, PBE0, $\omega$B97X-V, $\omega$B97M-V, CC-r$^2$SCAN, and CC-PBE0 yield $^1\Sigma_g^+$ as the lower-energy state, while the other methods yield $^3\Sigma_g^-$.

Table S1: Ground-state spectroscopic term symbols for the dimers.

| Dimer | Term Symbol | Dimer | Term Symbol | Dimer | Term Symbol | Dimer | Term Symbol | Dimer | Term Symbol |
|---|---|---|---|---|---|---|---|---|---|
| Sc–H | $^1\Sigma^+$ | Sc–Cl | - | Sc–O | $^2\Sigma^+$ | Sc–Sc | $^5\Sigma_u^-$ | Sc–H$^+$ | $^2\Delta$ |
| Ti–H | $^4\Phi$ | Ti–Cl | $^4\Phi$ | Ti–O | $^3\Delta$ | Ti–Ti | $^3\Delta_g$ | Ti–H$^+$ | $^3\Phi$ |
| V–H | $^5\Delta$ | V–Cl | $^5\Delta$ | V–O | $^4\Sigma^-$ | V–V | $^3\Sigma_g^-$ | V–H$^+$ | $^4\Delta$ |
| Cr–H | $^6\Sigma^+$ | Cr–Cl | $^6\Sigma^+$ | Cr–O | $^5\Pi$ | Cr–Cr | $^1\Sigma^+$ | Cr–H$^+$ | $^5\Sigma^+$ |
| Mn–H | $^7\Sigma^+$ | Mn–Cl | $^7\Sigma^+$ | Mn–O | $^6\Sigma^+$ | Mn–Mn | $^1\Sigma_g^+$ | Mn–H$^+$ | $^6\Sigma^+$ |
| Fe–H | $^4\Delta$ | Fe–Cl | $^6\Delta$ | Fe–O | $^5\Delta$ | Fe–Fe | $^9\Sigma_u^-$ | Fe–H$^+$ | $^5\Delta$ |
| Co–H | $^3\Phi$ | Co–Cl | $^3\Phi$ | Co–O | $^4\Delta$ | Co–Co | $^5\Delta_g$ | Co–H$^+$ | $^4\Phi$ |
| Ni–H | $^2\Delta$ | Ni–Cl | $^2\Pi$ | Ni–O | $^3\Sigma^-$ | Ni–Ni | $^1\Sigma_g^+$ or $^3\Sigma_g^-$ | Ni–H$^+$ | $^3\Delta$ |
| Cu–H | $^1\Sigma^+$ | Cu–Cl | $^1\Sigma^+$ | Cu–O | $^2\Pi$ | Cu–Cu | $^1\Sigma_g^+$ | Cu–H$^+$ | $^2\Sigma^+$ |
| Zn–H | $^2\Sigma^+$ | Zn–Cl | $^2\Sigma^+$ | Zn–O | $^1\Sigma^+$ | Zn–Zn | $^1\Sigma_g^+$ | Zn–H$^+$ | $^1\Sigma^+$ |

Table S2 lists the ground-state term symbols for each atom. We took metal cation ground states from Ref. 35, and neutral metal and ligand atom term symbols from Ref. 33.

Table S2: Ground-state spectroscopic term symbols for the atoms.

| Atom | Term Symbol | Atom | Term Symbol | Atom | Term Symbol |
|---|---|---|---|---|---|
| Sc | $^2D$ | Sc$^+$ | $^3D$ | H | $^2S$ |
| Ti | $^3F$ | Ti$^+$ | $^4F$ | O | $^3P$ |
| V | $^4F$ | V$^+$ | $^5D$ | Cl | $^2P$ |
| Cr | $^7S$ | Cr$^+$ | $^6S$ | | |
| Mn | $^6S$ | Mn$^+$ | $^7S$ | | |
| Fe | $^5D$ | Fe$^+$ | $^6D$ | | |
| Co | $^4F$ | Co$^+$ | $^3F$ | | |
| Ni | $^3F$ | Ni$^+$ | $^2D$ | | |
| Cu | $^2S$ | Cu$^+$ | $^1S$ | | |
| Zn | $^1S$ | Zn$^+$ | $^2S$ | | |

## S4 Reference Equilibrium Bond Distances and Dissociation Energies

Table S3 lists reference equilibrium bond distances ($r_e$) and dissociation energies ($D_e$) for each dimer. For several M–H and M–H$^+$ species, experimental $r_e$ values were either unavailable or inconsistent with CCSD(T)-optimized geometries reported in Ref. 33; in these cases, we used

B3LYP+X2C/TZVP optimized geometries for the CCSD(T) calculations (values in parentheses). We obtained reference $D_e$ values from experiment or *ab initio* calculations, as indicated. For M–H$^+$ dimers, we computed $D_e$ as $D_e = D_0 + E_{\text{ZPE}}$, where $D_0$ is the experimental bond dissociation energy at 0 K and $E_{\text{ZPE}}$ is the zero-point energy computed at the BP86/def2-TZVPP level of theory.[35] Table S4 lists the $D_0$ and $E_{\text{ZPE}}$ values for M–H$^+$.

Table S3: Reference equilibrium bond distances ($r_e$, in Å) and bond dissociation energies ($D_e$, in kcal/mol). Values in parentheses are $r_e$ from B3LYP+X2C/TZVP optimized geometries, used when experimental values were unavailable or inconsistent with DFT $r_e$.

| Dimer | $r_e$ | $D_e$ | Ref. | Dimer | $r_e$ | $D_e$ | Ref. | Dimer | $r_e$ | $D_e$ | Ref. | Dimer | $r_e$ | $D_e$ | Ref. | Dimer | $r_e$ | $D_e$ | Ref. |
|---|---|---|---|---|---|---|---|---|---|---|---|---|---|---|---|---|---|---|---|
| Sc–H | 1.77540(1.75011) | 50 ± 4 | 33 | Sc–Cl | - | - | - | Sc–O | 1.6661 | 161.0 ± 0.2 | 33 | Sc–Sc | 2.7000 | 11.00$^a$ | 43 | Sc–H$^+$ | (1.74314) | 57.68 ± 2.31 | 44 |
| Ti–H | 1.777 | 47.4 ± 1.4 | 33 | Ti–Cl | 2.2697 | 101.0 ± 2.0 | 33 | Ti–O | 1.6203 | 159.9 ± 1.6 | 33 | Ti–Ti | 1.9422 | 36.09 ± 4.15 | 45 | Ti–H$^+$ | (1.70419) | 57.58 ± 2.31 | 44 |
| V–H | 1.7300(1.6840) | 56.1 ± 1.5 | 33, 46 | V–Cl | 2.2145 | 102.5 ± 2.0 | 33 | V–O | 1.5893 | 150.9 ± 2.0 | 33 | V–V | 1.7660 | 64.25 | 32 | V–H$^+$ | (1.63734) | 50.30 ± 2.08 | 47 |
| Cr–H | 1.6554 | 51.1 ± 1.6 | 33 | Cr–Cl | 2.194 | 90.6 ± 1.3 | 33 | Cr–O | 1.615 | 104.8 ± 1.4 | 33 | Cr–Cr | 1.6788 | 35.98 ± 1.38 | 48 | Cr–H$^+$ | (1.59916) | 34.12 ± 2.08 | 49 |
| Mn–H | 1.7309 | 41.1 ± 1.4 | 33 | Mn–Cl | 2.2352 | 80.4 ± 1.3 | 33 | Mn–O | 1.6446 | 89.3 ± 1.8 | 33 | Mn–Mn | 3.4000 | 3.00 ± 2.31 | 39–41 | Mn–H$^+$ | (1.60674) | 50.06 ± 3.46 | 44 |
| Fe–H | 1.606(1.54080) | 44.8 ± 1.5 | 33, 46 | Fe–Cl | 2.1742 | 79.9 ± 1.3 | 33 | Fe–O | 1.6164 | 97.6 ± 0.2 | 33 | Fe–Fe | 2.0200 | 18.22$^b$ | 38 | Fe–H$^+$ | (1.55069) | 51.51 ± 1.38 | 44 |
| Co–H | 1.5327 | 54.6 ± 1.5 | 33, 46 | Co–Cl | 2.0656 | 81.9 ± 1.3 | 33 | Co–O | 1.6286 | 95.3 ± 2.1 | 33 | Co–Co | 2.1300 | 39.39 ± 6.00 | 45 | Co–H$^+$ | (1.50129) | 48.30 ± 1.38 | 44 |
| Ni–H | 1.4538 | 59.2 ± 1.9 | 33 | Ni–Cl | 2.0615 | 88.7 ± 1.3 | 33 | Ni–O | 1.6271 | 90.3 ± 0.7 | 33 | Ni–Ni | 2.1540 | 47.60 ± 0.20 | 31, 45 | Ni–H$^+$ | (1.47411) | 41.46 ± 1.84 | 44 |
| Cu–H | 1.4626(1.466) | 63.1 ± 4.8 | 33, 46 | Cu–Cl | 2.0512 | 90.2 ± 1.0 | 33 | Cu–O | 1.7246 | 71.0 ± 0.7 | 33 | Cu–Cu | 2.2193 | 47.92 ± 0.57 | 45, 50 | Cu–H$^+$ | (1.47738) | 31.48 | 51 |
| Zn–H | 1.5935 | 21.9 ± 0.2 | 33 | Zn–Cl | 2.1300 | 54.1 ± 1.0 | 33, 46 | Zn–O | 1.7047 | 38.2 ± 0.9 | 33 | Zn–Zn | 4.8000 | 0.78 ± 0.0001 | 52 | Zn–H$^+$ | (1.52109) | 60.76 | 53 |

$^a$ From $D_e \approx \frac{\omega_e}{4\omega_e x_e} - \Delta E \left[ \text{Sc} \left( ^2D \right) \to \text{Sc} \left( ^4F \right) \right]$ for $\text{Sc}_2 \to \text{Sc} \left( ^2D \right) + \text{Sc} \left( ^2D \right)$.
$^b$ RASPT2(16,12;2,16)/ANO-RCC-VQZVP[54–56] value from Ref. 38.

Table S4: Experimental bond dissociation energies and zero-point energies from Ref.35 for first-row transition metal hydride cations (kcal/mol). Table S3 lists the references for each dimer's $D_0$.

| Dimer | $D_0$ | $E_{\text{ZPE}}$ |
|---|---|---|
| ScH$^+$ | 55.34 ± 2.31 | 2.33 |
| TiH$^+$ | 55.11 ± 2.31 | 2.47 |
| VH$^+$ | 47.73 ± 2.08 | 2.56 |
| CrH$^+$ | 31.59 ± 2.08 | 2.53 |
| MnH$^+$ | 47.50 ± 3.46 | 2.55 |
| FeH$^+$ | 48.89 ± 1.38 | 2.62 |
| CoH$^+$ | 45.66 ± 1.38 | 2.65 |
| NiH$^+$ | 38.74 ± 1.84 | 2.72 |
| CuH$^+$ | 29.06 ± 0.00 | 2.42 |
| ZnH$^+$ | 58.11 ± 0.00 | 2.65 |

## S5 Reference Spin-Orbit Coupling Corrections

Table S5 lists spin-orbit coupling corrections ($\Delta E_{\text{SO}}$) for the M–H, M–Cl, M–O, M–M, and M–H$^+$ dimers, calculated according to Eq. S2. We took values from experimental data[31, 32, 46] or first-principles calculations.[33, 35]

Table S5: Spin-orbit coupling corrections ($\Delta E_{\text{SO}}$, in kcal/mol) for each dimer.

| Dimer | $\Delta E_{\text{SO}}$ | Ref. | Dimer | $\Delta E_{\text{SO}}$ | Ref. | Dimer | $\Delta E_{\text{SO}}$ | Ref. | Dimer | $\Delta E_{\text{SO}}$ | Ref. | Dimer | $\Delta E_{\text{SO}}$ | Ref. |
|---|---|---|---|---|---|---|---|---|---|---|---|---|---|---|
| Sc–H | −0.33 | 33 | Sc–Cl | - | - | Sc–O | −0.55 | 33 | Sc–Sc | 0.00 | - | Sc–H$^+$ | 0.05 | 35 |
| Ti–H | −0.19 | 33 | Ti–Cl | −0.88 | 33 | Ti–O | −0.57 | 33 | Ti–Ti | 0.00 | - | Ti–H$^+$ | 0.07 | 35 |
| V–H | −0.40 | 46 | V–Cl | −1.21 | 33 | V–O | −1.17 | 33 | V–V | −1.82 | 32 | V–H$^+$ | 0.07 | 35 |
| Cr–H | 0.00 | 33 | Cr–Cl | −0.76 | 33 | Cr–O | 0.10 | 33 | Cr–Cr | 0.00 | 48 | Cr–H$^+$ | 0.05 | 35 |
| Mn–H | 0.00 | 33 | Mn–Cl | −0.76 | 33 | Mn–O | −0.22 | 33 | Mn–Mn | 0.00 | - | Mn–H$^+$ | 0.10 | 35 |
| Fe–H | −0.02 | 32 | Fe–Cl | −0.60 | 33 | Fe–O | −0.17 | 33 | Fe–Fe | 0.00 | - | Fe–H$^+$ | 0.19 | 35 |
| Co–H | −0.22 | 46 | Co–Cl | −1.01 | 33 | Co–O | −0.94 | 33 | Co–Co | 0.00 | - | Co–H$^+$ | 0.07 | 35 |
| Ni–H | −1.09 | 33 | Ni–Cl | −3.00 | 33 | Ni–O | −3.10 | 33 | Ni–Ni | −5.56 | 31 | Ni–H$^+$ | −0.53 | 35 |
| Cu–H | 0.00 | 33 | Cu–Cl | −0.76 | 33 | Cu–O | 0.10 | 33 | Cu–Cu | 0.00 | - | Cu–H$^+$ | 0.60 | 35 |
| Zn–H | 0.00 | 33 | Zn–Cl | −0.76 | 46 | Zn–O | −0.22 | 33 | Zn–Zn | 0.00 | - | Zn–H$^+$ | 0.14 | 35 |

## S6 Potential Energy Curves for Cr$_2$

Figure S1 shows the potential energy surface of the antiferromagnetic Cr$_2$ dimer using spin-polarized KS-DFT or HF/def2-QZVPP and CCSD(T)/CBS with multiple reference densities. None of the DFT methods accurately describes the Cr$_2$ potential: USVWN5 massively overbinds ($-3.26$ eV), and UHF shows no binding minimum. However, UCCSD(T)/CBS with GGA orbitals recovers the correct physics. Notably, the quality of the KS-DFT potential does not predict UCCSD(T)/CBS performance: UPBE and UPW91 yield poor potentials yet provide excellent UCCSD(T)/CBS references. This indicates that the structure of the Fock matrix and the quality of the KS-DFT densities determine the accuracy of coupled-cluster calculations.

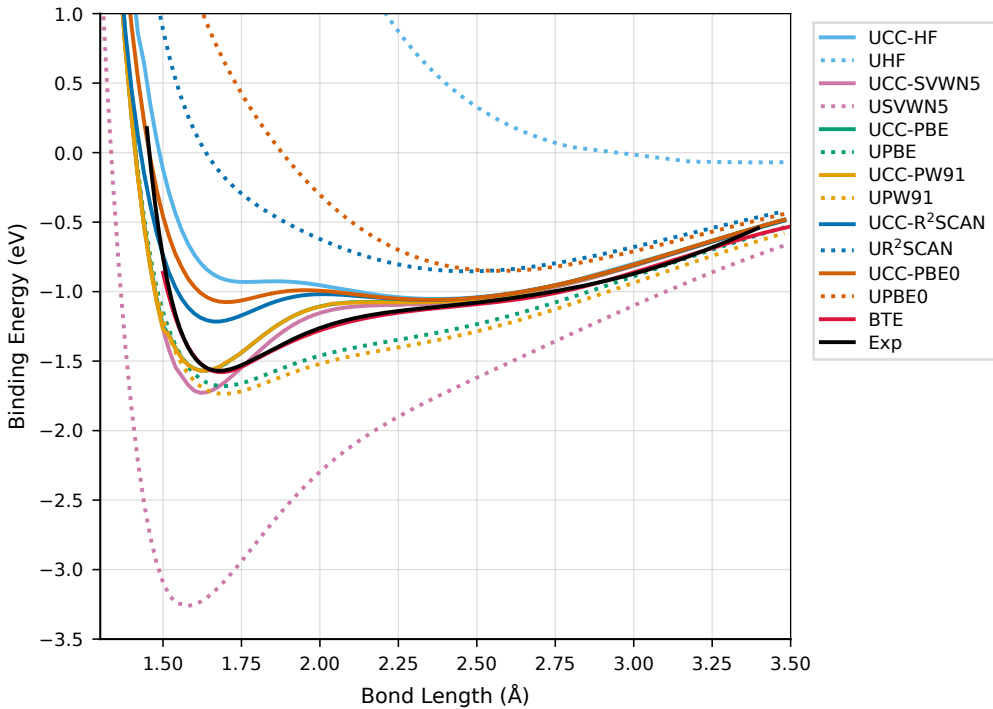

Figure S1: Potential energy curves for the Cr$_2$ dimer computed with UCCSD(T)/CBS (solid lines) and UKS-DFT/def2-QZVPP (dotted lines) using various reference methods. We took the experimental (Exp) and best theoretical estimate (BTE) curves from Ref. 48.

Table S6 presents calculated equilibrium bond lengths and dissociation energies for Cr$_2$ obtained from the minima of the potential energy curves (Figure S1) using HF, KS-DFT, and CCSD(T) methods. Among the DFT functionals, UPBE and UPW91 yield $r_e$ values in reasonable agreement with experiment, though they slightly overestimate $D_e$. In contrast, UHF and the hybrid functional UPBE0 severely overestimate the bond length and underestimate the dissociation energy, reflecting the well-known challenges of describing the multireference character of Cr$_2$. The CCSD(T)/CBS results show a strong dependence on the choice of reference orbitals: UCC-HF exhibits significant errors in both $r_e$ and $D_e$ accompanied by substantial spin contamination ($\langle S^2 \rangle = 5.494$), whereas UCC-PBE and UCC-PW91 provide excellent agreement with experiment, with deviations of only $-0.037$ Å in $r_e$ and $-0.001$ eV in $D_e$.

Table S6: Calculated equilibrium bond lengths ($r_e$) and dissociation energies ($D_e$) for $Cr_2$ using various electronic structure methods. Values correspond to the bottom of the potential energy curve; parentheses show deviations from experiment. We extrapolated CCSD(T) energies to the complete basis set (CBS) limit and computed $T_1$ diagnostic and $\langle S^2 \rangle$ values at the initial reference/def2-QZVPP theory.

| Method | $r_e$ (Å) | $D_e$ (eV) | $T_1$/QZ | $\langle S^2 \rangle$/QZ |
|---|---|---|---|---|
| HF and KS-DFT | | | | |
| UHF | 3.380 (+1.703) | 0.071 (−1.499) | | |
| USVWN5 | 1.580 (−0.097) | 3.262 (+1.692) | | |
| UPBE | 1.700 (+0.023) | 1.679 (+0.109) | | |
| UPW91 | 1.700 (+0.023) | 1.735 (+0.165) | | |
| UR$^2$SCAN | 2.480 (+0.803) | 0.854 (−0.716) | | |
| UPBE0 | 2.580 (+0.903) | 0.851 (−0.719) | | |
| CCSD(T)/CBS | | | | |
| UCC-HF | 2.380 (+0.703) | 1.053 (−0.517) | 0.0388 | 5.494 |
| UCC-SVWN5 | 1.620 (−0.057) | 1.729 (+0.159) | 0.0287 | 0.712 |
| UCC-PBE | 1.640 (−0.037) | 1.569 (−0.001) | 0.0234 | 1.291 |
| UCC-PW91 | 1.640 (−0.037) | 1.569 (−0.001) | 0.0235 | 1.329 |
| UCC-R$^2$SCAN | 1.680 (+0.003) | 1.215 (−0.355) | 0.0238 | 2.578 |
| UCC-PBE0 | 1.700 (+0.023) | 1.075 (−0.495) | 0.0221 | 3.099 |
| Multireference | | | | |
| BTE[a] | 1.686 (+0.009) | 1.580 (+0.010) | | |
| Experiment[a] | 1.677 | 1.570 | | |

[a]See Ref. 48.

## S7 Spark Plots of $D_e$ for Metal Dimers

Figure S2 presents a systematic comparison of bond dissociation energies (BDEs) across 49 first-row transition metal systems, contrasting CCSD(T)/CBS calculations performed with various reference determinants against standard Kohn-Sham DFT methods. The benchmark comprises a test set of 29 systems (M−H, M−Cl, and M−O bonds) for evaluating method performance, and a validation set of 20 systems (M−H$^+$ and M−M bonds) for assessing transferability across more diverse bonding situations.

**Test Set (M−H, M−Cl, M−O).** CCSD(T)/CBS with a conventional Hartree-Fock reference (CC-HF) yields an MAE of 0.30 eV with a maximum error of 2.37 eV (Cu−O). All DFT-referenced methods achieve comparable accuracy: CC-PBE (MAE = 0.14 eV, MAX = 0.40 eV), CC-PW91 (MAE = 0.14 eV, MAX = 0.39 eV), CC-r$^2$SCAN (MAE = 0.14 eV, MAX = 0.54 eV), and CC-PBE0 (MAE = 0.15 eV, MAX = 0.50 eV). The LDA-based CC-SVWN5 shows intermediate performance (MAE = 0.23 eV, MAX = 1.12 eV). Among individual bond types, metal oxides prove most challenging for CC-HF (MAE = 0.43 eV) due to failures on Fe−O (+0.85 eV) and Cu−O (−2.37 eV), while CC-PBE0 achieves the lowest MAE of 0.11 eV. Metal chlorides favor CC-r$^2$SCAN (MAE = 0.09 eV), and metal hydrides show similar performance across all DFT references (MAE = 0.17–0.19 eV). Catastrophic failures (|error| > 1 eV) in the test set occur for CC-HF (Cu−O, Fe−Cl) and CC-SVWN5 (Fe−O).

**Validation Set (M−H$^+$, M−M).** The validation set reveals stark differences between refer-

ence types. CC-HF exhibits an MAE of 0.95 eV with a maximum error of 10.69 eV, driven primarily by catastrophic failures on the homonuclear dimers. In contrast, CC-PW91 maintains excellent accuracy (MAE = 0.11 eV, MAX = 0.37 eV), as does CC-PBE (MAE = 0.14 eV, MAX = 0.51 eV). The meta-GGA reference CC-r$^2$SCAN shows degraded performance on the validation set (MAE = 0.41 eV, MAX = 3.76 eV) due to a catastrophic failure on Mn–Mn, while CC-PBE0 remains robust (MAE = 0.22 eV, MAX = 0.87 eV). For the M–H$^+$ series specifically, all methods perform comparably (MAE = 0.11–0.22 eV), though CC-HF fails for V–H$^+$ (−1.10 eV). The homonuclear dimers discriminate most strongly: CC-HF yields MAE = 1.68 eV with five systems exceeding 1 eV error (V–V, Mn–Mn, Fe–Fe, Co–Co), whereas CC-PW91 achieves MAE = 0.12 eV with no catastrophic failures. The exceptions are Cu–Cu and Zn–Zn, where CC-HF performs excellently (+0.09 eV and −0.009 eV error, respectively), reflecting the filled or nearly-filled $d^{10}$ configurations that eliminate the multireference character plaguing other homonuclear dimers. Additional validation set failures include CC-SVWN5 on Fe–Fe (−2.03 eV) and CC-r$^2$SCAN on Mn–Mn (−3.76 eV).

**Head-to-Head Comparison.** At least one CC-DFT method outperforms CC-HF in 62% of test set systems and 85% of validation set systems. By functional class, GGA references (PBE/PW91) provide the lowest error for 12 test set systems and 5 validation set systems, excelling for iron-containing species where CC-HF shows large errors. The hybrid CC-PBE0 reference proves optimal for 8 test set and 7 validation set systems, including several challenging dimers (Mn–Mn, Fe–Fe). The meta-GGA CC-r$^2$SCAN reference wins for 6 test set and 4 validation set systems, performing best for late transition metal chlorides and several homonuclear dimers (Sc–Sc, Ti–Ti, Ni–Ni).

**Special Cases: CC-SVWN5.** Despite exhibiting the highest overall MAE among DFT references, CC-SVWN5 achieves the lowest error for 3 test set systems (V–H, Co–H, Ti–Cl) and 4 validation set systems (V–H$^+$, Cu–H$^+$, V–V, Zn–Zn). The Co–H case is particularly striking: CC-HF, CC-PBE, CC-PW91, and CC-PBE0 all overbind by 0.34–0.37 eV, CC-r$^2$SCAN underbinds by 0.22 eV, yet CC-SVWN5 achieves −0.02 eV error. Similarly, for V–H$^+$ in the validation set, CC-HF fails catastrophically (−1.10 eV) while CC-SVWN5 yields only +0.07 eV error. These cases suggest that the extremely delocalized LDA density provides a superior reference for specific electronic configurations where GGA and hybrid functionals introduce systematic biases.

Notably, no GGA-referenced method produces errors exceeding 0.51 eV across either the test set or validation set, demonstrating the robust transferability of CC-PBE and CC-PW91 across diverse bonding environments for these metal dimers.

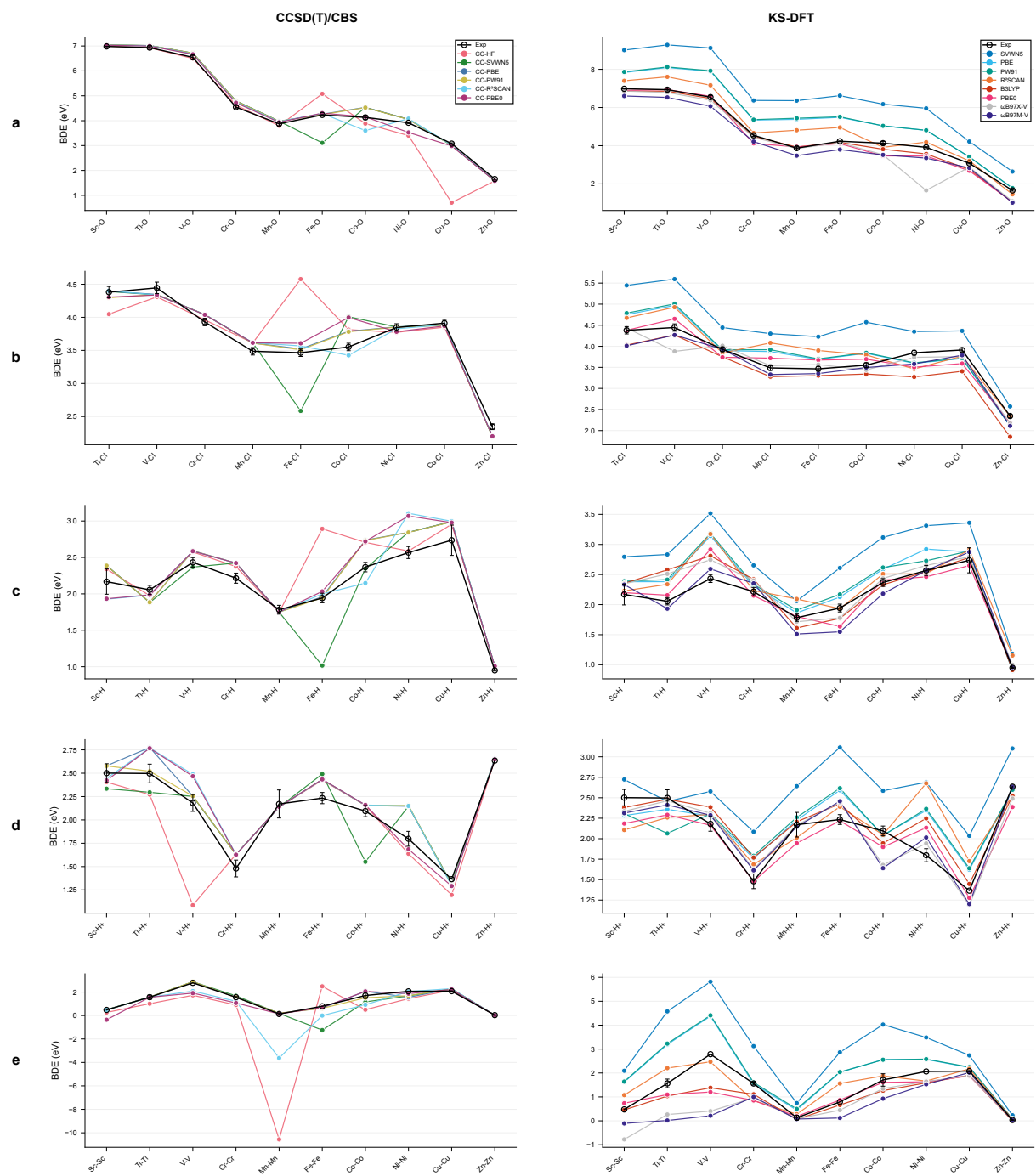

Figure S2: Calculated bond dissociation energies (BDEs) for first-row transition metal diatomics compared to experimental reference values. Left column: CCSD(T)/CBS with various DFT reference orbitals (CC-HF, CC-SVWN5, CC-PBE, CC-PW91, CC-R$^2$SCAN, CC-PBE0). Right column: KS-DFT methods (SVWN5, PBE, PW91, R$^2$SCAN, B3LYP, PBE0, $\omega$B97X-V, $\omega$B97M-V). Rows show different bond types: (**a**) metal oxides (M–O), (**b**) metal chlorides (M–Cl), (**c**) neutral metal hydrides (M–H), (**d**) metal hydride cations (M–H$^+$), and (**e**) homonuclear dimers (M–M). Open black circles with error bars represent experimental values and their reported uncertainties.

# S8 Varying HF Exchange and Its Effect on Orbital Energies

N$_2$ orbital energy changes with respect to proportional changes in HF and DFT exchange contributions (percent) are provided. The cc-pVTZ basis is used to obtain occupied MO eigenvalues whereas the aug-cc-pVTZ basis is used to obtain virtual MO eigenvalues. Basis functions are represented in the Cartesian angular form. The vertical dashed lines represent the standard percentage of HF exchange in each density functional approximation. Calculations using B3LYP, revPBE0, SCAN0, and variants using a semicanonicalization step were performed with a development version of Q-Chem v6.2.[57]

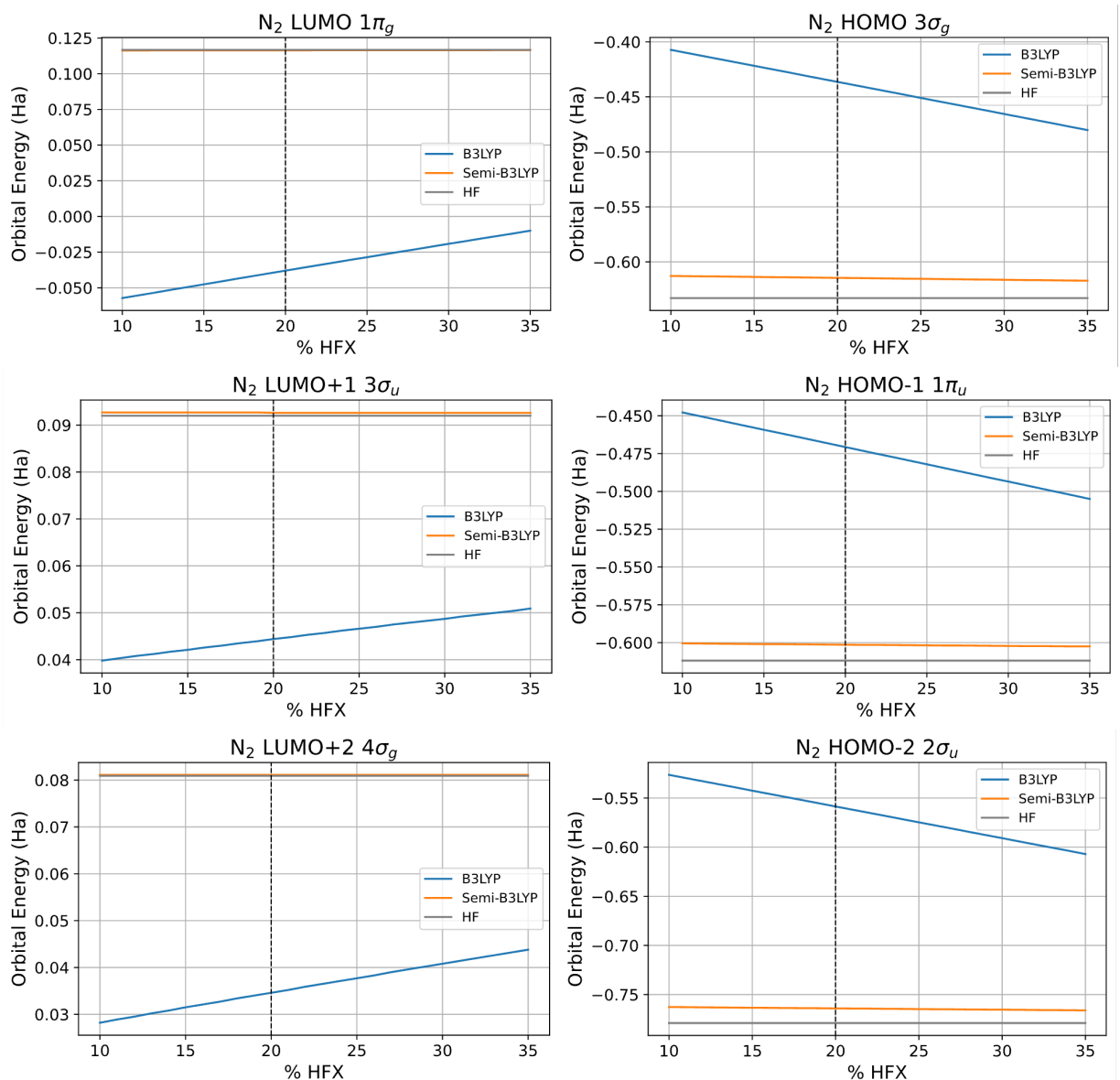

Figure S3: MO eigenvalue changes with respect to varying % HFX for B3LYP.

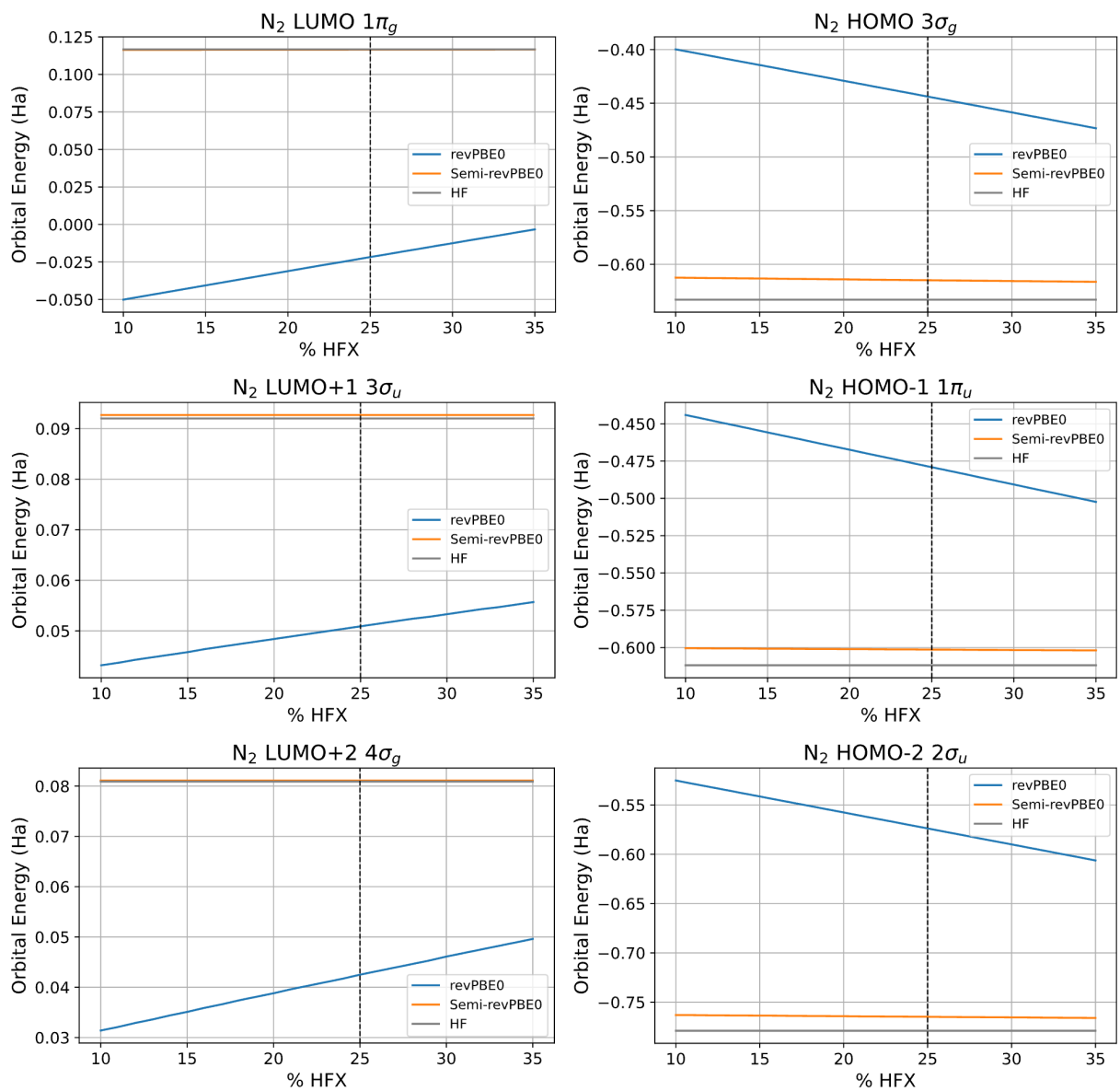

Figure S4: MO eigenvalue changes with respect to varying % HFX for revPBE0.

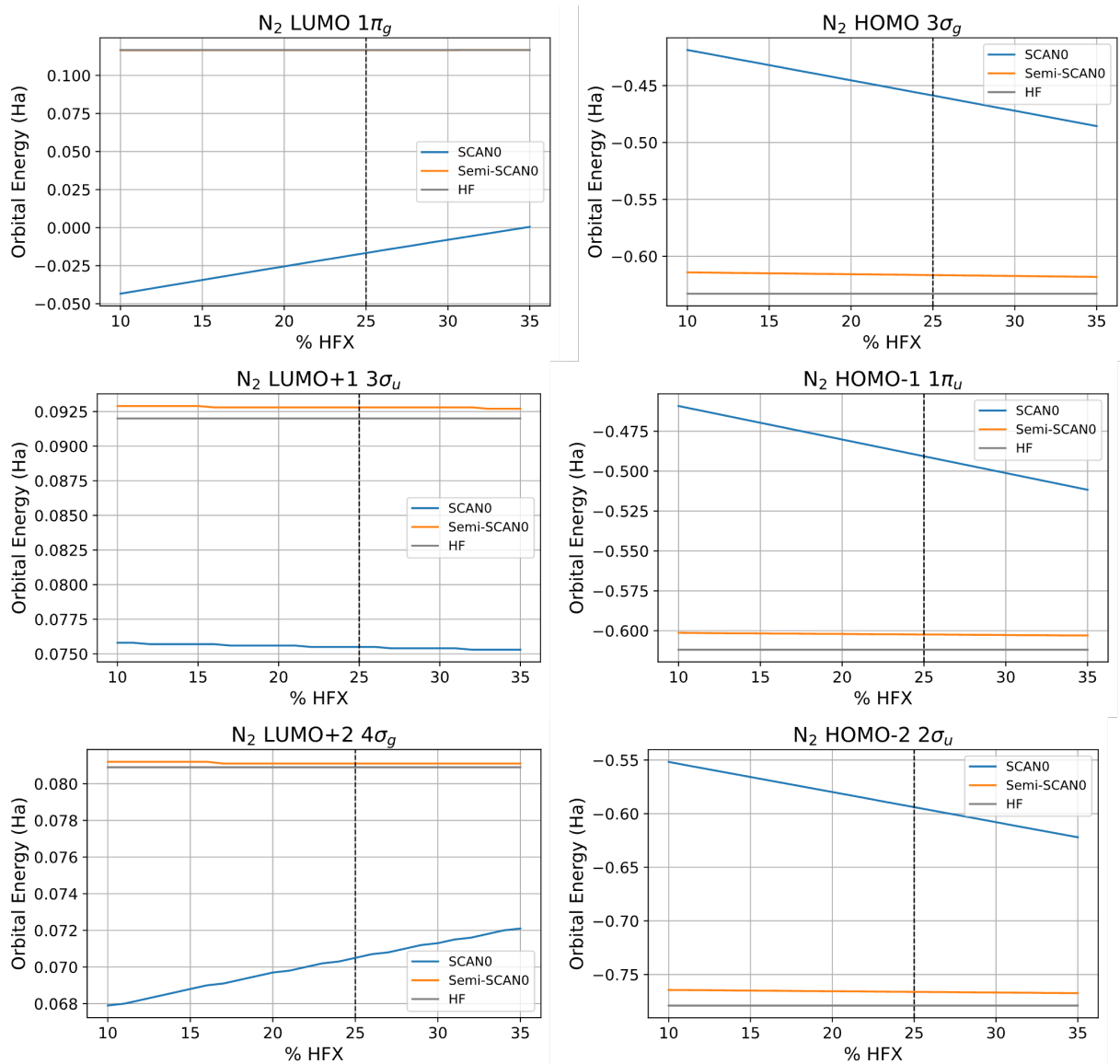

Figure S5: MO eigenvalue changes with respect to varying % HFX for SCAN0.

## S9  QUEST #1 Vertical Excitations Energies

EOM-CCSD/aug-cc-pVTZ, in the frozen-core approximation, is used to compute vertical excitation energies for molecules in the QUEST #1 database. When extrapolated FCI reference data is available, transitions with single-excitation character were considered. EOM-CC calculations and semicanonicalization were performed with a development version of Q-Chem v6.2. Basis functions are represented in the spherical harmonic (pure) form. Structures are obtained from Loos et al.[58]

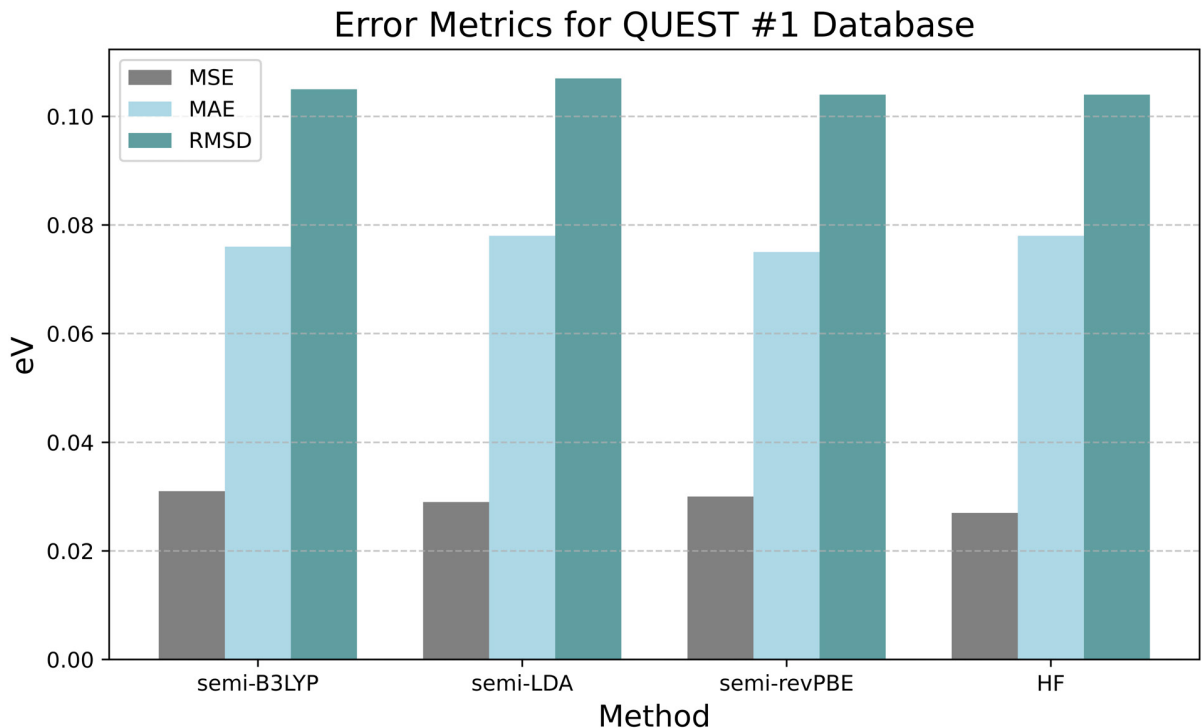

Figure S6: QUEST #1 excitation energies with HF-based and KS-based EOM-CCSD.

## S10 Correlated orbital energies with Kohn-Sham references

B3LYP, LDA, and revPBE references are used in tandem with electron propagator theory to compute vertical ionization potentials and electron affinities. Semicanonical orbitals and orbital energies are obtained from converged Kohn-Sham densities. These calculations are performed with a modified version of UQUANTCHEM.[59] All electrons are included in correlation corrections. Average pole strengths exceed 0.85 for the orbital detachments/attachments. The SG-1 grid is selected for the DFT SCF calculations.

### S10.1 Ionization Potentials

Principal ionization potentials (IPs) for 33 molecules are computed with Koopmans' theorem (KT), second-order (D2), partial third-order (P3), and approximately renormalized partial third-order (P3+) diagonal self-energy approximations using different wavefunction references. The cc-pVTZ basis is used and is represented in the Cartesian angular form. Second-order non-Brillouin singles contributions to the correlation corrections to the orbital energies are included. The results for IPs are compared to BD-T1 reference data.

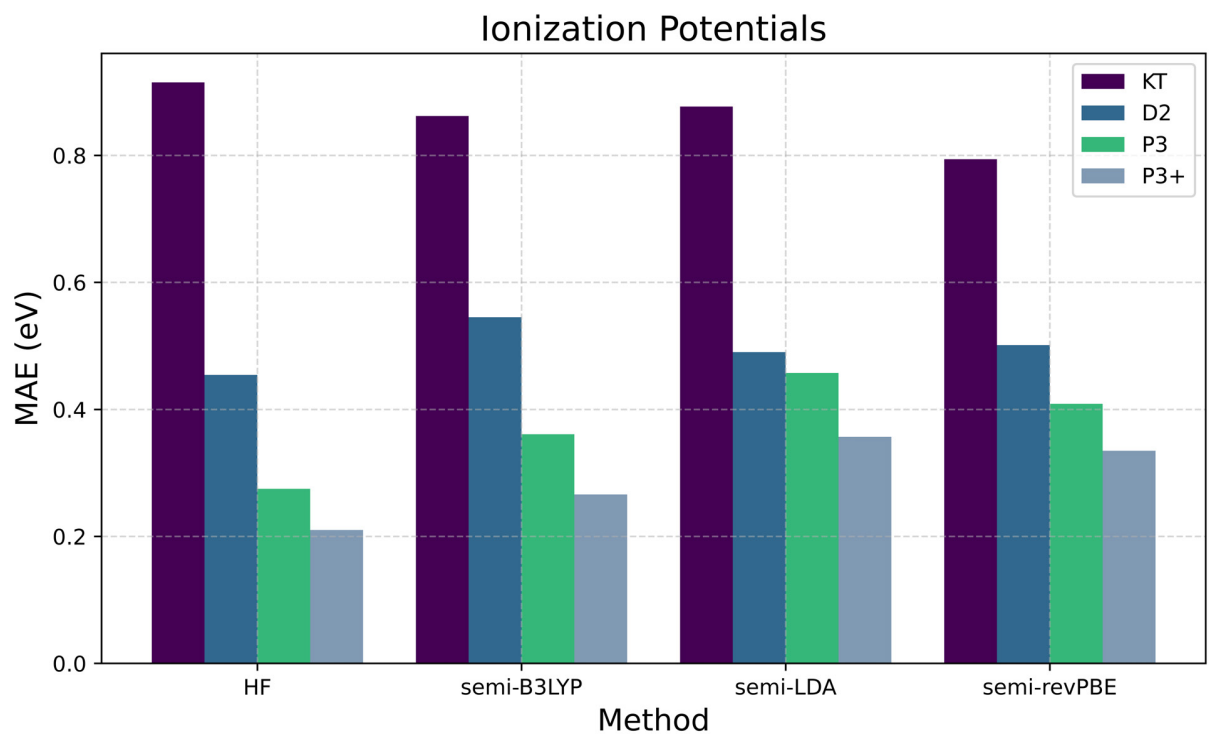

Figure S7: Ionization potentials obtained from electron propagator methods using HF and KS references.

Molecules are selected from the Chong-Gritsenko-Baerends (CGB) and GW100 datasets. Structures are obtained from literature.[60, 61]

**CGB**: $N_2$, $H_2O$, CO, $NH_3$, $F_2$, $O_3$, $H_2CO$, $C_2H_2$, $C_2H_4$, HCN, NNO, $CO_2$, HF, $HCONH_2$, $CH_4$, $C_2H_6$, $CH_3F$, LiH

**GW100**: $Li_2$, LiF, Ne, $BH_3$, BF, $HN_3$, $H_2O_2$, He, $N_2H_4$, $C_4$, $B_2H_6$, $H_2$, $PH_3$, $P_2$, PN

## S10.2 Electron Affinities

Electron affinities (EAs) for several molecules are computed with Koopmans' theorem (KT), second-order (D2), and third-order (L3) diagonal self-energy approximations. The aug-cc-pVTZ basis is used and is represented in the Cartesian angular form. The results for EAs are compared to BD-T1 reference data (NIST-EA-50).[62]

Table S7: Absolute deviations in eV from BD-T1 EA reference data.[a]

| Molecule | Method | KT | D2 | L3 |
|---|---|---|---|---|
| BeO | HF | 0.246 | 0.274 | 0.132 |
| | B3LYP | 0.506 | 0.066 | 0.172 |
| | LDA | 0.532 | 0.122 | 0.204 |
| | revPBE | 0.550 | 0.171 | 0.230 |
| BN | HF | 0.302 | 1.102 | 0.832 |
| | B3LYP | 0.051 | 0.342 | 0.280 |
| | LDA | nc | nc | nc |
| | revPBE | nc | nc | nc |
| $C_2$ | HF | 0.635 | 1.502 | 1.142 |
| | B3LYP | 0.337 | 0.736 | 0.538 |
| | LDA | nc | nc | nc |
| | revPBE | nc | nc | nc |
| $Li_2$ | HF | 0.344 | 0.043 | 0.022 |
| | B3LYP | 0.372 | 0.086 | 0.066 |
| | LDA | 0.361 | 0.065 | 0.046 |
| | revPBE | 0.366 | 0.075 | 0.055 |
| NaH | HF | 0.061 | 0.014 | 0.020 |
| | B3LYP | 0.141 | 0.056 | 0.059 |
| | LDA | 0.150 | 0.058 | 0.062 |
| | revPBE | 0.142 | 0.055 | 0.059 |
| LiH | HF | 0.083 | 0.007 | 0.004 |
| | B3LYP | 0.114 | 0.034 | 0.025 |
| | LDA | 0.116 | 0.031 | 0.021 |
| | revPBE | 0.119 | 0.036 | 0.028 |
| NaLi | HF | 0.313 | 0.012 | 0.001 |
| | B3LYP | 0.353 | 0.061 | 0.049 |
| | LDA | 0.349 | 0.049 | 0.038 |
| | revPBE | 0.344 | 0.052 | 0.041 |

[a] Entries with "nc" indicate that the SCF did not converge with a conventional DIIS solver and SAD guess. Second-order non-Brillouin singles are included.

## S11 Large Molecules

Orbital energies are obtained from SCF calculations using Hartree-Fock and the semi-empirical functional PBEh-3c with the def2-mSVP basis in spherical harmonic (pure) form. The SG-3 grid is selected. Structures are optimized with PBEh-3c/def2-mSVP. EOM-MP2 is used to obtain vertical IPs for metal-free phthalocyanine ($H_2Pc$), free base porphyrin ($H_2P$), coronene, and Buckminsterfullerene ($C_{60}$).[63–66] The frozen core approximation is employed along with frozen natural orbitals with a 99% natural occupation threshold. These calculations were performed with a development version of Q-Chem v6.2.

Table S8: Orbital Energies and Ionization Potentials for Large Molecules

| Molecule | Orbital | PBEh-3c (Ha) | semi-PBEh-3c (Ha) | HF (Ha) | IP@ semi-PBEh-3c (eV) | IP@ HF (eV) | Exp (eV) |
|---|---|---|---|---|---|---|---|
| $H_2PC$ | HOMO | -0.210 | -0.186 | -0.194 | 6.37 | 6.65 | 6.41 |
|  | HOMO - 1 | -0.287 | -0.315 | -0.328 |  |  |  |
|  | HOMO - 2 | -0.289 | -0.319 | -0.331 |  |  |  |
|  | LUMO | -0.100 | -0.003 | -0.011 |  |  |  |
|  | LUMO + 1 | -0.098 | 0.000 | -0.009 |  |  |  |
|  | LUMO + 2 | -0.025 | 0.097 | 0.087 |  |  |  |
| $H_2P$ | HOMO | -0.226 | -0.220 | -0.227 | 7.20 | 7.06 | 6.9 |
|  | HOMO - 1 | -0.228 | -0.231 | -0.245 |  |  |  |
|  | HOMO - 2 | -0.285 | -0.328 | -0.338 |  |  |  |
|  | LUMO | -0.078 | 0.018 | 0.009 |  |  |  |
|  | LUMO + 1 | -0.077 | 0.020 | 0.011 |  |  |  |
|  | LUMO + 2 | -0.005 | 0.123 | 0.114 |  |  |  |
| Coronene | HOMO | -0.240 | -0.247 | -0.261 | 7.54 | 7.40 | 7.21 |
|  | HOMO - 1 | -0.240 | -0.247 | -0.261 |  |  |  |
|  | HOMO - 2 | -0.295 | -0.323 | -0.335 |  |  |  |
|  | LUMO | -0.043 | 0.068 | 0.060 |  |  |  |
|  | LUMO + 1 | -0.043 | 0.068 | 0.060 |  |  |  |
|  | LUMO + 2 | 0.002 | 0.128 | 0.117 |  |  |  |
| $C_{60}$ | HOMO | -0.272 | -0.279 | -0.294 | 8.41 | 8.29 | 7.64 |
|  | HOMO - 1 | -0.272 | -0.279 | -0.294 |  |  |  |
|  | HOMO - 2 | -0.272 | -0.279 | -0.294 |  |  |  |
|  | LUMO | -0.126 | -0.023 | -0.031 |  |  |  |
|  | LUMO + 1 | -0.126 | -0.023 | -0.031 |  |  |  |
|  | LUMO + 2 | -0.126 | -0.023 | -0.031 |  |  |  |

## S12 NNED Metrics

NNED metrics are obtained for the TinySpins25 (abbreviated as ST25) and BDE20 datasets.
**TinySpins25 (ST25)**: AgCl, AgF, AgH, AuCl, AuF, AuH, CdO, CdSe, CdS, CuBr, CuCl, CuF, CuH, HgO, HgSe, HgS, PtC, RuC, ScBr, ScCl, ScF, ScH, ZnO, ZnSe, ZnS
**BDE20**: $Ti_2$, $V_2$, $Mn_2$, $Fe_2$, FeO, CoO, NiO, CuO, TiCl, FeCl, CoCl, NiCl, TiH, VH, FeH, CoH, $VH^+$, $CrH^+$, $MnH^+$, $NiH^+$

NNED-based MR thresholds, $x$, for a specific functional used in a specific dataset are determined through $\frac{\text{AVE } T_1}{\text{AVE NNED}} = \frac{\text{REF } T_1}{x}$, where the conventional reference $T_1$ limit REF $T_1$ is either 0.02 or 0.05.

| **ST25** | **LDA** | **B3LYP** | **PBE** | **PBE0** | **SCAN** | **SCAN0** |
|---|---|---|---|---|---|---|
| T1 = 0.02 equiv. | 0.020 | 0.026 | 0.022 | 0.030 | 0.022 | 0.027 |
| T1 = 0.05 equiv. | 0.051 | 0.066 | 0.054 | 0.076 | 0.055 | 0.069 |

Table S9: Functional specific NNED limits with def2-QZVPPD. Entries are averages of singlet and triplet NNED limits.

| **BDE20** | **LDA** | **PW91** | **r$^2$SCAN** | **PBE** | **PBE0** |
|---|---|---|---|---|---|
| T1 = 0.02 equiv. | 0.012 | 0.012 | 0.011 | 0.011 | 0.008 |
| T1 = 0.05 equiv. | 0.031 | 0.031 | 0.028 | 0.028 | 0.021 |

Table S10: Functional specific NNED limits with def2-QZVPP. Entries are the NNED limits of the ground state dimers.

The TinySpins25 (ST25) NNED averages for each functional range from 0.018-0.035 across singlets and triplets. The BDE20 NNED averages for each functional range from 0.035-0.051. The value 0.035 is chosen as the demarcation point for MR classification that bifurcates into rules based on strong or moderate NNED limits. We determine a general-use set of NNED thresholds based on those designated for different types of density functional and chemical species. The general upper and lower NNED threshold is based on the average of the highest and lowest NNEDs equivalents to $T_1 = 0.05$ for the ST25 and BDE20 data sets, respectively. When possible, $T_1$ metrics used to obtain the NNED equivalents consider all electrons. The following decision trees illustrate the general classification process:

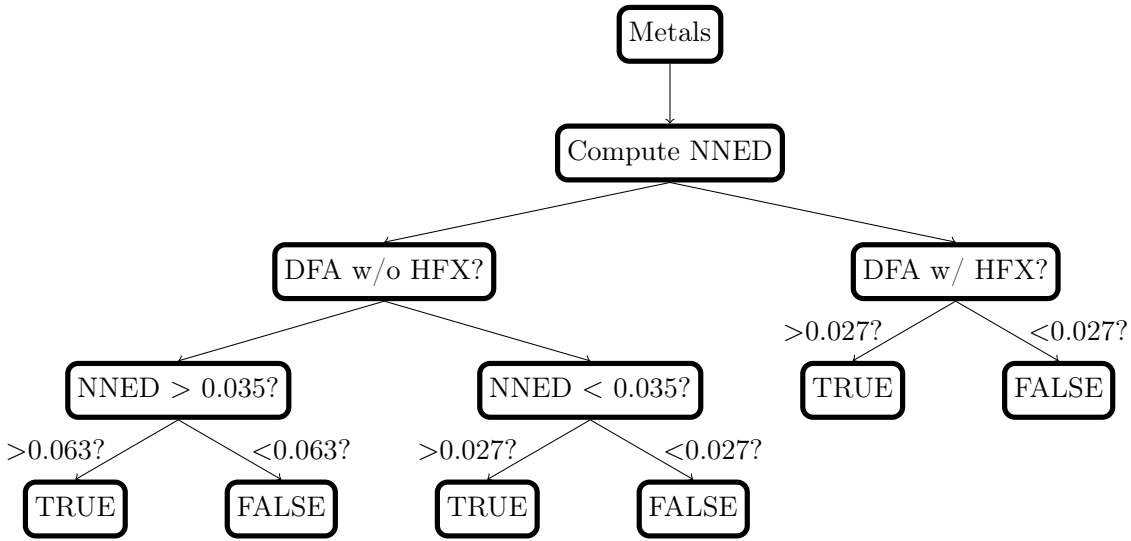

Figure S8: MR decision tree for metal species based on general NNED limits.

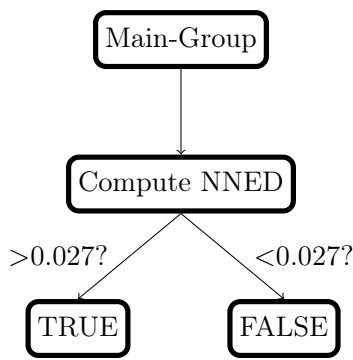

Figure S9: MR decision tree for main-group species based on general NNED limits.

For Figure S8, metal systems are filtered based on the inclusion of HFX in the DFA. Average NNED and $||\Delta_P||_F$ values are smaller with hybrid functionals, warranting an additional bifurcation based on the DFA type. The systems are further binned according to the overall magnitudes of the NNED values and the upper and lower thresholds defined by the strongly and weakly correlated test sets (see Tables S9 and S10). A simpler MR classification scheme is performed for main-group species as shown in Figure S9.

## S12.1 Features and Criteria

Qualitative features for the BDE20 dataset are examined. The following are MR designations based on functional specific NNED limits. Results below use the strong $T_1 = 0.05$ NNED limit equivalents. $T_1$, $\Delta_P$, and NNED metrics are also provided.

| Species | HF-T1 | LDA-NNED | PW91-NNED | r²SCAN-NNED | PBE-NNED | PBE0-NNED |
|---|---|---|---|---|---|---|
| CoCl | yes | no | no | no | yes | no |
| CoH | yes | yes | yes | no | yes | yes |
| CoO | yes | yes | yes | yes | yes | yes |
| CrH$^+$ | yes | yes | yes | yes | yes | yes |
| CuO | yes | yes | yes | yes | yes | yes |
| Fe$_2$ | yes | yes | yes | yes | yes | yes |
| FeCl | no | yes | no | no | yes | no |
| FeH | yes | yes | yes | no | yes | no |
| FeO | yes | yes | yes | yes | yes | yes |
| Mn$_2$ | no | yes | yes | yes | yes | yes |
| MnH$^+$ | no | yes | yes | yes | yes | yes |
| NiCl | no | no | no | no | yes | no |
| NiH$^+$ | yes | yes | yes | yes | yes | yes |
| NiO | yes | yes | yes | yes | yes | yes |
| Ti$_2$ | yes | yes | yes | yes | yes | yes |
| TiCl | yes | yes | yes | yes | yes | yes |
| TiH | yes | yes | yes | yes | yes | yes |
| V$_2$ | yes | yes | yes | yes | yes | yes |
| VH | yes | yes | yes | yes | yes | yes |
| VH$^+$ | no | yes | yes | no | yes | yes |

BDE20 Dataset: Threshold-based designations for MR Character (Strong)

Figure S10: MR character designations of BDE20 given the strong NNED limits.

17

The following are MR designations based off <u>general</u> NNED limits.

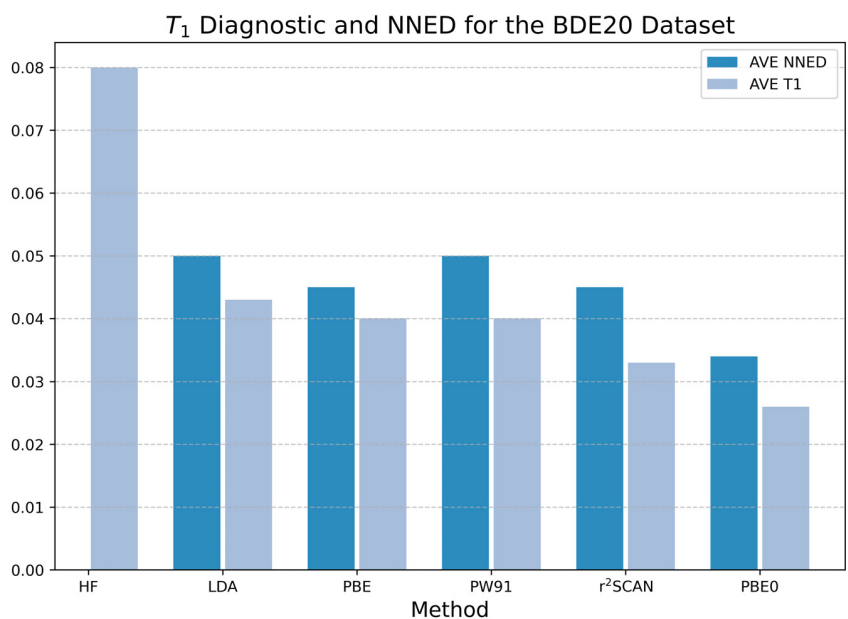

Figure S11: MR character designations of BDE20 given the general NNED limits.

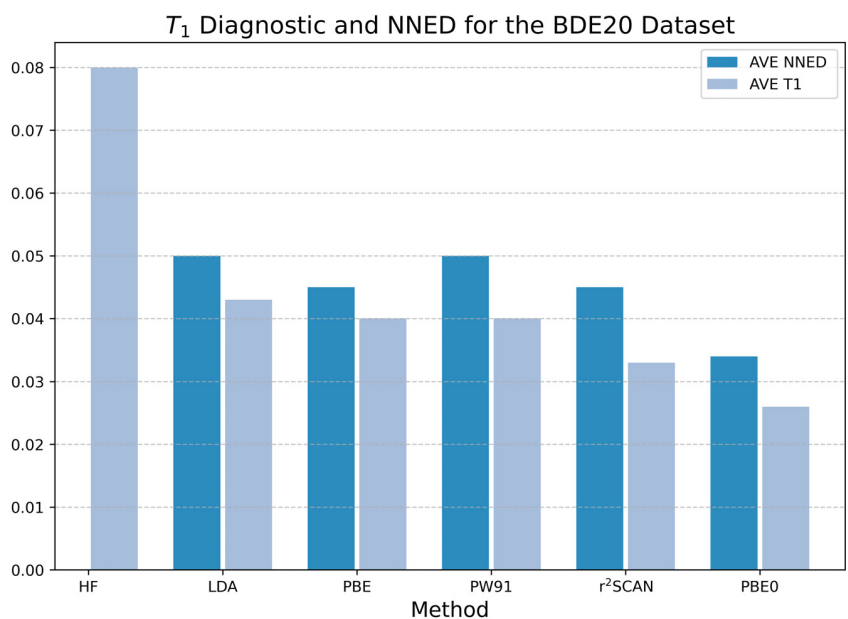

Figure S12: Average $T_1$ diagnostics and NNED metrics from various KS references for species within the BDE20 dataset.

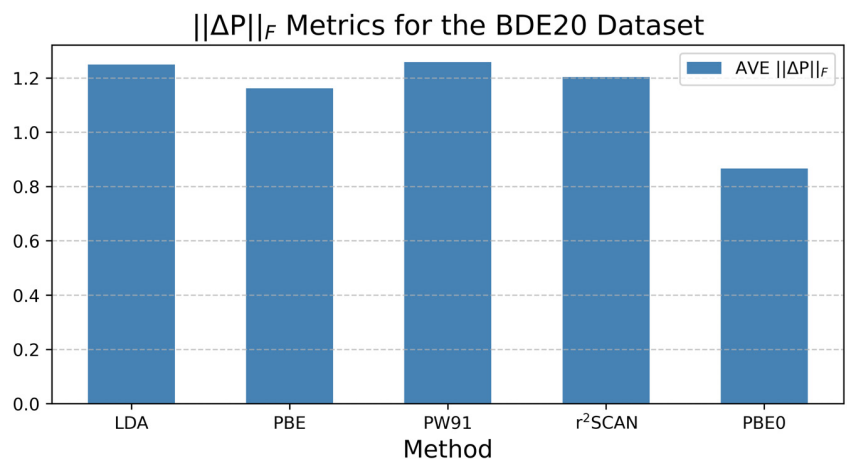

Figure S13: Average Frobenius norm metrics ($||\Delta_P||_F$) using def2-QZVPP.

## S13 Difference Density Natural Orbitals

Difference density natural orbitals (DDNOs) and electron displacement eigenpairs ($\delta_\pm$) for select main-group species are displayed below.

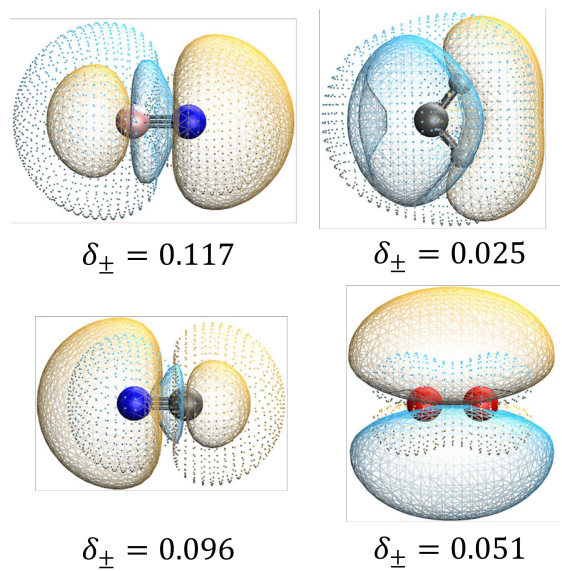

Figure S14: DDNOs for singlets BN, CN$^+$, CH$_2$, and O$_2$ computed between HF and SCAN with def2-QZVPP. The largest $\delta_\pm$ values are reported. The orbitals corresponding to $\delta_-$ and $\delta_+$ are represented by radial dots and line meshes, respectively. An isosurface value of 0.02 is used. For BN, CN$^+$, CH$_2$, and O$_2$, the NNED values are 0.035, 0.033, 0.017, and 0.019, respectively. Similarly, the $||\Delta_P||_F$ values are 0.361, 0.311, 0.103, and 0.193, respectively.

## S14 Singlet-Triplet Gaps: Metal Diatomics

Vertical singlet-triplet (ST) gaps are computed with CCSD(T)/def2-QZVPPD. Effective core potentials for the corresponding basis set are applied to neutral species containing 4d and 5d elements (lanthanides excluded). RHF and ROHF solutions are obtained for the singlets and triplets respectively. $T_1$ diagnostic metrics and the Frobenius norms of the occupied-virtual block of the semicanonical Fock operator $F_{ov}$ are computed. Hartree-Fock and various density functional references are used. Semicanonicalization and CC calculations were performed with a development version of Q-Chem v6.2.

Structures are optimized in Q-Chem v6.2 with $\omega$B9M-V/def2-TZVPP. Reference data used as best theoretical estimates are computed with CBS extrapolated (def2-SVPD/def2-TZVPPD) CCSDT(Q)$_\Lambda$ starting from a (RO)HF/def2-QZVPPD reference. CCSDT(Q)$_\Lambda$ calculations were performed with the MRCC v25.1.1 program.[67]

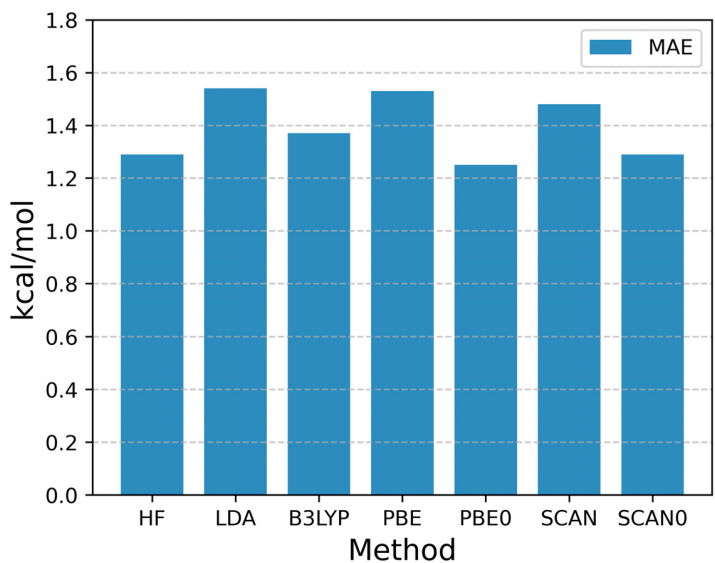

Figure S15: Average ST gap errors for the TinySpins25 dataset. ST gaps obtained with CCSD(T)/def2-QZVPPD with various SCF references are compared to CCSDT(Q)$_\Lambda$/CBS results.

### S14.1 Features and Criteria

Quantitative and qualitative features for the TinySpins25 dataset are examined.

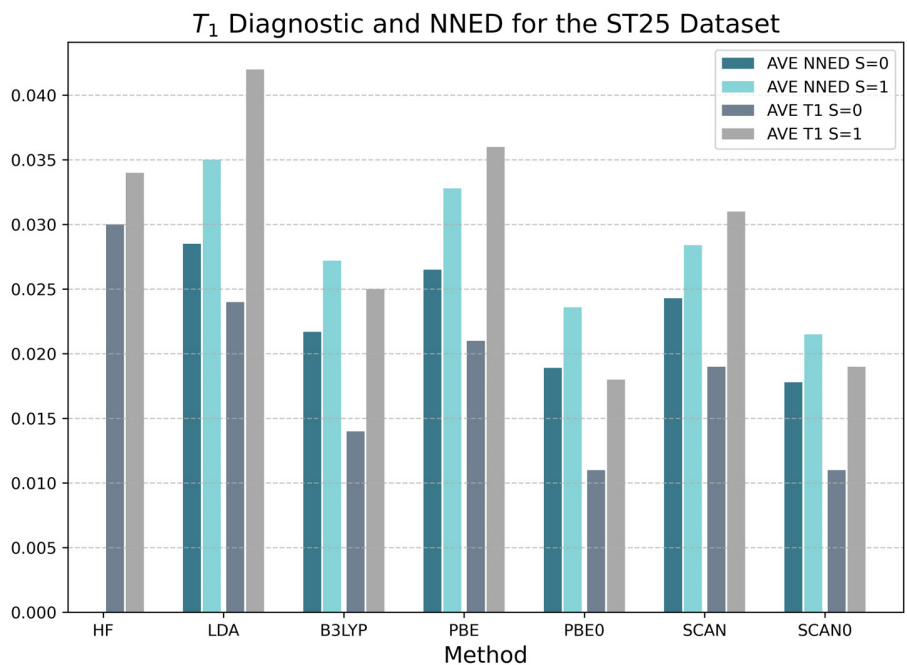

Figure S16: Average $T_1$ diagnostics and NNED metrics from various KS references for singlet and triplet species within the TinySpins25 dataset.

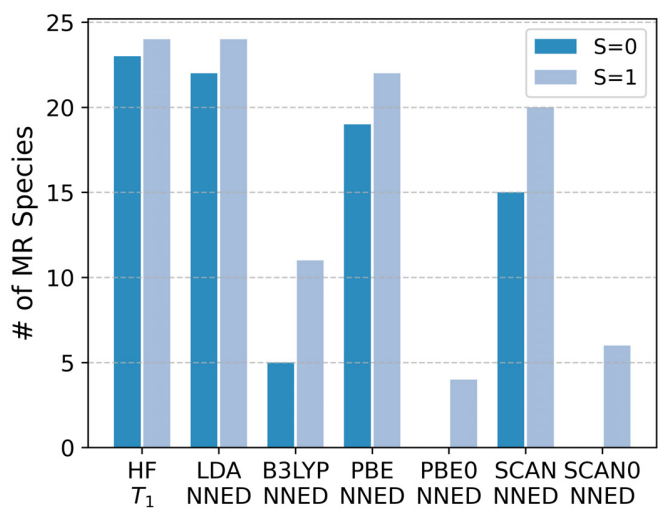

Figure S17: Number of singlet and triplet species within the TinySpins25 dataset predicted to be of moderate MR character based on HF-$T_1$ ($> 0.02$) and functional-specific NNED metrics.

For this data set, singlet and triplet NNEDs follow the trends of their $T_1$ counterparts and, interestingly, the decrease in MR counts appears to coincide with ascending rungs of functionals according to Jacob's Ladder. NNED thresholds based on the higher $T_1$ limit ($> 0.05$) conclude that no singlet or triplet metal diatomic in the TinySpins25 (ST25) set is considered strongly MR within KS-CC. The Frobenius norms of the difference density ($||\Delta_P||_F$) are also computed for the ST25

21

set. SCAN0 provides the smallest average $||\Delta_P||_F$ for singlet and triplets with B3LYP and PBE0 following suit, indicating the smaller distances between the $P_{\text{KS}}$ of these DFAs and $P_{\text{HF}}$ compared to LDA or revPBE. This captures, qualitatively, the trends of the observed ST25 MAEs with the selected DFAs.

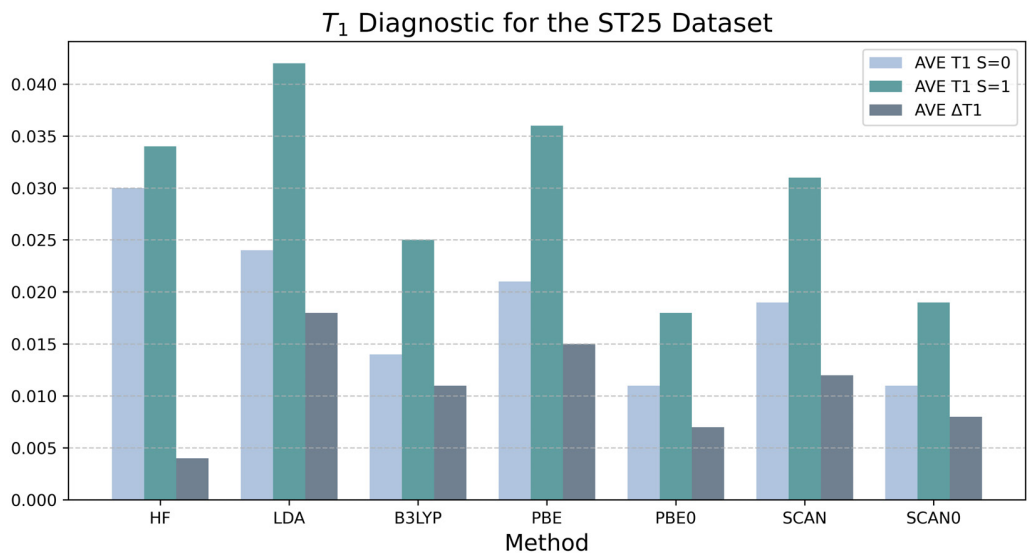

Figure S18: Average $T_1$ diagnostics using def2-QZVPPD.

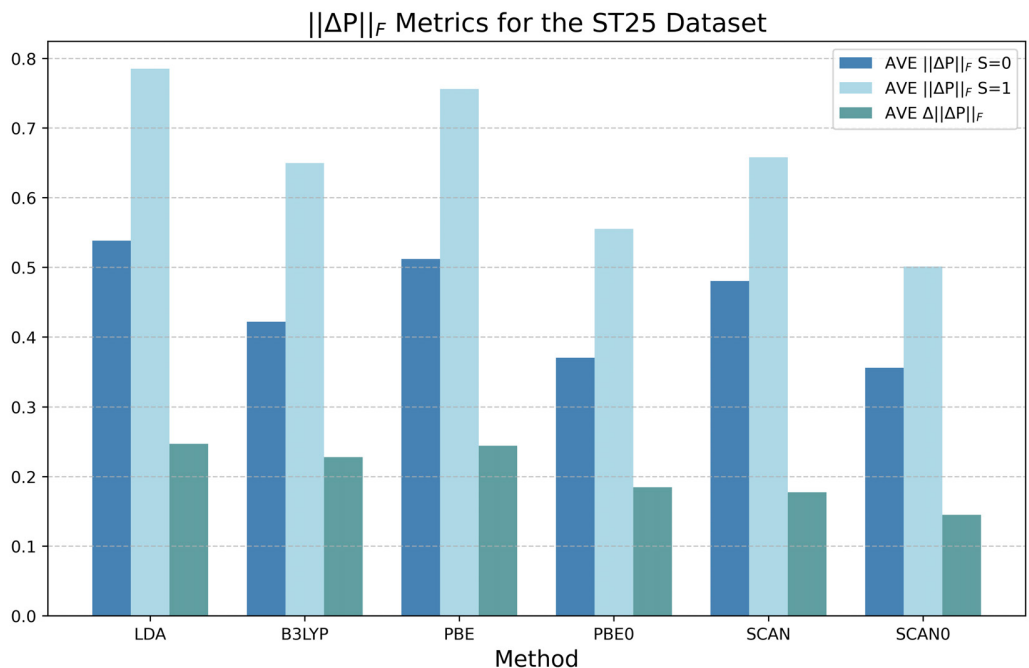

Figure S19: Average Frobenius norm metrics ($||\Delta_P||_F$) using def2-QZVPPD. $\Delta||\Delta_P||_F$ is the absolute difference between average singlet and triplet $||\Delta_P||_F$.

22

Figure S20: Identifying improvements over HF-CCSD(T) with respect to CCSDT(Q)$_\Lambda$. "Small gap" refers to values < 11.5 kcal/mol.

The following are MR designations based off functional specific NNED limits. No TinySpins25 species was deemed MR using the $T_1 = 0.05$ NNED equivalents, so results below are based the loose $T_1 = 0.02$ NNED equivalents.

## ST25 Dataset: Threshold-based designations for MR Character in Singlets

| Species | HF-T1 | LDA-NNED | B3LYP-NNED | PBE-NNED | PBE0-NNED | SCAN-NNED | SCAN0-NNED |
|---|---|---|---|---|---|---|---|
| AgCl | yes | yes | no | yes | no | yes | no |
| AgF | yes | yes | yes | yes | no | yes | no |
| AgH | yes | yes | yes | yes | no | yes | no |
| AuCl | yes | yes | no | yes | no | yes | no |
| AuF | yes | yes | no | yes | no | yes | no |
| AuH | yes | yes | yes | yes | no | yes | no |
| CdO | yes | yes | yes | yes | no | yes | no |
| CdS | yes | yes | no | yes | no | no | no |
| CdSe | yes | yes | no | no | no | no | no |
| CuBr | yes | no | no | no | no | no | no |
| CuCl | yes | yes | no | yes | no | no | no |
| CuF | yes | yes | no | yes | no | yes | no |
| CuH | yes | yes | no | yes | no | yes | no |
| HgO | yes | yes | no | yes | no | yes | no |
| HgS | yes | yes | no | yes | no | no | no |
| HgSe | no | no | no | no | no | no | no |
| PtC | yes | yes | yes | yes | no | yes | no |
| RuC | yes | yes | no | yes | no | yes | no |
| ScBr | yes | yes | no | no | no | no | no |
| ScCl | yes | yes | no | yes | no | no | no |
| ScF | yes | yes | no | yes | no | yes | no |
| ScH | yes | yes | no | yes | no | yes | no |
| ZnO | yes | yes | no | yes | no | yes | no |
| ZnS | yes | yes | no | no | no | no | no |
| ZnSe | no | no | no | no | no | no | no |

Figure S21: MR character designations of TinySpins25 singlets given the weak NNED limits.

## ST25 Dataset: Threshold-based designations for MR Character in Triplets

| Species | HF-T1 | LDA-NNED | B3LYP-NNED | PBE-NNED | PBE0-NNED | SCAN-NNED | SCAN0-NNED |
|---|---|---|---|---|---|---|---|
| AgCl | yes | yes | yes | yes | yes | yes | yes |
| AgF | yes | yes | yes | yes | yes | yes | yes |
| AgH | yes | yes | yes | yes | no | yes | yes |
| AuCl | yes | yes | yes | yes | yes | yes | yes |
| AuF | yes | yes | yes | yes | no | yes | no |
| AuH | yes | yes | no | yes | no | yes | no |
| CdO | yes | yes | yes | yes | no | yes | no |
| CdS | yes | yes | no | yes | no | yes | no |
| CdSe | yes | yes | no | no | no | no | no |
| CuBr | yes | yes | yes | yes | no | yes | no |
| CuCl | yes | yes | no | yes | no | yes | no |
| CuF | yes | yes | no | yes | no | yes | no |
| CuH | yes | yes | yes | yes | yes | yes | yes |
| HgO | yes | yes | yes | yes | no | yes | no |
| HgS | yes | yes | no | yes | no | yes | no |
| HgSe | yes | yes | no | no | no | no | no |
| PtC | yes | yes | yes | yes | no | yes | yes |
| RuC | yes | yes | no | yes | no | yes | no |
| ScBr | yes | yes | no | yes | no | no | no |
| ScCl | yes | yes | no | yes | no | yes | no |
| ScF | yes | yes | no | yes | no | yes | no |
| ScH | yes | yes | yes | yes | no | yes | no |
| ZnO | yes | yes | no | yes | no | yes | no |
| ZnS | yes | yes | no | yes | no | no | no |
| ZnSe | no | no | no | no | no | no | no |

Figure S22: MR character designations of TinySpins25 triplets given the weak NNED limits.

The following are MR designations based off <u>general</u> NNED limits.

**ST25 Dataset: Threshold-based designations for MR Character in Singlets (General)**

| Species | HF-T1 | LDA-NNED | B3LYP-NNED | PBE-NNED | PBE0-NNED | SCAN-NNED | SCAN0-NNED |
|---|---|---|---|---|---|---|---|
| AgCl | no | yes | no | no | no | no | no |
| AgF | no | yes | no | yes | no | yes | no |
| AgH | no | no | yes | no | no | no | no |
| AuCl | no | yes | no | no | no | no | no |
| AuF | no | yes | no | yes | no | yes | no |
| AuH | no | no | yes | yes | no | yes | no |
| CdO | no | no | no | yes | no | yes | no |
| CdS | no | no | no | no | no | no | no |
| CdSe | no | no | no | no | no | no | no |
| CuBr | no | no | no | no | no | no | no |
| CuCl | no | no | no | no | no | no | no |
| CuF | no | yes | no | yes | no | no | no |
| CuH | no | yes | no | yes | no | yes | no |
| HgO | no | no | no | yes | no | yes | no |
| HgS | no | no | no | no | no | no | no |
| HgSe | no | no | no | no | no | no | no |
| PtC | no | no | yes | no | yes | yes | no |
| RuC | no | yes | no | yes | no | no | no |
| ScBr | no | no | no | no | no | no | no |
| ScCl | no | yes | no | no | no | no | no |
| ScF | no | yes | no | yes | no | no | no |
| ScH | no | yes | no | no | no | no | no |
| ZnO | no | no | no | no | no | no | no |
| ZnS | no | no | no | no | no | no | no |
| ZnSe | no | no | no | no | no | no | no |

Figure S23: MR character designations of TinySpins25 singlets given the general NNED limits.

**ST25 Dataset: Threshold-based designations for MR Character in Triplets (General)**

| Species | HF-T1 | LDA-NNED | B3LYP-NNED | PBE-NNED | PBE0-NNED | SCAN-NNED | SCAN0-NNED |
|---|---|---|---|---|---|---|---|
| AgCl | no | no | yes | no | yes | no | yes |
| AgF | no | no | yes | no | yes | no | yes |
| AgH | no | no | yes | yes | yes | no | yes |
| AuCl | no | no | yes | no | yes | no | yes |
| AuF | no | no | yes | no | no | yes | no |
| AuH | no | yes | no | yes | no | no | no |
| CdO | no | no | yes | yes | no | yes | no |
| CdS | no | yes | no | no | no | no | no |
| CdSe | no | no | no | no | no | no | no |
| CuBr | no | yes | yes | yes | no | yes | no |
| CuCl | no | no | no | no | no | yes | no |
| CuF | no | no | no | yes | no | no | no |
| CuH | yes | no | yes | no | yes | no | yes |
| HgO | no | no | yes | no | no | yes | no |
| HgS | no | yes | no | no | no | no | no |
| HgSe | no | no | no | no | no | no | no |
| PtC | yes | no | yes | no | yes | no | yes |
| RuC | yes | yes | no | yes | no | yes | no |
| ScBr | no | no | no | no | no | no | no |
| ScCl | no | yes | no | yes | no | no | no |
| ScF | no | yes | no | yes | no | no | no |
| ScH | no | no | yes | no | yes | yes | no |
| ZnO | no | yes | no | yes | no | no | no |
| ZnS | no | no | no | no | no | no | no |
| ZnSe | no | no | no | no | no | no | no |

Figure S24: MR character designations of TinySpins25 triplets given the general NNED limits.

## S15  Singlet-Triplet Gaps: Main-Group

Singlet-triplet (ST) or triplet-singlet (TS) gaps are computed with restricted (R) and unrestricted (U) CCSD(T)/cc-pVTZ with the frozen-core approximation. Symmetry preserved and broken symmetry (BS) singlet solutions are examined. $T_1$ diagnostic metrics and Frobenius norms of the occupied-virtual block of the semicanonical Fock operator $F_{ov}$ are computed. Total spin-squared expectation values $\langle S^2 \rangle$ are provided for unrestricted self-consistent-field (USCF) and coupled cluster (UCCSD) solutions. Hartree-Fock and various density functional references are used. Semicanonicalization and CC calculations were performed with a development version of Q-Chem v6.2.

The approximate spin projection scheme of Yamaguchi and Noodleman is applied to the broken symmetry singlet UCC calculations. In some cases, the accuracy of AP spin-state energetics may be affected if $\langle S^2 \rangle$ exceeds 1, signaling spin state mixing beyond the triplet. Spin-state structures and reference data used as best theoretical estimates are taken from the literature.[68–73] NF and $CH_2$ bond lengths are specifically taken from the CCCBDB at the CCSD(T)/cc-pVTZ level of theory.[74] These molecules can be separated into two classes. BN, $C_2$, $BO^+$, and $CN^+$ are isoelectronic (Class I). Apart from BN, these molecules are assumed to have singlet ground states. $CH_2$, NF, $O_2$, and NH have triplet ground states (Class II). Class I ST gaps are generally smaller than Class II's.

The following are individual STGs for $C_2$, $BO^+$, $CN^+$ NF, $O_2$, and NH computed with CCSD(T)/cc-pVTZ with HF and KS-DFT determinants. The solid and dashed horizontal orange line represent the reference theoretical estimate and experimental value (if available), respectively. When applicable, broken-symmetry references are used.

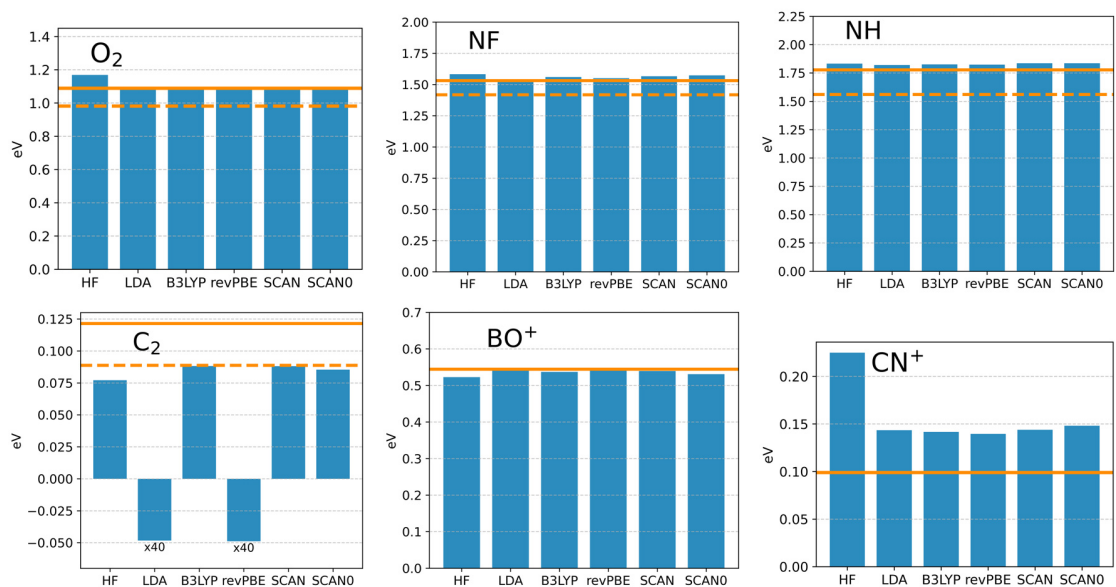

Figure S25: Singlet-triplet gaps computed with HF-CCSD(T) and KS-CCSD(T).

The STG of $C_2$ is corrected slightly towards the baseline theoretical estimate, except for when the LDA and revPBE densities are selected, as they are the most distorted away from the HF density. Results similar to $CH_2$ are observed in $BO^+$ since it is largely dominated by the HF configuration, more-so than its isoelectronic kin.[70] The $O_2$ molecule also sees some refinements with all DFAs. Errors for KS-CC STGs in NF and NH are also marginally reduced or left unchanged in comparison to HF-CC. For the diradical singlets that require unrestricted orbitals to be properly described by

26

a single determinant, we remark that residual spin contamination at the SCF level often persists in the CC reference. For this reason, the approximate projection (AP) technique is applied to the broken symmetry state.[75–77]

$T_1$, $\Delta_P$, and NNED metrics are also provided. Some percentage of HFX appears to be useful in obtaining more conservative designation of MR character. $||\Delta_P||_F$ and the NNED metric are also complementary in gauging the distance from $P_{\text{HF}}$; we note that SCAN or SCAN0 densities are typically closer to $P_{\text{HF}}$.

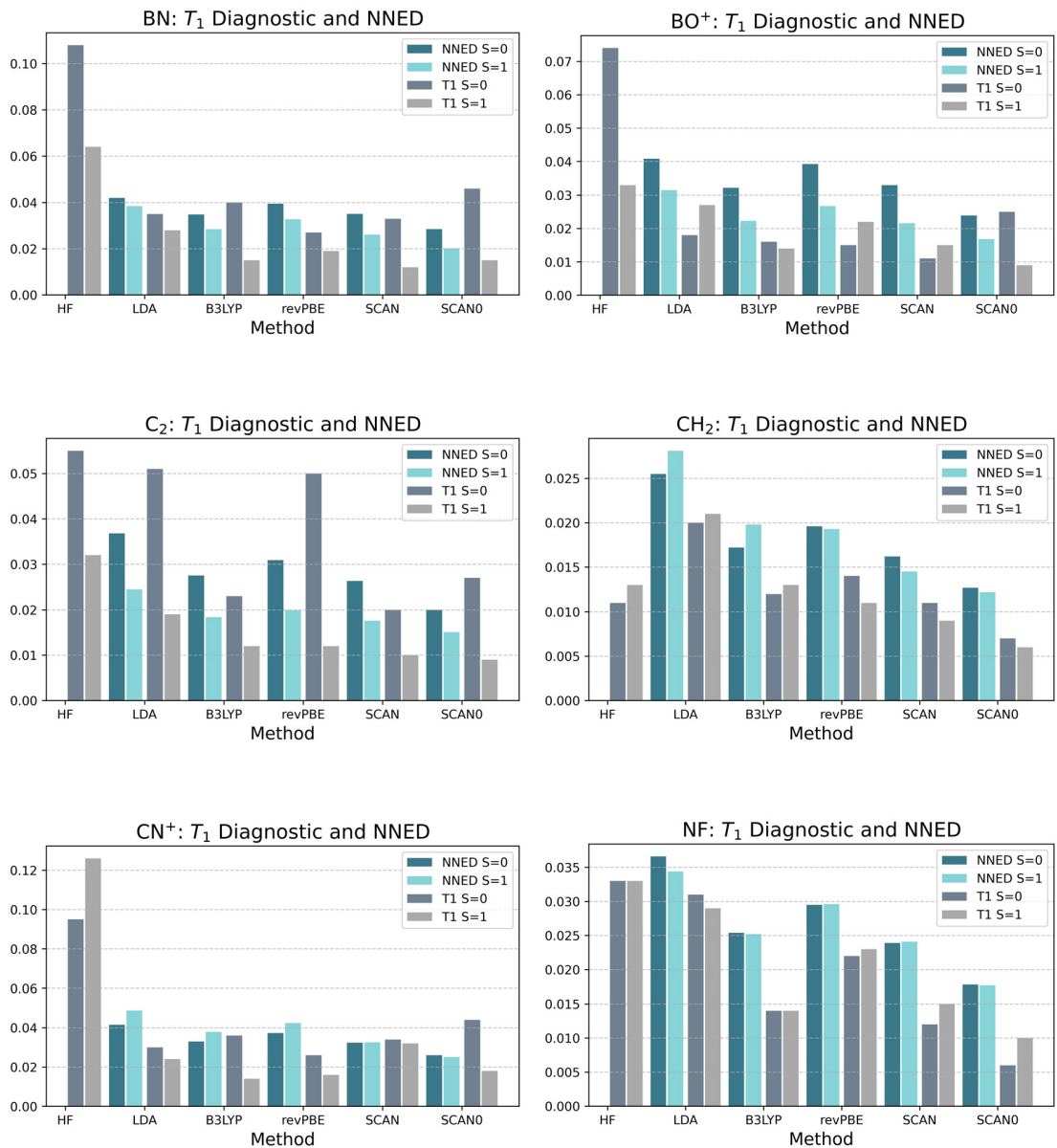

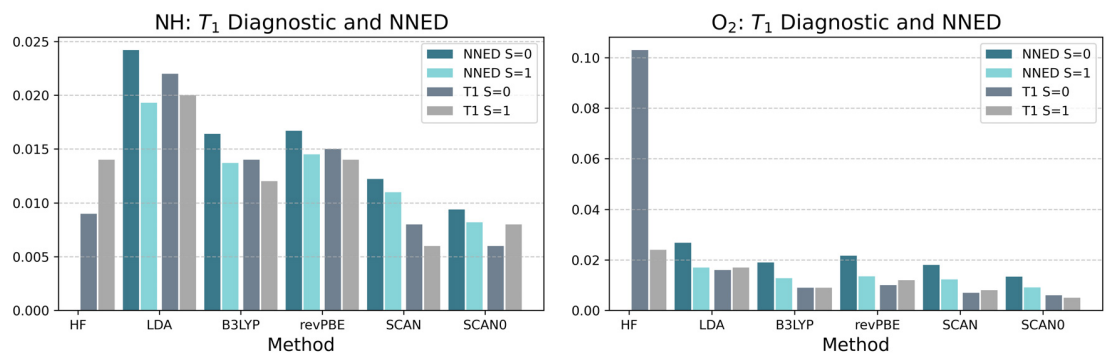

Figure S26: Comparison of $T_1$ diagnostics and NNED values for various main-group species. Broken-symmetry references are used when applicable. SCF and CCSD references are obtained using the cc-pVTZ basis set.

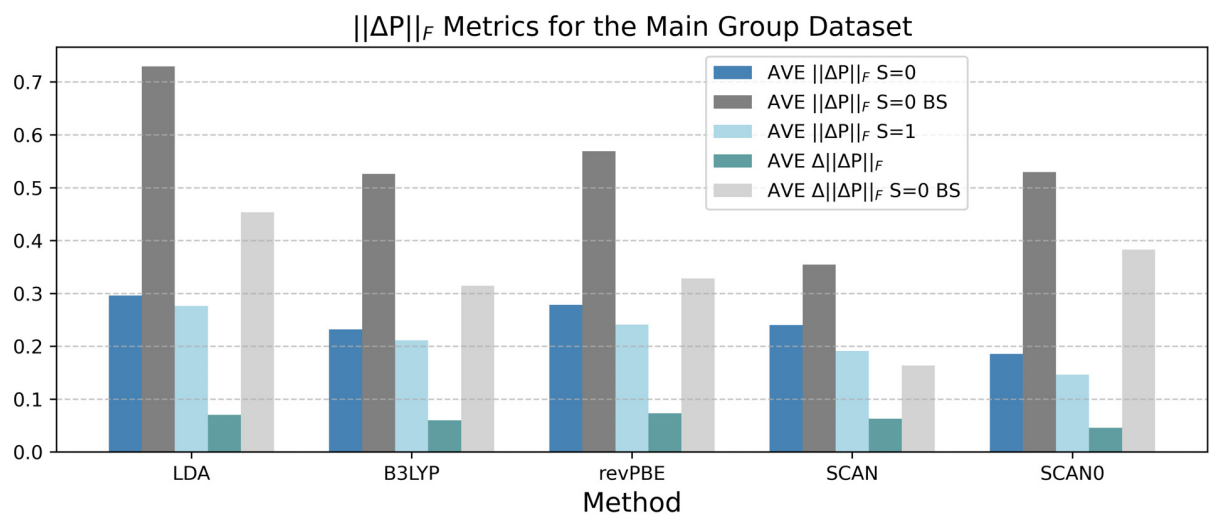

Figure S27: Average Frobenius norm metrics ($||\Delta_P||_F$) using cc-pVTZ.

## S15.1  Closed-shell Singlets: CCSD(T)

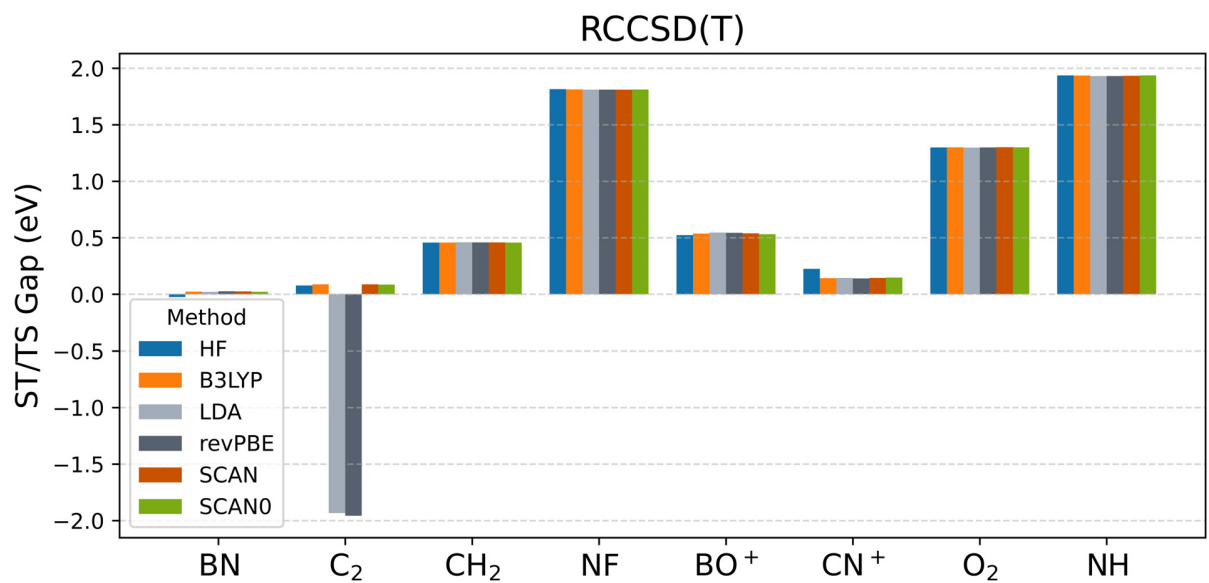

Figure S28: Singlet-triplet gaps with closed-shell singlets.

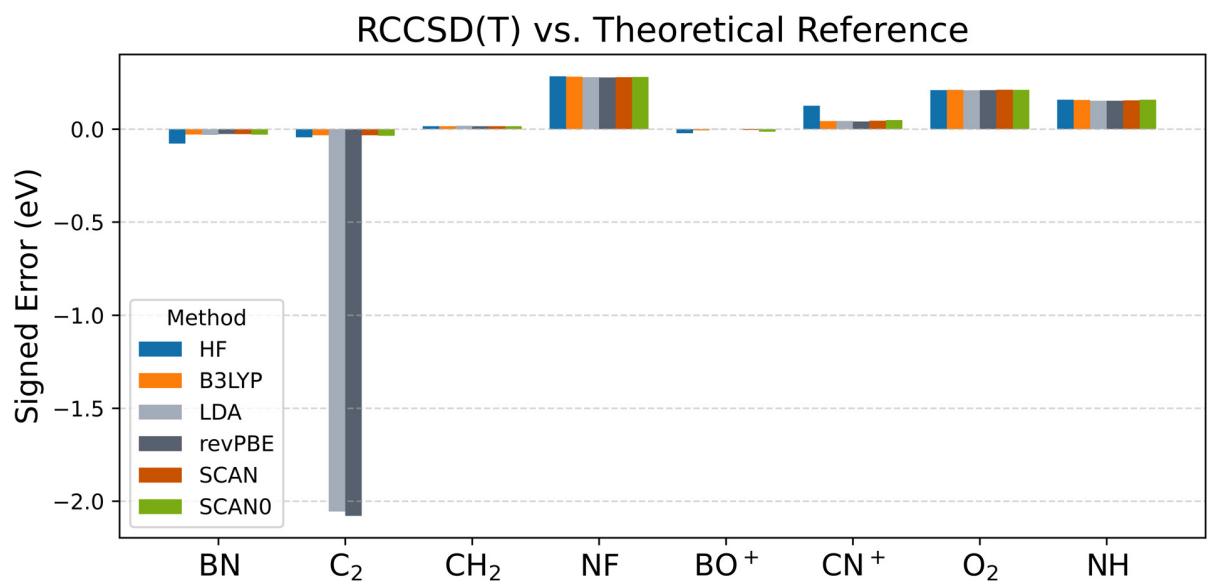

Figure S29: Signed errors of RCCSD(T) against theoretical references.

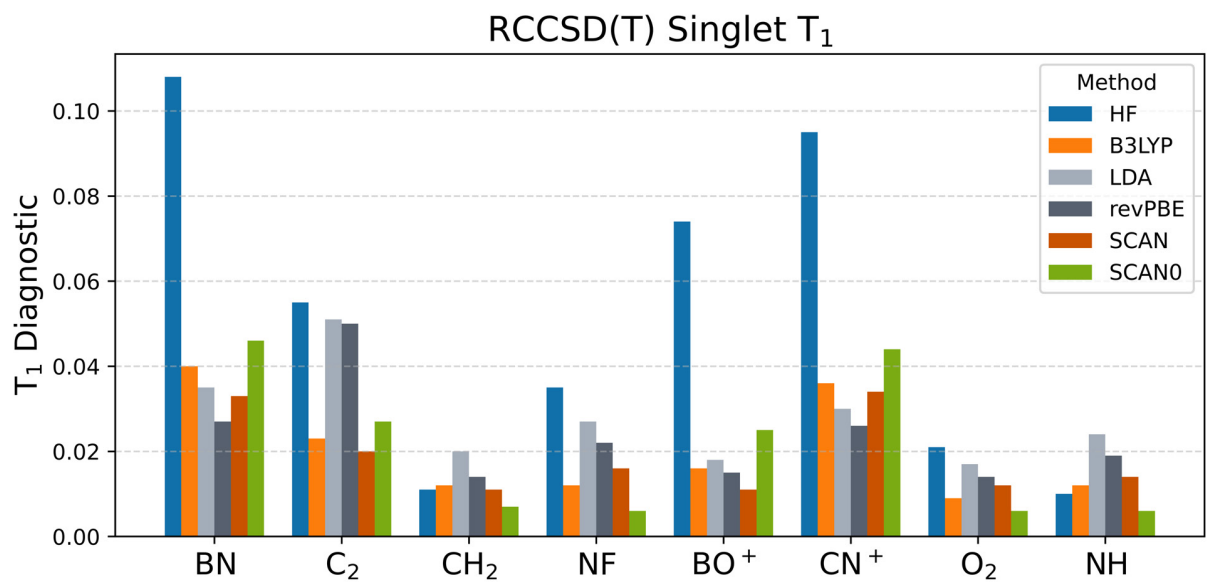

Figure S30: Closed-shell singlet $T_1$ with various reference determinants.

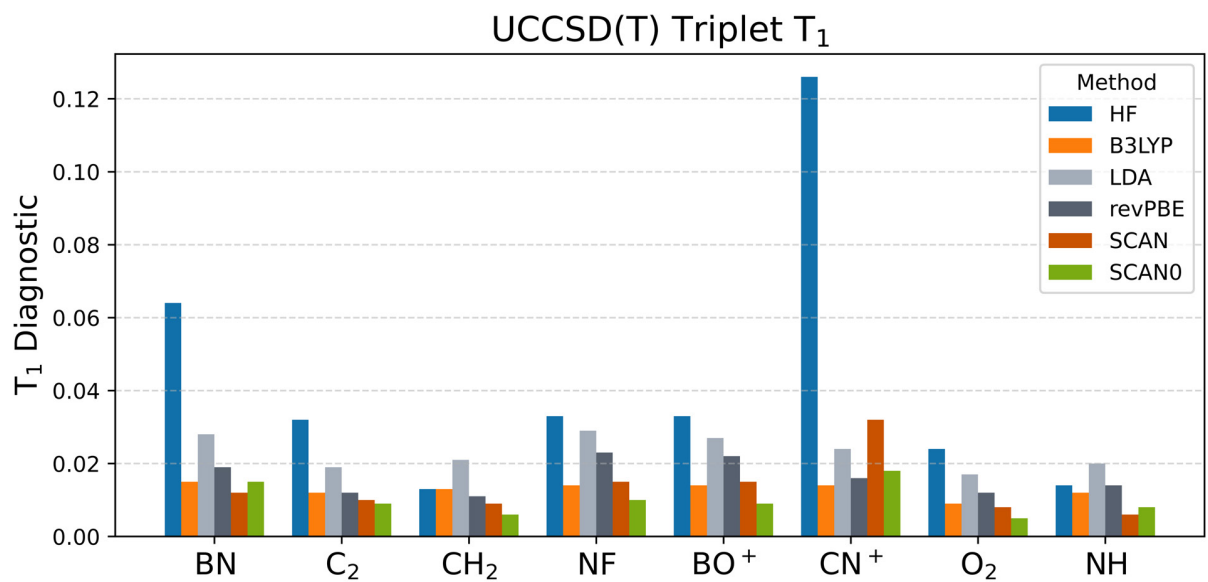

Figure S31: Unrestricted triplet $T_1$ with various reference determinants.

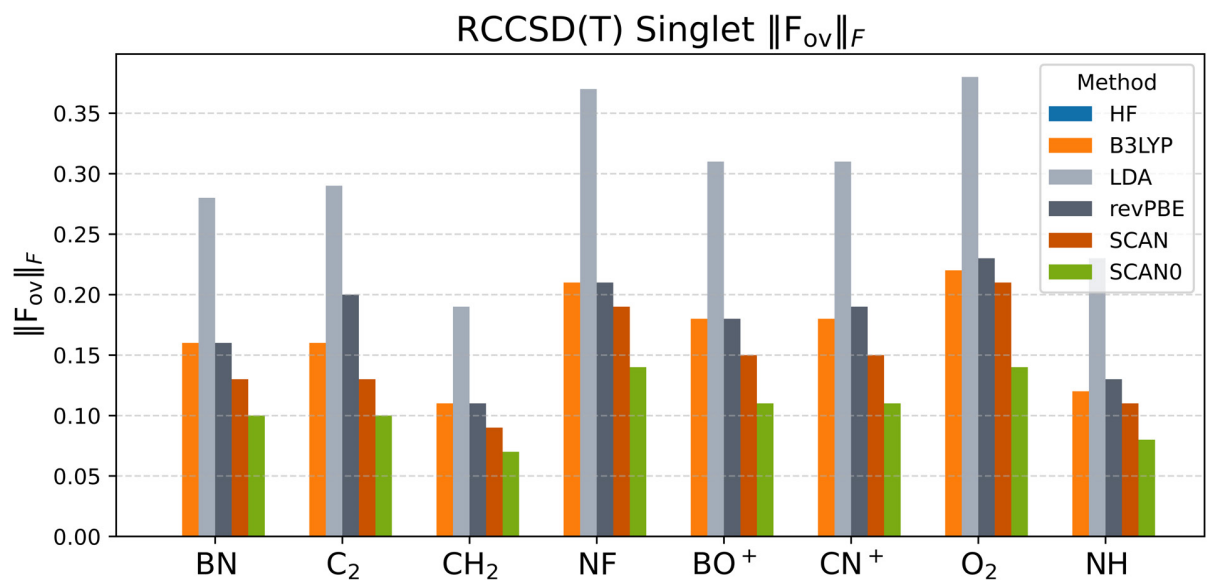

Figure S32: Closed-shell singlet occupied-virtual Fock matrix Frobenius norms with various reference determinants.

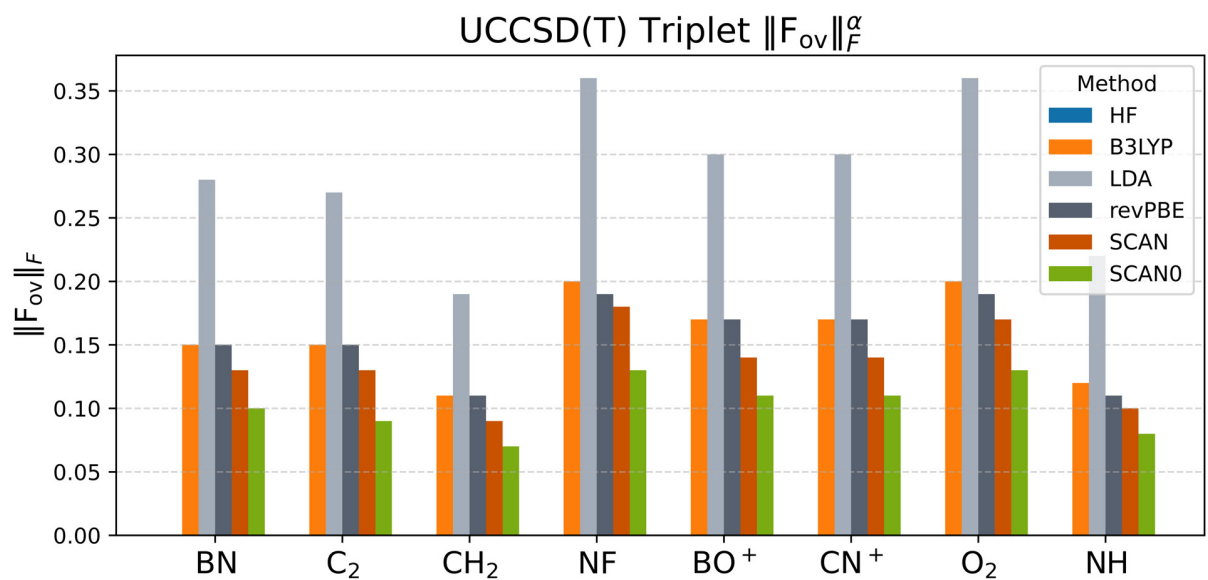

Figure S33: Unrestricted triplet occupied-virtual Fock matrix (alpha) Frobenius norms with various reference determinants.

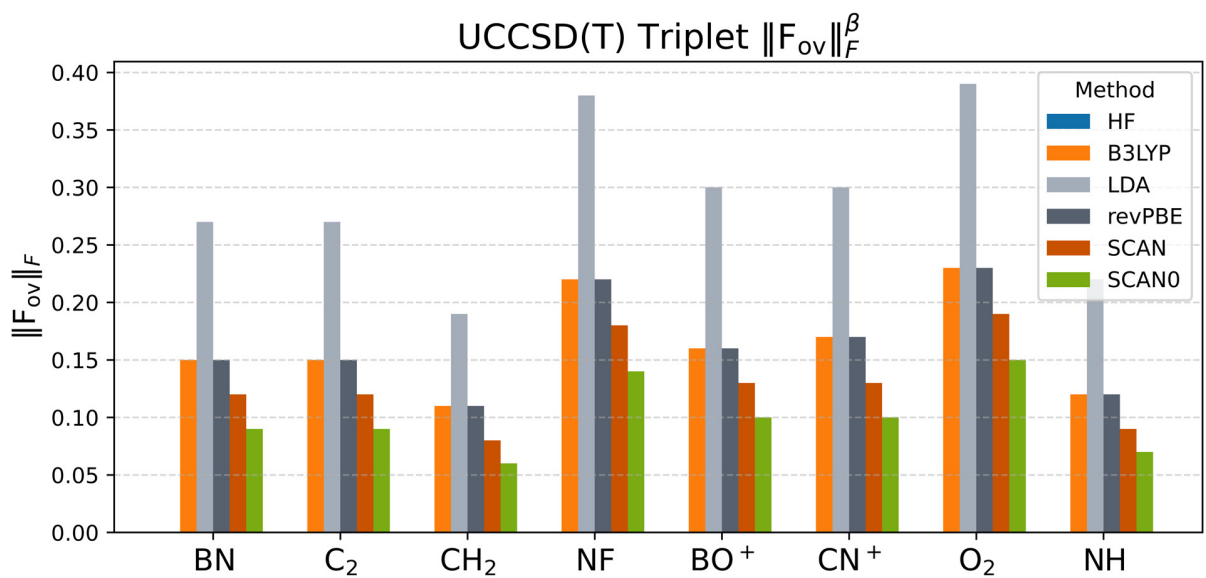

Figure S34: Unrestricted triplet occupied-virtual Fock matrix (beta) Frobenius norms with various reference determinants.

## S15.2 Open-shell Singlets: CCSD(T)

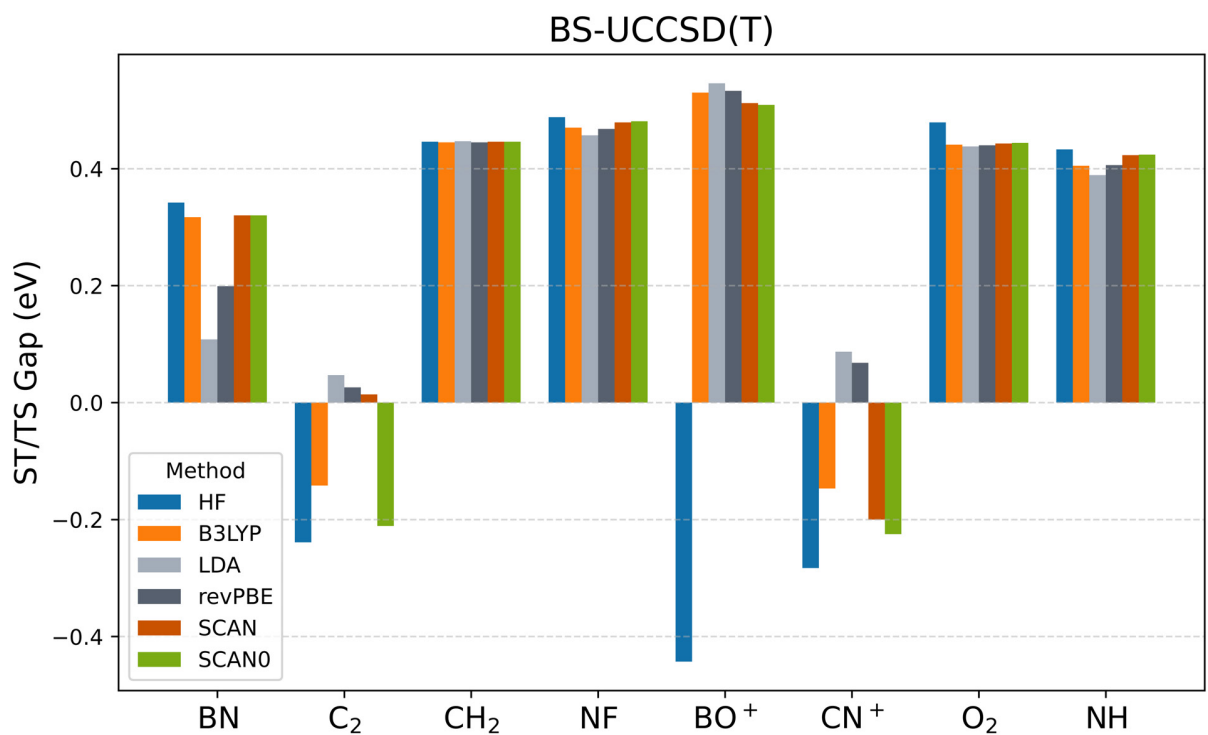

Figure S35: Singlet-triplet gaps with unrestricted broken-symmetry singlets.

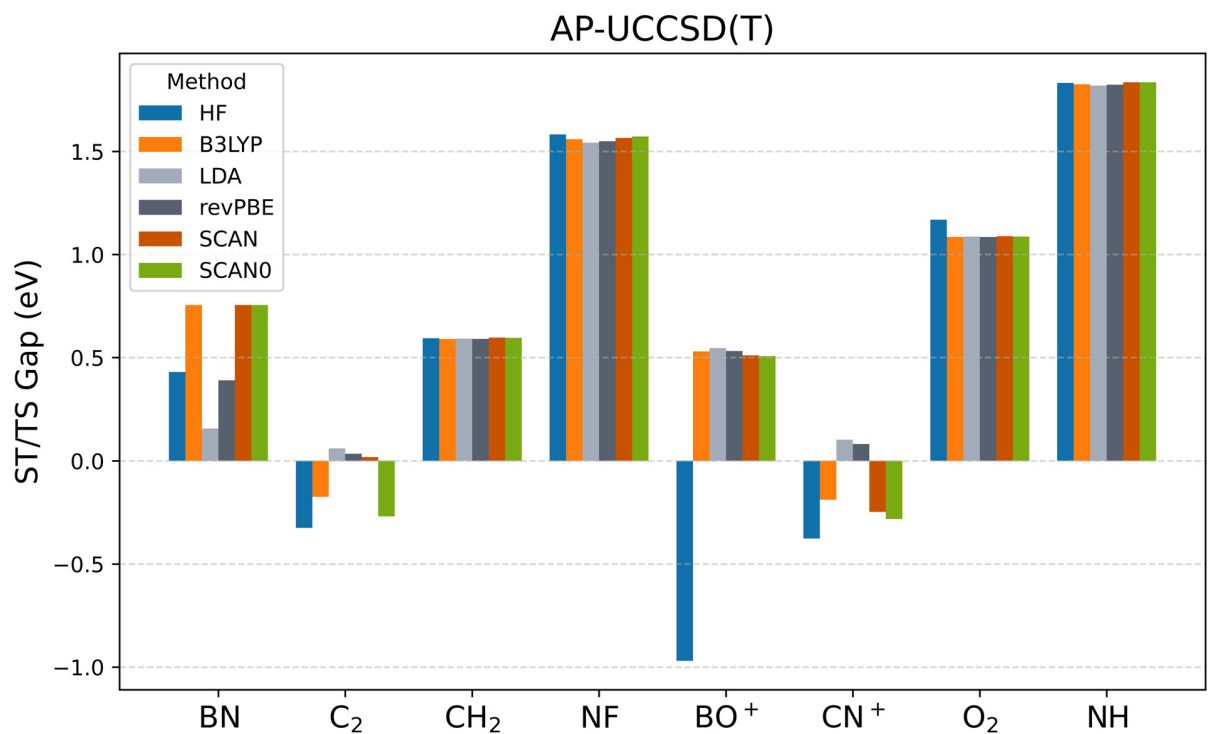

Figure S36: Singlet-triplet gaps corrected by approximate spin projection with unrestricted broken-symmetry singlets.

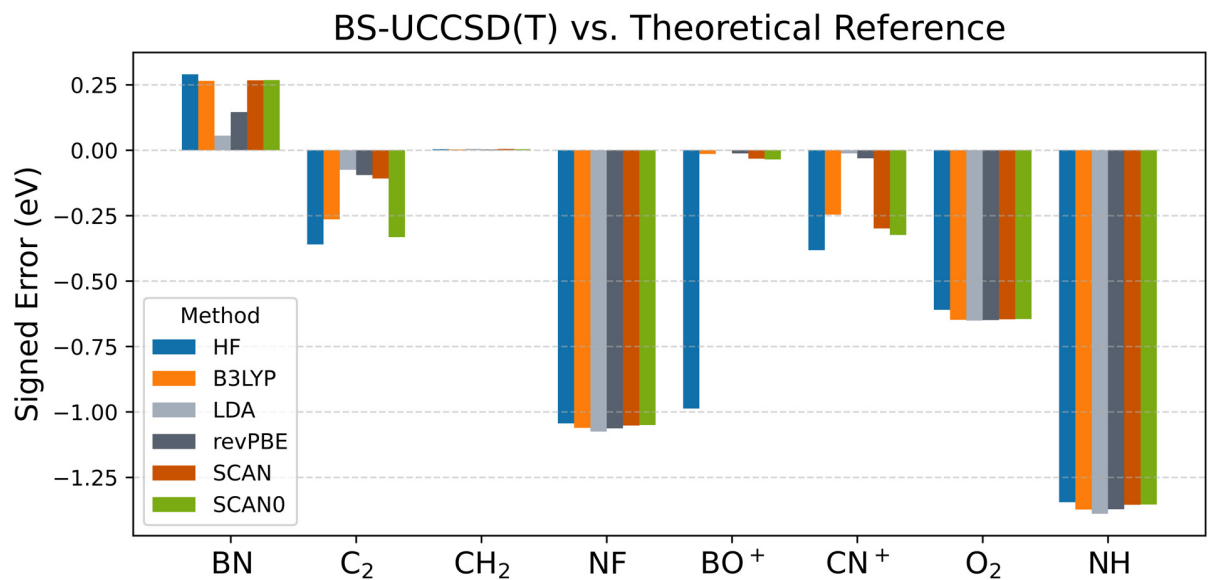

Figure S37: Signed errors of UCCSD(T) against theoretical references.

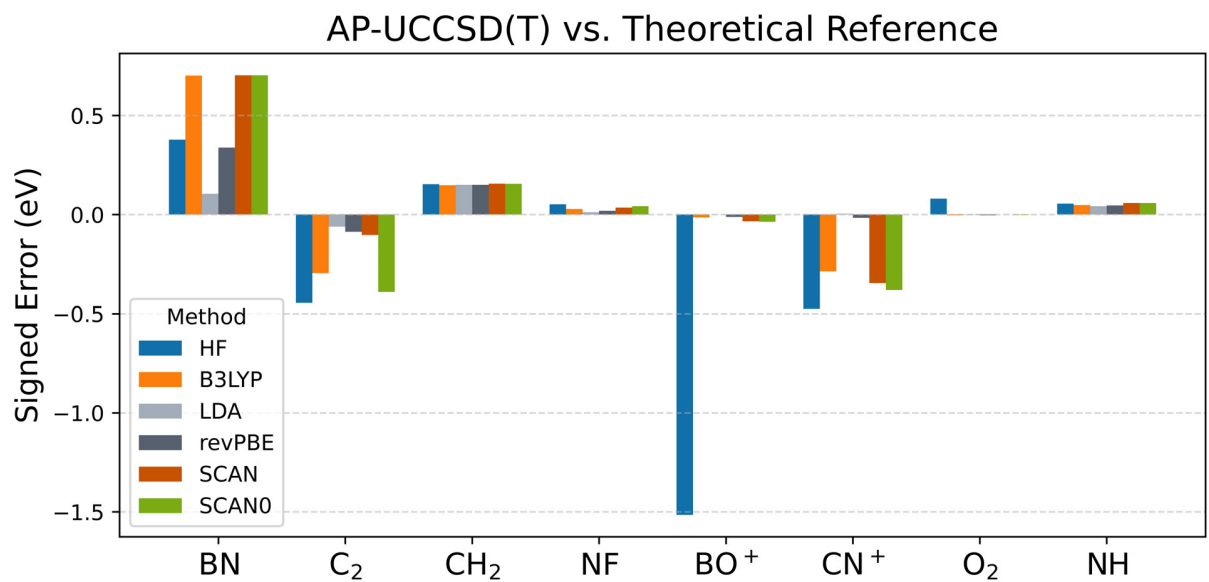

Figure S38: Signed errors of UCCSD(T) with approximate spin projection against theoretical references.

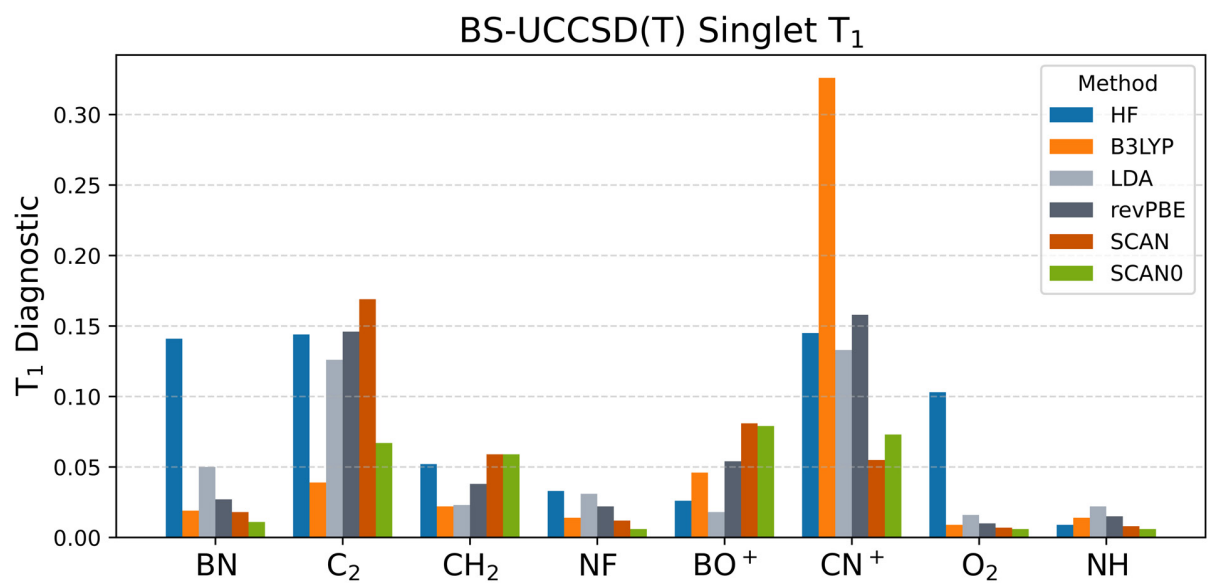

Figure S39: Unrestricted broken-symmetry singlet $T_1$ with various reference determinants.

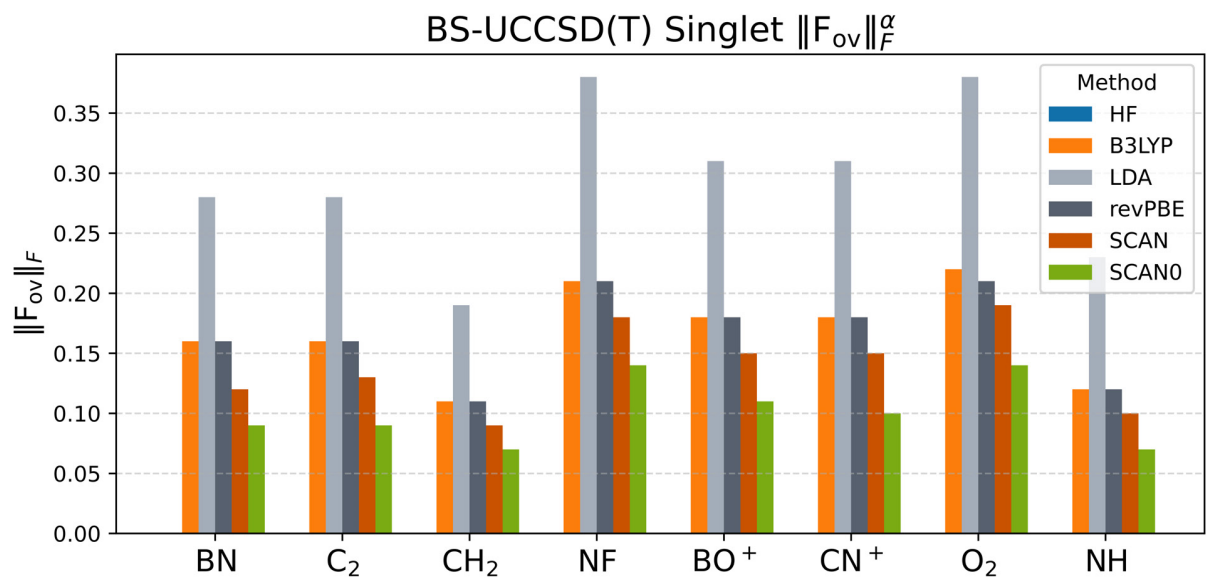

Figure S40: Unrestricted singlet occupied-virtual Fock matrix (alpha) Frobenius norms with various reference determinants.

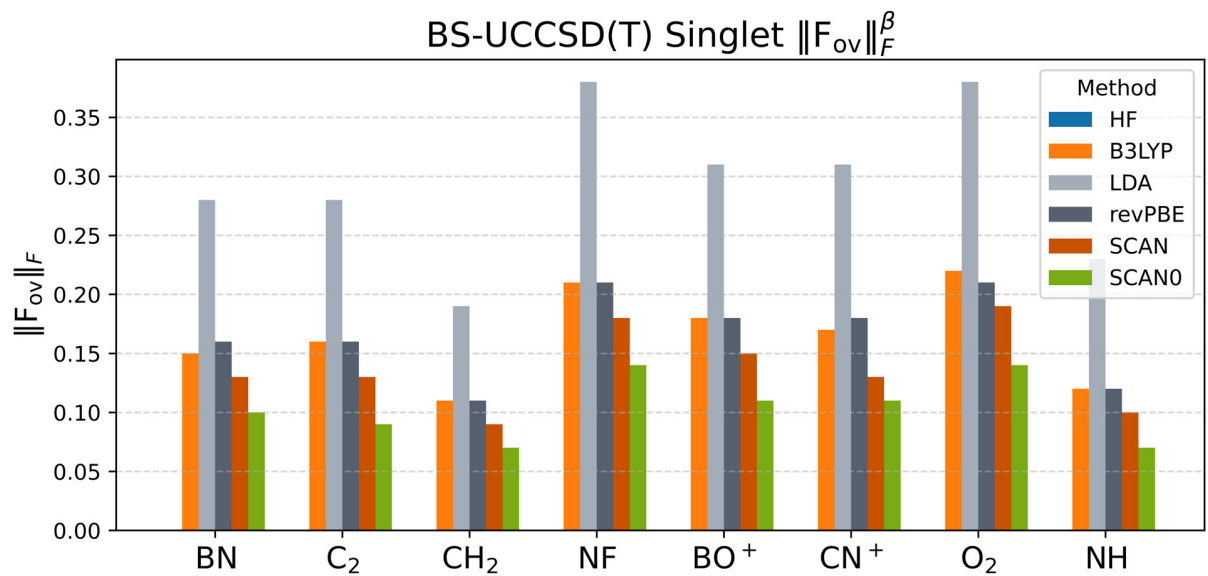

Figure S41: Unrestricted singlet occupied-virtual Fock matrix (beta) Frobenius norms with various reference determinants.

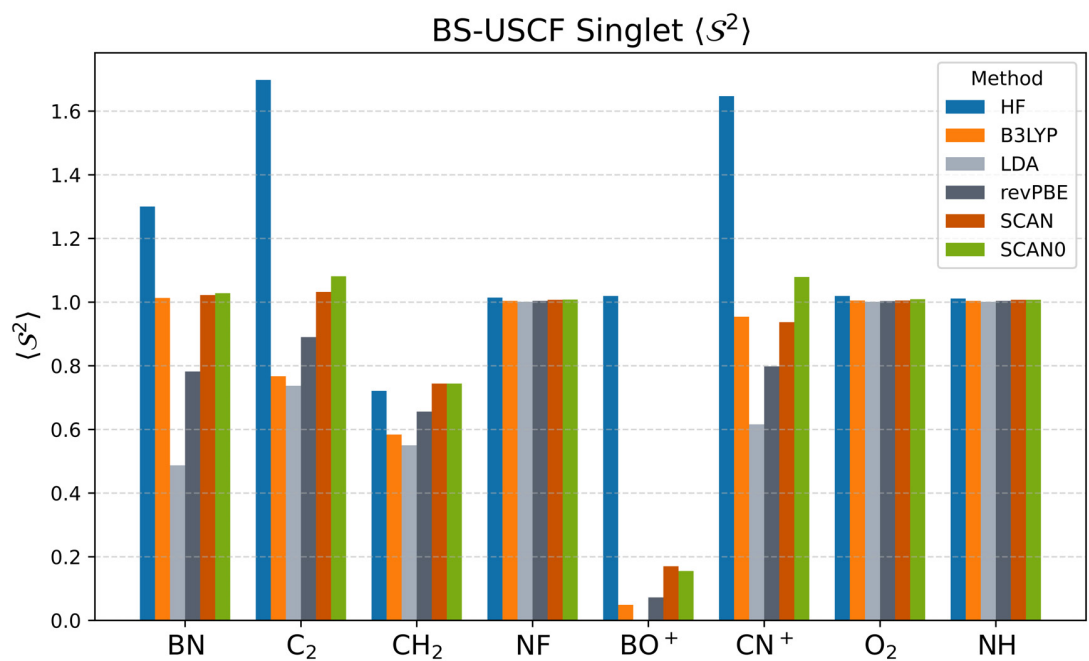

Figure S42: Unrestricted singlet SCF total spin-squared expectation values with various reference determinants.

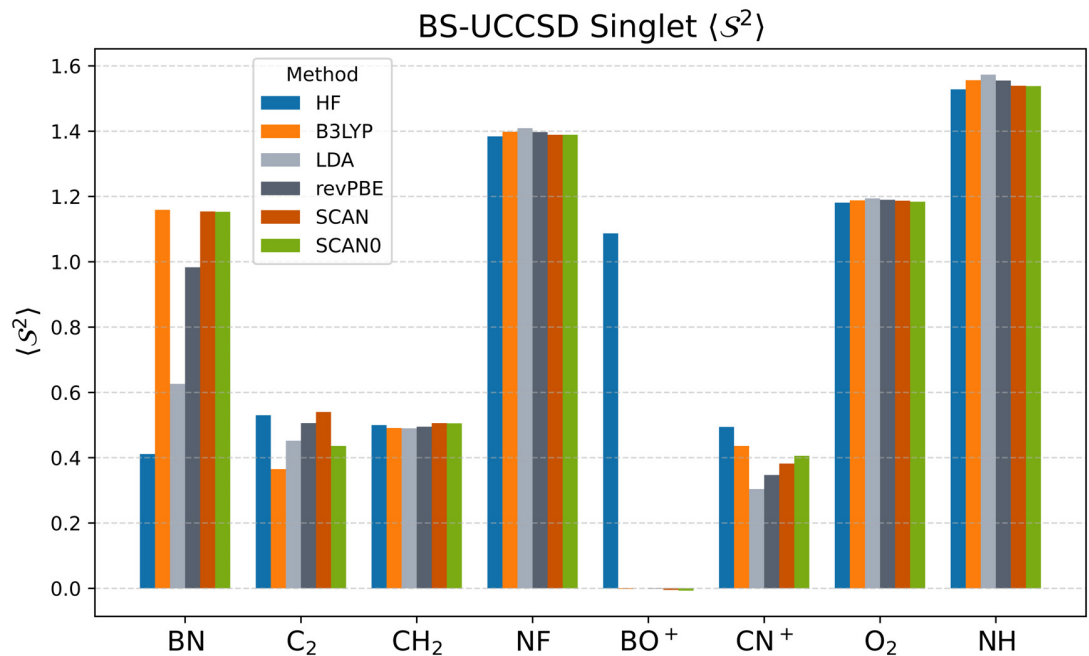

Figure S43: Unrestricted singlet CCSD total spin-squared expectation values with various reference determinants.

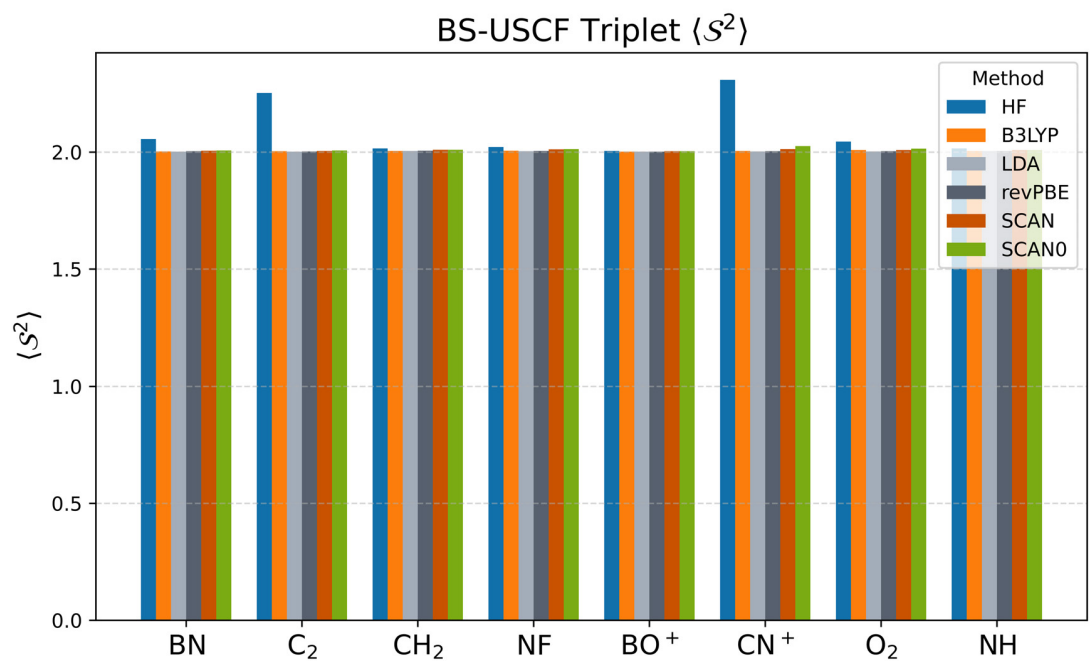

Figure S44: Unrestricted triplet SCF total spin-squared expectation values with various reference determinants.

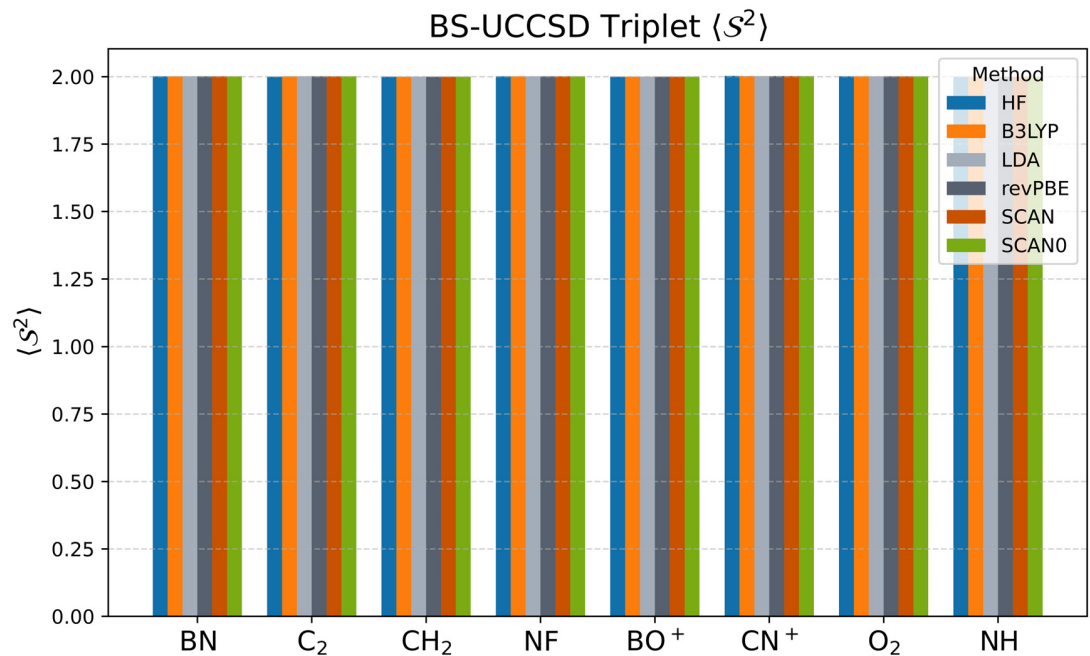

Figure S45: Unrestricted triplet CCSD total spin-squared expectation values with various reference determinants.

## S15.3 Features and Criteria

Quantitative and qualitative features are mapped out for the main-group singlet-triplet gap results. Gap sign, errors with respect to CCSD(T) with a HF reference (2 sig. figs., rounded), magnitude of Kohn-Sham $T_1$ diagnostics versus HF, and differences of the absolute value of the smallest orbital energy denominators ( $\Delta_{ia}/\Delta_{ijab}$ ) with respect to HF are examined.

### RKS Features

| Molecule-Functional | Correct Gap Sign? | KS-Gap Err < HF? | S=0 T1-KS < T1-HF? | S=1 T1-KS < T1-HF? | $|\delta HF|_{ia} \geq 0.01$ au? | $|\delta HF|_{ijab} \geq 0.01$ au? |
|---|---|---|---|---|---|---|
| BO$^+$-B3LYP | yes | yes | yes | yes | yes | yes |
| BO$^+$-LDA | yes | yes | yes | yes | yes | yes |
| BO$^+$-SCAN | yes | yes | yes | yes | yes | yes |
| BO$^+$-SCAN0 | yes | yes | yes | yes | yes | yes |
| BO$^+$-revPBE | yes | yes | yes | yes | yes | yes |
| CH$_2$-B3LYP | yes | no | no | no | no | yes |
| CH$_2$-LDA | yes | no | no | no | no | yes |
| CH$_2$-SCAN | yes | no | no | yes | no | yes |
| CH$_2$-SCAN0 | yes | no | yes | yes | no | no |
| CH$_2$-revPBE | yes | no | no | yes | no | yes |
| CN$^+$-B3LYP | yes | yes | yes | yes | no | no |
| CN$^+$-LDA | yes | yes | yes | yes | no | yes |
| CN$^+$-SCAN | yes | yes | yes | yes | no | no |
| CN$^+$-SCAN0 | yes | yes | yes | yes | no | no |
| CN$^+$-revPBE | yes | yes | yes | yes | no | no |
| C$_2$-B3LYP | yes | yes | yes | yes | yes | yes |
| C$_2$-LDA | no | no | yes | yes | yes | yes |
| C$_2$-SCAN | yes | yes | yes | yes | yes | yes |
| C$_2$-SCAN0 | yes | yes | yes | yes | yes | yes |
| C$_2$-revPBE | no | no | yes | yes | no | yes |
| O$_2$-B3LYP | yes | no | yes | yes | yes | yes |
| O$_2$-LDA | yes | no | yes | yes | yes | yes |
| O$_2$-SCAN | yes | no | yes | yes | yes | yes |
| O$_2$-SCAN0 | yes | no | yes | yes | yes | yes |
| O$_2$-revPBE | yes | no | yes | yes | yes | yes |
| BN-B3LYP | yes | yes | yes | yes | yes | yes |
| BN-LDA | yes | yes | yes | yes | yes | yes |
| BN-SCAN | yes | yes | yes | yes | yes | yes |
| BN-SCAN0 | yes | yes | yes | yes | yes | yes |
| BN-revPBE | yes | yes | yes | yes | yes | yes |
| NF-B3LYP | yes | no | yes | yes | yes | yes |
| NF-LDA | yes | no | yes | yes | yes | yes |
| NF-SCAN | yes | no | yes | yes | yes | yes |
| NF-SCAN0 | yes | no | yes | yes | yes | yes |
| NF-revPBE | yes | no | yes | yes | yes | yes |
| NH-B3LYP | yes | no | no | yes | no | no |
| NH-LDA | yes | yes | no | no | no | no |
| NH-SCAN | yes | no | no | yes | no | yes |
| NH-SCAN0 | yes | no | yes | yes | no | no |
| NH-revPBE | yes | yes | no | no | no | no |

Figure S46: Features of main-group singlet-triplet gap results with closed-shell singlet references and various reference determinants.

| Molecule-Functional | Correct Gap Sign? | KS-Gap Err < HF? | S=0 T1-KS < T1-HF? | S=1 T1-KS < T1-HF? | UKS improves RKS? | $|\delta HF|ia \geq 0.01$ au? | $|\delta HF|ijab \geq 0.01$ au? |
|---|---|---|---|---|---|---|---|
| BN-B3LYP | yes | no | yes | yes | no | yes | yes |
| BN-LDA | yes | yes | yes | yes | no | yes | yes |
| BN-SCAN | yes | no | yes | yes | no | yes | yes |
| BN-SCAN0 | yes | no | yes | yes | no | yes | yes |
| BN-revPBE | yes | yes | yes | yes | no | yes | yes |
| $BO^+$-B3LYP | yes | yes | no | yes | no | yes | yes |
| $BO^+$-LDA | yes | yes | yes | yes | no | yes | yes |
| $BO^+$-SCAN | yes | yes | no | yes | no | yes | yes |
| $BO^+$-SCAN0 | yes | yes | no | yes | no | yes | yes |
| $BO^+$-revPBE | yes | yes | no | yes | no | yes | yes |
| $CH_2$-B3LYP | yes | yes | yes | no | no | yes | yes |
| $CH_2$-LDA | yes | no | yes | no | no | yes | yes |
| $CH_2$-SCAN | yes | no | no | yes | no | no | yes |
| $CH_2$-SCAN0 | yes | no | no | yes | no | no | yes |
| $CH_2$-revPBE | yes | no | yes | yes | no | yes | yes |
| $CN^+$-B3LYP | no | yes | no | yes | no | yes | yes |
| $CN^+$-LDA | yes | yes | yes | yes | yes | yes | yes |
| $CN^+$-SCAN | no | yes | yes | yes | no | yes | yes |
| $CN^+$-SCAN0 | no | yes | yes | yes | no | yes | yes |
| $CN^+$-revPBE | yes | yes | no | yes | yes | yes | yes |
| $C_2$-B3LYP | no | yes | yes | yes | no | yes | yes |
| $C_2$-LDA | yes | yes | yes | yes | no | yes | yes |
| $C_2$-SCAN | yes | yes | no | yes | yes | yes | yes |
| $C_2$-SCAN0 | no | yes | yes | yes | no | yes | yes |
| $C_2$-revPBE | yes | yes | no | yes | yes | yes | yes |
| NF-B3LYP | yes | yes | yes | yes | yes | yes | yes |
| NF-LDA | yes | yes | yes | yes | yes | yes | yes |
| NF-SCAN | yes | yes | yes | yes | yes | yes | yes |
| NF-SCAN0 | yes | yes | yes | yes | yes | yes | yes |
| NF-revPBE | yes | yes | yes | yes | yes | yes | yes |
| NH-B3LYP | yes | yes | no | yes | yes | yes | yes |
| NH-LDA | yes | yes | no | no | yes | yes | yes |
| NH-SCAN | yes | no | yes | yes | yes | no | yes |
| NH-SCAN0 | yes | no | yes | yes | yes | no | yes |
| NH-revPBE | yes | yes | no | no | yes | yes | yes |
| $O_2$-B3LYP | yes | yes | yes | yes | yes | yes | yes |
| $O_2$-LDA | yes | yes | yes | yes | yes | yes | yes |
| $O_2$-SCAN | yes | yes | yes | yes | yes | yes | yes |
| $O_2$-SCAN0 | yes | yes | yes | yes | yes | yes | yes |
| $O_2$-revPBE | yes | yes | yes | yes | yes | yes | yes |

UKS Features (AP)

Criteria

Figure S47: Features of main-group singlet-triplet gap results corrected by approximate spin projection with unrestricted broken-symmetry singlet references and various reference determinants.

## S16   $T_1$ and Absolute Error Correlations for Metal Hydrides

The relationship between $T_1^2$ and absolute errors with respect to experiment is examined for metal-hydrides (M-H) and metal-hydride cations (M-$H^+$). Alternative BDE calculations using CCSD(T)/def2-QZVPP with different functionals are performed. All electrons are considered. Structures, relativistic corrections, and vibrational corrections from the literature are included.[35] We also examine the qualitative feature of improvements upon HF-CC using KS-CC with the selected functionals. There is no consistent correlation between the magnitude of $T_1^2$ and absolute errors.

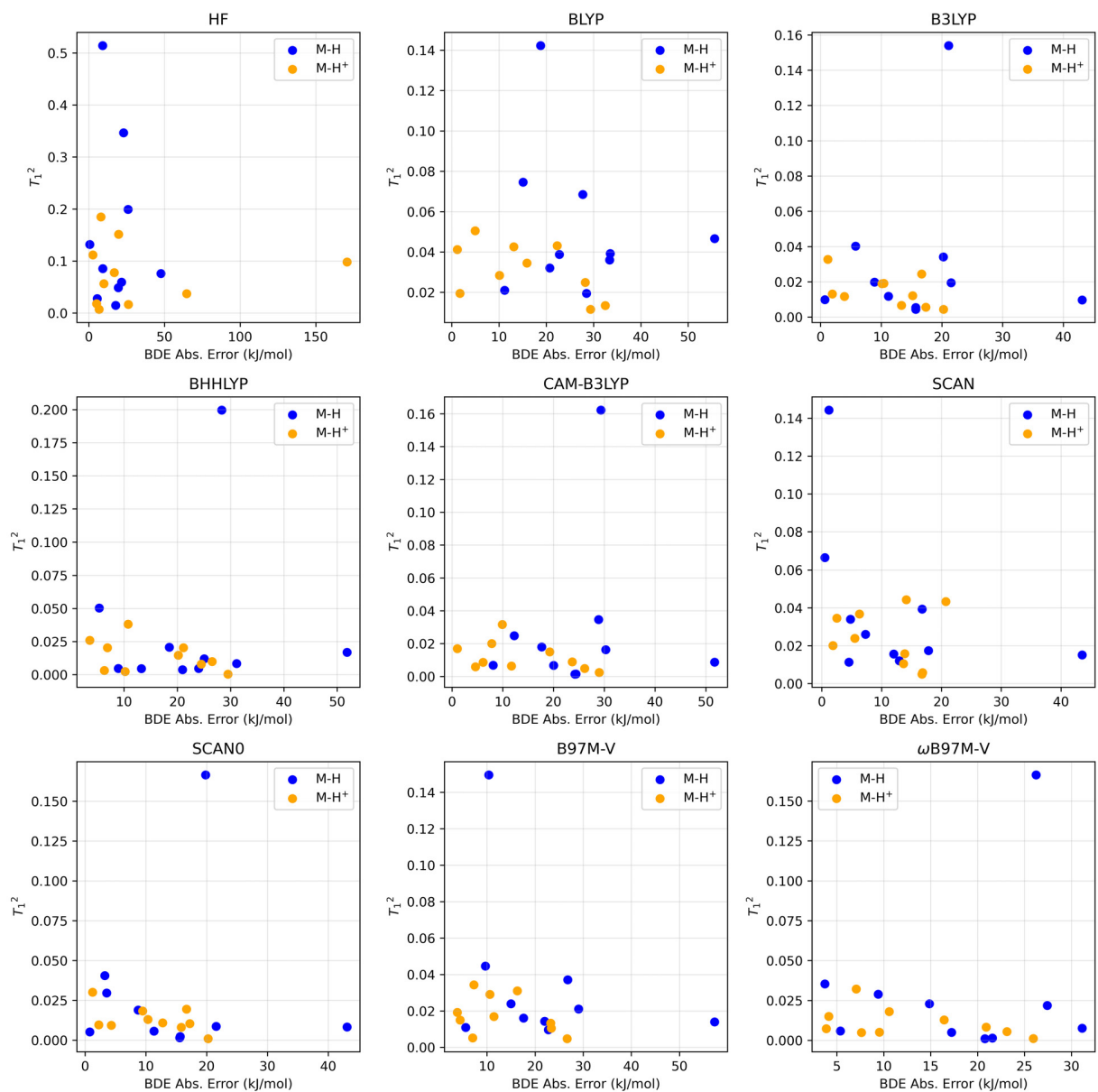

Figure S48: Scatter plots comparing the magnitudes of $T_1^2$ and absolute errors with respect to experimental BDEs. HF-based and KS-based CCSD(T)/def2-QZVPP calculations, with various functionals, are performed.

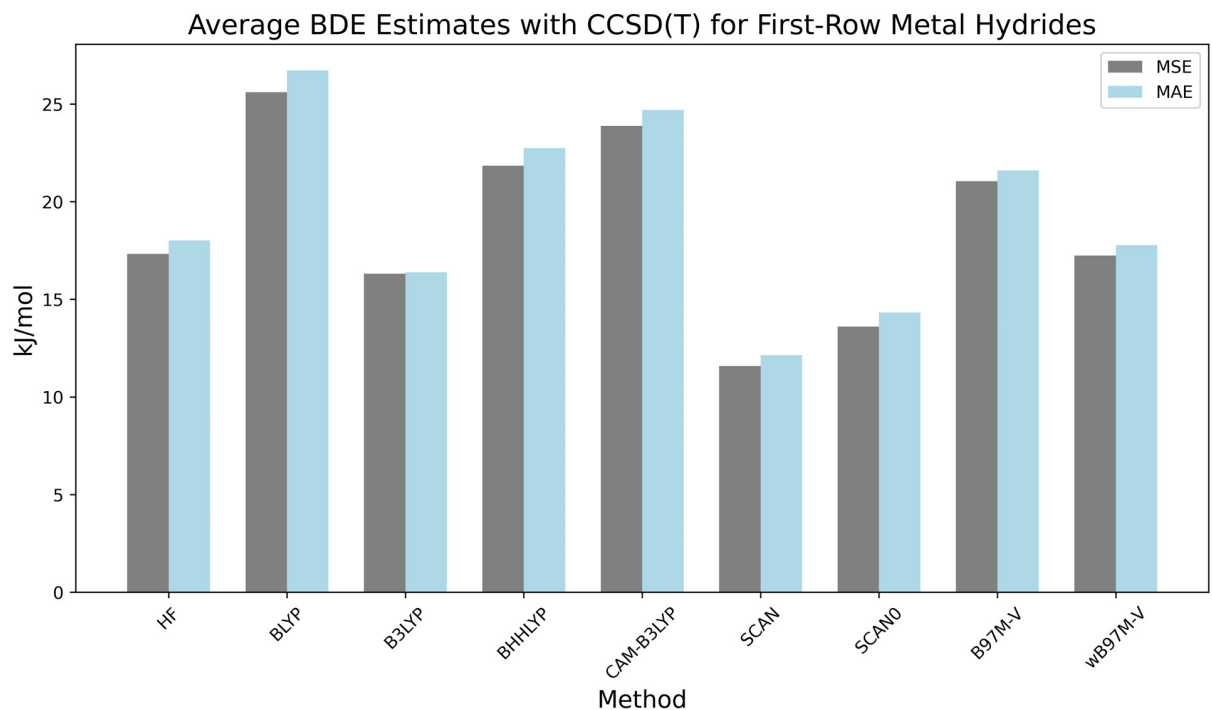

Figure S49: Signed and absolute errors with respect to experiment for metal-hydride BDEs computed with CCSD(T) with HF and various DFAs.

Figure S50: Qualitative features indicating absolute error reductions with respect to HF-CCSD(T) for metal-hydrides.

41

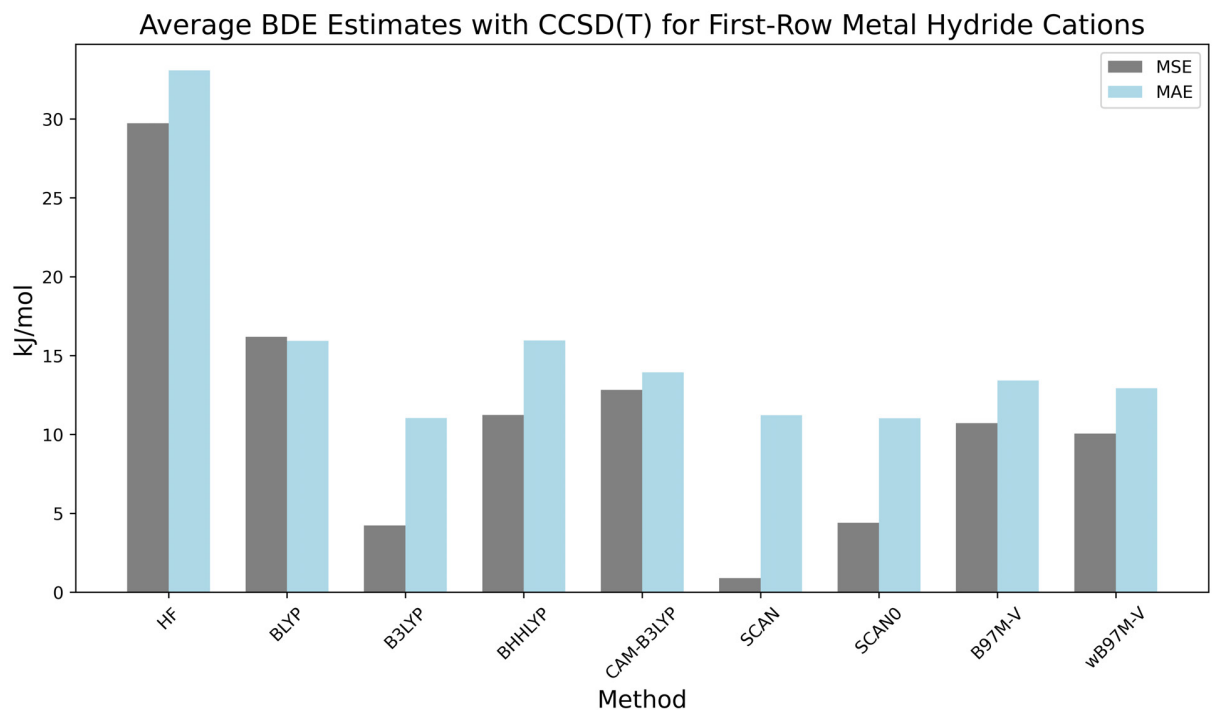

Figure S51: Signed and absolute errors with respect to experiment for metal-hydride cation BDEs computed with CCSD(T) with HF and various DFAs.

Figure S52: Qualitative features indicating absolute error reductions with respect to HF-CCSD(T) for metal-hydride cations

# References

[1] Frank Neese. The ORCA program system. *WIREs Comput. Mol. Sci.*, 2(1):73–78, 1 2012. ISSN 1759-0876. doi: 10.1002/wcms.81.

[2] Frank Neese. Software update: the ORCA program system, version 4.0. *WIREs Comput. Mol. Sci.*, 8(1), 1 2018. ISSN 1759-0876. doi: 10.1002/wcms.1327.

[3] Frank Neese, Frank Wennmohs, Ute Becker, and Christoph Riplinger. The ORCA quantum chemistry program package. *J. Chem. Phys.*, 152(22), 6 2020. ISSN 0021-9606. doi: 10.1063/5.0004608.

[4] Frank Neese. The SHARK integral generation and digestion system. *J. Comp. Chem.*, 44(3): 381–396, 1 2023. ISSN 0192-8651. doi: 10.1002/jcc.26942.

[5] Frank Neese. Software Update: The ORCA Program System—Version 6.0. *WIREs Comput. Mol. Sci.*, 15(2), 3 2025. ISSN 1759-0876. doi: 10.1002/wcms.70019.

[6] S. H. Vosko, L. Wilk, and M. Nusair. Accurate spin-dependent electron liquid correlation energies for local spin density calculations: a critical analysis. *Can. J. Phys.*, 58(8):1200–1211, 8 1980. ISSN 0008-4204. doi: 10.1139/p80-159.

[7] John P. Perdew, Kieron Burke, and Matthias Ernzerhof. Generalized Gradient Approximation Made Simple. *Phys. Rev. Lett.*, 77(18):3865–3868, 10 1996. ISSN 0031-9007. doi: 10.1103/PhysRevLett.77.3865.

[8] John P Perdew, Kieron Burke, and Matthias Ernzerhof. Generalized Gradient Approximation Made Simple [Phys. Rev. Lett. 77, 3865 (1996)]. *Phys. Rev. Lett.*, 78(7):1396–1396, 1997. doi: 10.1103/PhysRevLett.78.1396. URL https://link.aps.org/doi/10.1103/PhysRevLett.78.1396.

[9] John P. Perdew and Yue Wang. Accurate and simple analytic representation of the electron-gas correlation energy. *Phys. Rev. B*, 45(23):13244–13249, 6 1992. ISSN 0163-1829. doi: 10.1103/PhysRevB.45.13244.

[10] John P. Perdew and Yue Wang. Erratum: Accurate and simple analytic representation of the electron-gas correlation energy [Phys. Rev. B 45, 13244 (1992)]. *Phys. Rev. B*, 98(7):079904, 8 2018. ISSN 2469-9950. doi: 10.1103/PhysRevB.98.079904.

[11] James W. Furness, Aaron D. Kaplan, Jinliang Ning, John P. Perdew, and Jianwei Sun. Accurate and Numerically Efficient r$^{2}$ SCAN Meta-Generalized Gradient Approximation. *J. Phys. Chem. Lett.*, 11(19):8208–8215, 10 2020. ISSN 1948-7185. doi: 10.1021/acs.jpclett.0c02405.

[12] James W. Furness, Aaron D. Kaplan, Jinliang Ning, John P. Perdew, and Jianwei Sun. Correction to "Accurate and Numerically Efficient r$^{2}$ SCAN Meta-Generalized Gradient Approximation". *J. Phys. Chem. Lett.*, 11(21):9248–9248, 11 2020. ISSN 1948-7185. doi: 10.1021/acs.jpclett.0c03077.

[13] Carlo Adamo and Vincenzo Barone. Toward reliable density functional methods without adjustable parameters: The PBE0 model. *J. Chem. Phys.*, 110(13):6158–6170, 4 1999. ISSN 00219606. doi: 10.1063/1.478522.

[14] Axel D. Becke. Density-functional thermochemistry. III. The role of exact exchange. *J. Chem. Phys.*, 98(7):5648–5652, 1993. ISSN 00219606. doi: 10.1063/1.464913.

[15] Narbe Mardirossian and Martin Head-Gordon. ωB97X-V: A 10-parameter, range-separated hybrid, generalized gradient approximation density functional with nonlocal correlation, designed by a survival-of-the-fittest strategy. *Phys. Chem. Chem. Phys.*, 16(21):9904, 2014. ISSN 1463-9076. doi: 10.1039/c3cp54374a.

[16] Narbe Mardirossian and Martin Head-Gordon. ωB97M-V: A combinatorially optimized, range-separated hybrid, meta-GGA density functional with VV10 nonlocal correlation. *J. Chem. Phys.*, 144(21), 6 2016. ISSN 0021-9606. doi: 10.1063/1.4952647.

[17] Daoling Peng, Nils Middendorf, Florian Weigend, and Markus Reiher. An efficient implementation of two-component relativistic exact-decoupling methods for large molecules. *J. Chem. Phys.*, 138(18), 5 2013. ISSN 0021-9606. doi: 10.1063/1.4803693.

[18] Yannick J. Franzke, Nils Middendorf, and Florian Weigend. Efficient implementation of one- and two-component analytical energy gradients in exact two-component theory. *J. Chem. Phys.*, 148(10), 3 2018. ISSN 0021-9606. doi: 10.1063/1.5022153.

[19] Patrik Pollak and Florian Weigend. Segmented Contracted Error-Consistent Basis Sets of Double- and Triple-ζ Valence Quality for One- and Two-Component Relativistic All-Electron Calculations. *J. Chem. Theory Comput.*, 13(8):3696–3705, 8 2017. ISSN 1549-9618. doi: 10.1021/acs.jctc.7b00593.

[20] Amir Karton and Jan M. L. Martin. Comment on: "Estimating the Hartree–Fock limit from finite basis set calculations" [Jensen F (2005) Theor Chem Acc 113:267]. *Theor. Chem. Acc.*, 115(4):330–333, 4 2006. ISSN 1432-881X. doi: 10.1007/s00214-005-0028-6.

[21] Shijun Zhong, Ericka C. Barnes, and George A. Petersson. Uniformly convergent n-tuple-ζ augmented polarized (nZaP) basis sets for complete basis set extrapolations. I. Self-consistent field energies. *J. Chem. Phys.*, 129(18), 11 2008. ISSN 0021-9606. doi: 10.1063/1.3009651.

[22] Donald G. Truhlar. Basis-set extrapolation. *Chem. Phys. Lett.*, 294(1-3):45–48, 9 1998. ISSN 00092614. doi: 10.1016/S0009-2614(98)00866-5.

[23] Frank Neese and Edward F. Valeev. Revisiting the Atomic Natural Orbital Approach for Basis Sets: Robust Systematic Basis Sets for Explicitly Correlated and Conventional Correlated <i>ab initio</i> Methods? *J. Chem. Theory Comput.*, 7(1):33–43, 1 2011. ISSN 1549-9618. doi: 10.1021/ct100396y.

[24] Florian Weigend and Reinhart Ahlrichs. Balanced basis sets of split valence, triple zeta valence and quadruple zeta valence quality for H to Rn: Design and assessment of accuracy. *Phys. Chem. Chem. Phys.*, 7(18):3297, 2005. ISSN 1463-9076. doi: 10.1039/b508541a.

[25] Florian Weigend, Filipp Furche, and Reinhart Ahlrichs. Gaussian basis sets of quadruple zeta valence quality for atoms H–Kr. *J. Chem. Phys.*, 119(24):12753–12762, 12 2003. ISSN 0021-9606. doi: 10.1063/1.1627293.

[26] Florian Weigend. Accurate Coulomb-fitting basis sets for H to Rn. *Phys. Chem. Chem. Phys.*, 8(9):1057, 2006. ISSN 1463-9076. doi: 10.1039/b515623h.

[27] Arnim Hellweg, Christof Hättig, Sebastian Höfener, and Wim Klopper. Optimized accurate auxiliary basis sets for RI-MP2 and RI-CC2 calculations for the atoms Rb to Rn. *Theor. Chem. Acc.*, 117(4):587–597, 3 2007. ISSN 1432-881X. doi: 10.1007/s00214-007-0250-5.

[28] E. van Lenthe, E. J. Baerends, and J. G. Snijders. Relativistic regular two-component Hamiltonians. *J. Chem. Phys.*, 99(6):4597–4610, 9 1993. ISSN 0021-9606. doi: 10.1063/1.466059.

[29] Christoph van Wüllen. Molecular density functional calculations in the regular relativistic approximation: Method, application to coinage metal diatomics, hydrides, fluorides and chlorides, and comparison with first-order relativistic calculations. *J. Chem. Phys.*, 109(2):392–399, 7 1998. ISSN 0021-9606. doi: 10.1063/1.476576.

[30] L. Visscher and K.G. Dyall. Dirac–Fock Atomic Electronic Structure Calculations Using Different Nuclear Charge Distributions. *Atom. Data Nucl. Data Tabl.*, 67(2):207–224, 11 1997. ISSN 0092640X. doi: 10.1006/adnd.1997.0751.

[31] Nathan E. Schultz, Yan Zhao, and Donald G. Truhlar. Databases for transition element bonding: Metal-metal bond energies and bond lengths and their use to test hybrid, hybrid meta, and meta density functionals and generalized gradient approximations. *J. Phys. Chem. A*, 109(19):4388–4403, 2005. doi: 10.1021/jp0504468.

[32] Junwei Lucas Bao, Samuel O. Odoh, Laura Gagliardi, and Donald G. Truhlar. Predicting Bond Dissociation Energies of Transition-Metal Compounds by Multiconfiguration Pair-Density Functional Theory and Second-Order Perturbation Theory Based on Correlated Participating Orbitals and Separated Pairs. *J. Chem. Theory Comput.*, 13(2):616–626, 2 2017. ISSN 1549-9618. doi: 10.1021/acs.jctc.6b01102.

[33] Yuri A. Aoto, Ana Paula de Lima Batista, Andreas Köhn, and Antonio G. S. de Oliveira-Filho. How To Arrive at Accurate Benchmark Values for Transition Metal Compounds: Computation or Experiment? *J. Chem. Theory Comput.*, 13(11):5291–5316, 11 2017. ISSN 1549-9618. doi: 10.1021/acs.jctc.7b00688.

[34] J. Bruce. Schilling, William A. Goddard, and J. L. Beauchamp. Theoretical studies of transition-metal hydrides. 2. Calcium monohydride(1+) through zinc monohydride(1+). *J. Phys. Chem.*, 91(22):5616–5623, 10 1987. ISSN 0022-3654. doi: 10.1021/j100306a024.

[35] Klaus A. Moltved and Kasper P. Kepp. The Metal Hydride Problem of Computational Chemistry: Origins and Consequences. *J. Phys. Chem. A*, 123(13):2888–2900, 4 2019. ISSN 1089-5639. doi: 10.1021/acs.jpca.9b02367.

[36] Wanyi Jiang, Nathan J. DeYonker, and Angela K. Wilson. Multireference Character for 3d Transition-Metal-Containing Molecules. *J. Chem. Theory Comput.*, 8(2):460–468, 2 2012. ISSN 1549-9618. doi: 10.1021/ct2006852.

[37] Apostolos Kalemos and Aristides Mavridis. The electronic structure of Ti2 and Ti2+. *J. Chem. Phys.*, 135(13), 10 2011. ISSN 0021-9606. doi: 10.1063/1.3643380.

[38] Chad E. Hoyer, Giovanni Li Manni, Donald G. Truhlar, and Laura Gagliardi. Controversial electronic structures and energies of Fe2, ${\rm Fe}_2^+$ Fe 2+, and ${\rm Fe}_2^-$ Fe 2- resolved by RASPT2 calculations. *J. Chem. Phys.*, 141(20), 11 2014. ISSN 0021-9606. doi: 10.1063/1.4901718.

[39] Arthur Kant, Sin-Shong Lin, and Bernard Strauss. Dissociation Energy of Mn2. *J. Chem. Phys.*, 49(4):1983–1985, 8 1968. ISSN 0021-9606. doi: 10.1063/1.1670350.

[40] C. A. Baumann, R. J. Van Zee, S. V. Bhat, and W. Weltner. ESR of diatomic manganese and pentaatomic manganese molecules in rare gas matrices. *J. Chem. Phys.*, 78(1):190–199, 1 1983. ISSN 0021-9606. doi: 10.1063/1.444540.

[41] M. Cheeseman, R. J. Van Zee, H. L. Flanagan, and W. Weltner. Transition-metal diatomics: Mn2, Mn+2, CrMn. *J. Chem. Phys.*, 92(3):1553–1559, 2 1990. ISSN 0021-9606. doi: 10.1063/1.458086.

[42] Rosendo Pou-Amerigo, Manuela Merchan, Ignacio Nebot-Gil, Per-Ake Malmqvist, and Bjorn O. Roos. The chemical bonds in CuH, Cu2, NiH, and Ni2 studied with multiconfigurational second order perturbation theory. *J. Chem. Phys.*, 101(6):4893–4902, 9 1994. ISSN 0021-9606. doi: 10.1063/1.467411.

[43] Apostolos Kalemos, Ilya G. Kaplan, and Aristides Mavridis. The Sc2 dimer revisited. *J. Chem. Phys.*, 132(2), 1 2010. ISSN 0021-9606. doi: 10.1063/1.3290951.

[44] J. L. Elkind and P. B. Armentrout. State-specific reactions of atomic transition-metal ions with molecular hydrogen, hydrogen deuteride, and molecular deuterium: effects of d orbitals on chemistry. *J. Phys. Chem.*, 91(8):2037–2045, 4 1987. ISSN 0022-3654. doi: 10.1021/j100292a012.

[45] Gennady L. Gutsev and Charles W. Bauschlicher. Chemical Bonding, Electron Affinity, and Ionization Energies of the Homonuclear 3d Metal Dimers. *J. Phys. Chem. A*, 107(23):4755–4767, 6 2003. ISSN 1089-5639. doi: 10.1021/jp030146v.

[46] Zongtang Fang, Monica Vasiliu, Kirk A. Peterson, and David A. Dixon. Prediction of Bond Dissociation Energies/Heats of Formation for Diatomic Transition Metal Compounds: CCSD(T) Works. *J. Chem. Theory Comput.*, 13(3):1057–1066, 3 2017. ISSN 1549-9618. doi: 10.1021/acs.jctc.6b00971.

[47] P. B. Armentrout, Yih-Chung Chang, and Cheuk-Yiu Ng. What is the Bond Dissociation Energy of the Vanadium Hydride Cation? *J. Phys. Chem. A*, 124(26):5306–5313, 7 2020. ISSN 1089-5639. doi: 10.1021/acs.jpca.0c04517.

[48] Henrik R. Larsson, Huanchen Zhai, C. J. Umrigar, and Garnet Kin-Lic Chan. The Chromium Dimer: Closing a Chapter of Quantum Chemistry. *J. Am. Chem. Soc.*, 144(35):15932–15937, 9 2022. ISSN 0002-7863. doi: 10.1021/jacs.2c06357.

[49] J. L. Elkind and P. B. Armentrout. Effect of kinetic and electronic energy on the reactions of Cr+ with H2, HD, and D2. *J. Chem. Phys.*, 86(4):1868–1877, 2 1987. ISSN 0021-9606. doi: 10.1063/1.452138.

[50] Eric A. Rohlfing and James J. Valentini. UV laser excited fluorescence spectroscopy of the jet-cooled copper dimer. *J. Chem. Phys.*, 84(12):6560–6566, 6 1986. ISSN 0021-9606. doi: 10.1063/1.450708.

[51] J. L. Elkind and P. B. Armentrout. Transition-metal hydride bond energies: first and second row. *Inorg. Chem.*, 25(8):1078–1080, 4 1986. ISSN 0020-1669. doi: 10.1021/ic00228a004.

[52] M. Czajkowski, R. Bobkowski, and L. Krause. transitions in Zn2 excited in crossed molecular and laser beams. *Phys. Rev. A*, 41(1):277–282, 1 1990. ISSN 1050-2947. doi: 10.1103/PhysRevA.41.277.

[53] K. P. Huber and G. Herzberg. *Molecular Spectra and Molecular Structure*. Springer US, Boston, MA, 1 edition, 1979. ISBN 978-1-4757-0963-6. doi: 10.1007/978-1-4757-0961-2.

[54] Jeppe Olsen, Björn O. Roos, Poul Jørgensen, and Hans Jörgen Aa. Jensen. Determinant based configuration interaction algorithms for complete and restricted configuration interaction spaces. *J. Chem. Phys.*, 89(4):2185–2192, 8 1988. ISSN 0021-9606. doi: 10.1063/1.455063.

[55] Per Ake Malmqvist, Kristine Pierloot, Abdul Rehaman Moughal Shahi, Christopher J. Cramer, and Laura Gagliardi. The restricted active space followed by second-order perturbation theory method: Theory and application to the study of CuO2 and Cu2O2 systems. *J. Chem. Phys.*, 128(20), 5 2008. ISSN 0021-9606. doi: 10.1063/1.2920188.

[56] Björn O. Roos, Roland Lindh, Per-Ake Malmqvist, Valera Veryazov, and Per-Olof Widmark. New Relativistic ANO Basis Sets for Transition Metal Atoms. *J. Phys. Chem. A*, 109(29): 6575–6579, 7 2005. ISSN 1089-5639. doi: 10.1021/jp0581126.

[57] Evgeny Epifanovsky, Andrew T. B. Gilbert, Xintian Feng, Joonho Lee, Yuezhi Mao, Narbe Mardirossian, Pavel Pokhilko, Alec F. White, Marc P. Coons, Adrian L. Dempwolff, Zhengting Gan, Diptarka Hait, Paul R. Horn, Leif D. Jacobson, Ilya Kaliman, Jörg Kussmann, Adrian W. Lange, Ka Un Lao, Daniel S. Levine, Jie Liu, Simon C. McKenzie, Adrian F. Morrison, Kaushik D. Nanda, Felix Plasser, Dirk R. Rehn, Marta L. Vidal, Zhi-Qiang You, Ying Zhu, Bushra Alam, Benjamin J. Albrecht, Abdulrahman Aldossary, Ethan Alguire, Josefine H. Andersen, Vishikh Athavale, Dennis Barton, Khadiza Begam, Andrew Behn, Nicole Bellonzi, Yves A. Bernard, Eric J. Berquist, Hugh G. A. Burton, Abel Carreras, Kevin Carter-Fenk, Romit Chakraborty, Alan D. Chien, Kristina D. Closser, Vale Cofer-Shabica, Saswata Dasgupta, Marc de Wergifosse, Jia Deng, Michael Diedenhofen, Hainam Do, Sebastian Ehlert, Po-Tung Fang, Shervin Fatehi, Qingguo Feng, Triet Friedhoff, James Gayvert, Qinghui Ge, Gergely Gidofalvi, Matthew Goldey, Joe Gomes, Cristina E. González-Espinoza, Sahil Gulania, Anastasia O. Gunina, Magnus W. D. Hanson-Heine, Phillip H. P. Harbach, Andreas Hauser, Michael F. Herbst, Mario Hernández Vera, Manuel Hodecker, Zachary C. Holden, Shannon Houck, Xunkun Huang, Kerwin Hui, Bang C. Huynh, Maxim Ivanov, Ádám Jász, Hyunjun Ji, Hanjie Jiang, Benjamin Kaduk, Sven Kähler, Kirill Khistyaev, Jaehoon Kim, Gergely Kis, Phil Klunzinger, Zsuzsanna Koczor-Benda, Joong Hoon Koh, Dimitri Kosenkov, Laura Koulias, Tim Kowalczyk, Caroline M. Krauter, Karl Kue, Alexander Kunitsa, Thomas Kus, István Ladjánszki, Arie Landau, Keith V. Lawler, Daniel Lefrancois, Susi Lehtola, Run R. Li, Yi-Pei Li, Jiashu Liang, Marcus Liebenthal, Hung-Hsuan Lin, You-Sheng Lin, Fenglai Liu, Kuan-Yu Liu, Matthias Loipersberger, Arne Luenser, Aaditya Manjanath, Prashant Manohar, Erum Mansoor, Sam F. Manzer, Shan-Ping Mao, Aleksandr V. Marenich, Thomas Markovich, Stephen Mason, Simon A. Maurer, Peter F. McLaughlin, Maximilian F. S. J. Menger, Jan-Michael Mewes, Stefanie A. Mewes, Pierpaolo Morgante, J. Wayne Mullinax, Katherine J. Oosterbaan, Garrette Paran, Alexander C. Paul, Suranjan K. Paul, Fabijan Pavošević, Zheng Pei, Stefan Prager, Emil I. Proynov, Ádám Rák, Eloy Ramos-Cordoba, Bhaskar Rana, Alan E. Rask, Adam Rettig, Ryan M. Richard, Fazle Rob, Elliot Rossomme, Tarek Scheele, Maximilian Scheurer, Matthias Schneider, Nickolai Sergueev, Shaama M. Sharada, Wojciech Skomorowski, David W. Small, Christopher J. Stein, Yu-Chuan Su, Eric J. Sundstrom, Zhen Tao, Jonathan Thirman,

47

Gábor J. Tornai, Takashi Tsuchimochi, Norm M. Tubman, Srimukh Prasad Veccham, Oleg Vydrov, Jan Wenzel, Jon Witte, Atsushi Yamada, Kun Yao, Sina Yeganeh, Shane R. Yost, Alexander Zech, Igor Ying Zhang, Xing Zhang, Yu Zhang, Dmitry Zuev, Alán Aspuru-Guzik, Alexis T. Bell, Nicholas A. Besley, Ksenia B. Bravaya, Bernard R. Brooks, David Casanova, Jeng-Da Chai, Sonia Coriani, Christopher J. Cramer, György Cserey, A. Eugene DePrince, Robert A. DiStasio, Andreas Dreuw, Barry D. Dunietz, Thomas R. Furlani, William A. Goddard, Sharon Hammes-Schiffer, Teresa Head-Gordon, Warren J. Hehre, Chao-Ping Hsu, Thomas-C. Jagau, Yousung Jung, Andreas Klamt, Jing Kong, Daniel S. Lambrecht, WanZhen Liang, Nicholas J. Mayhall, C. William McCurdy, Jeffrey B. Neaton, Christian Ochsenfeld, John A. Parkhill, Roberto Peverati, Vitaly A. Rassolov, Yihan Shao, Lyudmila V. Slipchenko, Tim Stauch, Ryan P. Steele, Joseph E. Subotnik, Alex J. W. Thom, Alexandre Tkatchenko, Donald G. Truhlar, Troy Van Voorhis, Tomasz A. Wesolowski, K. Birgitta Whaley, H. Lee Woodcock, Paul M. Zimmerman, Shirin Faraji, Peter M. W. Gill, Martin Head-Gordon, John M. Herbert, and Anna I. Krylov. Software for the frontiers of quantum chemistry: An overview of developments in the Q-Chem 5 package. *J. Chem. Phys.*, 155(8), August 2021. ISSN 1089-7690. doi: 10.1063/5.0055522.

[58] Pierre-François Loos, Anthony Scemama, Aymeric Blondel, Yann Garniron, Michel Caffarel, and Denis Jacquemin. A mountaineering strategy to excited states: Highly accurate reference energies and benchmarks. *J. Chem. Theory Comput.*, 14(8):4360–4379, 2018. doi: 10.1021/acs.jctc.8b00406.

[59] Petros Souvatzis. Uquantchem: A versatile and easy to use quantum chemistry computational software. *Comput. Phys. Commun.*, 185(1):415–421, 2014. ISSN 0010-4655. doi: https://doi.org/10.1016/j.cpc.2013.09.014.

[60] Duminda S. Ranasinghe, Johannes T. Margraf, Ajith Perera, and Rodney J. Bartlett. Vertical valence ionization potential benchmarks from equation-of-motion coupled cluster theory and QTP functionals. *J. Chem. Phys.*, 150(7):074108, 02 2019. ISSN 0021-9606. doi: 10.1063/1.5084728.

[61] Ernest Opoku, Filip Pawłowski, and J. V. Ortiz. Electron propagator self-energies versus improved GW100 vertical ionization energies. *J. Chem. Theory Comput.*, 18(8):4927–4944, 2022. doi: 10.1021/acs.jctc.2c00502.

[62] Ernest Opoku, Filip Pawłowski, and J. V. Ortiz. Ab initio electron propagators with an hermitian, intermediately normalized superoperator metric applied to vertical electron affinities. *J. Phys. Chem. A*, 128(23):4730–4749, 2024. doi: 10.1021/acs.jpca.4c02050.

[63] E. Ortí and J. L. Brédas. Electronic structure of metal-free phthalocyanine: A valence effective hamiltonian theoretical study. *J. Chem. Phys.*, 89(2):1009–1016, July 1988. ISSN 1089-7690. doi: 10.1063/1.455251.

[64] O. Dolgounitcheva, V. G. Zakrzewski, and J. V. Ortiz. Ab initio electron propagator calculations on the ionization energies of free base porphine, magnesium porphyrin, and zinc porphyrin. *J. Phys. Chem. A*, 109(50):11596–11601, October 2005. ISSN 1520-5215. doi: 10.1021/jp0538060.

[65] Detlef Schröder, Jessica Loos, Helmut Schwarz, Roland Thissen, Dorin V. Preda, Lawrence T. Scott, Doina Caraiman, Maxim V. Frach, and Diethard K. Böhme. Single and double ionization

of corannulene and coronene. *Helv. Chim. Acta*, 84(6):1625–1634, June 2001. ISSN 1522-2675. doi: 10.1002/1522-2675(20010613)84:6<1625::aid-hlca1625>3.0.co;2-0.

[66] Manuel Díaz-Tinoco, O. Dolgounitcheva, V. G. Zakrzewski, and J. V. Ortiz. Composite electron propagator methods for calculating ionization energies. *J. Chem. Phys.*, 144(22):224110, 2016. ISSN 0021-9606. doi: 10.1063/1.4953666.

[67] Dávid Mester, Péter R. Nagy, József Csóka, László Gyevi-Nagy, P. Bernát Szabó, Réka A. Horváth, Klára Petrov, Bence Hégely, Bence Ladóczki, Gyula Samu, Balázs D. Lőrincz, and Mihály Kállay. Overview of developments in the MRCC program system. *J. Phys. Chem. A*, 129(8):2086–2107, 2025. doi: 10.1021/acs.jpca.4c07807.

[68] Ajith Perera, Robert W. Molt, Victor F. Lotrich, and Rodney J. Bartlett. *Singlet–triplet separations of di-radicals treated by the DEA/DIP-EOM-CCSD methods*, page 153–165. Springer Berlin Heidelberg, June 2014. ISBN 9783662481486. doi: $10.1007/978-3-662-48148-6_14$.

[69] Xiangzhu Li and Josef Paldus. Singlet–triplet separation in BN and $C_2$: Simple yet exceptional systems for advanced correlated methods. *Chem. Phys. Lett.*, 431(1–3):179–184, November 2006. ISSN 0009-2614. doi: 10.1016/j.cplett.2006.09.053.

[70] Kirk A. Peterson. Accurate multireference configuration interaction calculations on the lowest $^1\sigma^+$ and $^3\pi$ electronic states of $C_2$, $CN^+$, BN, and $BO^+$. *J. Chem. Phys.*, 102(1):262–277, January 1995. ISSN 1089-7690. doi: 10.1063/1.469399.

[71] Barry J. Moss and William A. Goddard. Configuration interaction studies on low-lying states of $O_2$. *J. Chem. Phys.*, 63(8):3523–3531, October 1975. ISSN 1089-7690. doi: 10.1063/1.431791.

[72] Takayuki Fueno, Vlasta Bonacic-Koutecky, and Jaroslav Koutecky. Ab initio CI study of chemical reactions of singlet and triplet imidogen (NH) radicals. *J. Am. Chem. Soc.*, 105(17):5547–5557, August 1983. ISSN 1520-5126. doi: 10.1021/ja00355a004.

[73] Stavros Kardahakis, Jiří Pittner, Petr Čársky, and Aristides Mavridis. Multireference configuration interaction and coupled-cluster calculations on the $x^3\sigma^-$, $a^1\delta$, and $b^1\sigma^+$ states of the NF molecule. *Int. J. Quantum Chem.*, 104(4):458–467, April 2005. ISSN 1097-461X. doi: 10.1002/qua.20618. URL http://dx.doi.org/10.1002/qua.20618.

[74] Russel D. Johnson III. NIST computational chemistry comparison and benchmark. database NIST standard reference database number 101. release 22. https://cccbdb.nist.gov/, 2022. URL https://cccbdb.nist.gov/.

[75] K. Yamaguchi, F. Jensen, A. Dorigo, and K.N. Houk. A spin correction procedure for unrestricted hartree-fock and møller-plesset wavefunctions for singlet diradicals and polyradicals. *Chem. Phys. Lett.*, 149(5):537–542, 1988. ISSN 0009-2614. doi: https://doi.org/10.1016/0009-2614(88)80378-6. URL https://www.sciencedirect.com/science/article/pii/0009261488803786.

[76] L. Noodleman, C.Y. Peng, D.A. Case, and J.-M. Mouesca. Orbital interactions, electron delocalization and spin coupling in iron-sulfur clusters. *Coord. Chem. Rev*, 144:199–244, 1995. ISSN 0010-8545. doi: https://doi.org/10.1016/0010-8545(95)07011-L.

[77] Toru Saito, Satomichi Nishihara, Yusuke Kataoka, Yasuyuki Nakanishi, Yasutaka Kitagawa, Takashi Kawakami, Shusuke Yamanaka, Mitsutaka Okumura, and Kizashi Yamaguchi. Reinvestigation of the reaction of ethylene and singlet oxygen by the approximate spin projection method. comparison with multireference coupled-cluster calculations. *J. Phys. Chem. A*, 114 (30):7967–7974, 2010. doi: 10.1021/jp102635s.



\end{document}